\documentclass[acmsmall,screen,natbib=false]{acmart}
\usepackage[T1]{fontenc}
\usepackage[utf8]{inputenc}
\usepackage{graphicx,xcolor}
\usepackage{cleveref}
\usepackage[maxbibnames=99,giveninits=true]{biblatex}
\usepackage{stmaryrd}
\usepackage{soul}
\usepackage{mathpartir}
\usepackage{listings}
\usepackage{subcaption}
\usepackage{rrhpf}
\usepackage{tikz}
\usepackage{calrsfs}
\usepackage{paralist}
\usepackage{wrapfig}
\usepackage{needspace}
\newtheorem{definition}{Definition}

\newtheorem*{example*}{Example}
\newtheorem*{theorem*}{Theorem}
\newcommand{\basicCodeStyle}{\ttfamily\small}
\newcommand{\keywordCodeStyle}{\bfseries}

\newcommand\thelanguage{Notac}

\newcommand\code[1]{\ensuremath{\mathtt{#1}}}
\newcommand{\codeparens}[1]{\ensuremath{\code{(}#1\code{)}}}
\newcommand{\codeeq}{\ensuremath{\mathop{\mbox{\texttt{=}}}}}

\newcommand{\deref}[1]{\ensuremath{\code{*}#1}}
\newcommand{\addrof}[1]{\ensuremath{\mbox{\texttt{\&}}#1}}
\newcommand{\cast}[1]{\ensuremath{\code{cast(}#1\code{)}}}
\newcommand{\binop}[3]{\ensuremath{#1\mathop{#2}#3}}

\newcommand{\assg}[2]{\ensuremath{#1\codeeq#2}}
\newcommand{\ptrassg}[2]{\ensuremath{\code{*}\codeparens{#1}\codeeq#2}}
\newcommand{\malloc}[2]{\ensuremath{#1\codeeq\code{malloc(}#2\code{)}}}
\newcommand{\free}[1]{\ensuremath{\code{free(}#1\code{)}}}
\newcommand{\sequence}[2]{\ensuremath{#1 \code{;} #2}}
\newcommand{\cond}[3]{\code{if}\codeparens{#1}\; #2 \;\code{else}\; #3}
\newcommand{\cnd}[2]{\code{if}\codeparens{#1}\; #2}
\newcommand{\while}[2]{\code{while}\codeparens{#1}\; #2}
\newcommand{\obs}[1]{\code{observe}\codeparens{#1}}
\newcommand{\cmdskip}{\code{skip}}

\newcommand{\upd}[3]{\ensuremath{{#1}\lbrack{#2}\mapsto{#3}\rbrack}}
\newcommand\dom[1]{\mathsf{dom}({#1})}
\newcommand\img[1]{\mathsf{img}({#1})}
\newcommand{\domain}[1]{\ensuremath{\mathsf{#1}}}
\newcommand{\MALLOC}[1]{\ensuremath{\mathbf{malloc}(#1)}}
\newcommand{\FREE}[1]{\ensuremath{\mathbf{free}(#1)}}
\newcommand{\NULL}{\ensuremath{\mathbf{null}}}
\newcommand{\INIT}[1]{\ensuremath{\mathbf{init}(#1)}}
\newcommand{\conf}[1]{\ensuremath{\langle #1 \rangle}}
\newcommand{\finconf}[1]{\ensuremath{\langle #1 \rangle}}
\newcommand{\expeval}{\Downarrow}
\newcommand\lveval{\downarrow}
\newcommand{\ejudg}[3][E,H,\allocstrategy]{#1\vdash #2 \expeval #3}
\newcommand{\ljudg}[3][E,H,\allocstrategy]{#1\vdash #2 \lveval #3}
\newcommand{\cjudg}[3][E,H,\allocstrategy]{\conf{#2; #1} \to \finconf{#3}}
\newcommand{\cjudgtr}[4][E,H,\allocstrategy]{\conf{#2; #1} \to_{#3} \finconf{#4}}
\newcommand{\cjudgtrext}[5][E,H,\allocstrategy]{\conf{#2; #1} \to_{#3}
\conf{#4;#5}}
\newcommand{\nullexpr}{\ensuremath{\code{NULL}}}

\newcommand{\tr}{\ensuremath{t}}
\newcommand{\emptytr}{\ensuremath{\epsilon}}

\newcommand{\event}{\ensuremath{\eta}}

\newcommand{\evobs}[1]{\mathsf{obs}(#1)}
\newcommand{\evmalloc}[2]{\mathsf{malloc}(#1,#2)}
\newcommand{\evmallocfail}[1]{\mathsf{mfail}(#1)}

\newcommand{\evfree}[1]{\mathsf{free}(#1)}
\newcommand{\evcast}[1]{\mathsf{cast}(#1)}

\newcommand{\symbmalloc}[1]{\mathrm{m}(#1)}
\newcommand{\symbfail}[1]{\mathrm{n}(#1)}
\newcommand{\symbfree}[1]{\mathrm{f}\langle #1 \rangle}
\newcommand\symbevent{\sigma}
\newcommand\symbseq{\vec\symbevent}
\newcommand\symbseqlookup[2]{#1[#2]}
\newcommand{\symbseqemp}{\epsilon}
\newcommand{\symbcons}{\cdot}
\newcommand\freeindex[2]{\operatorname{idx}(#1 | #2)}
\newcommand\mallocfreerel[1]{\mathrel{\curvearrowleft_{#1}}}
\newcommand\clientupdmeta{u}
\newcommand\clientupd[3][\clientupdmeta]{#1 (#2, #3)}
\newcommand\clientupdseq{\vec\clientupdmeta}
\newcommand\allocsymmap{\Phi}

\newcommand{\isbot}[2]{\ensuremath{#1_{\bot}(#2)}}

\newcommand{\intv}[1]{\ensuremath{{[}#1{]}}}
\newcommand{\intvL}[1]{\ensuremath{{[}#1)}}

\newcommand{\heapeq}[1]{\ensuremath{\mathrel{=}_{#1}}}
\newcommand\lenof[1]{| {#1} |}
\newcommand\addressesof[1]{\llbracket {#1} \rrbracket}

\newcommand\symfilter[2]{ \lfloor #1 \rfloor_{#2}}
\newcommand\stepcorresponds[5]{#1, #2 \vdash #3 / #4 \hookrightarrow #5 }

\newcommand\hexlit[1]{0\mathrm{x}#1}

\newcommand\tracesim[2]{#1 \simeq #2}
\newcommand\tracecons{\cdot}
\newcommand\allocatorimpactmeta{\mathbb{A}}
\newcommand\progclass{\ensuremath{P}}
\newcommand\allocatorprogressimpactmeta{\mathbb{P}}
\newcommand\allocatorimpactproggen{\allocatorprogressimpactmeta}

\newcommand\eventclassalloc[1]{\progclass^{+#1}}
\newcommand{\eventclasscast}{\progclass^{\code{cast}}}

\newcommand\downcharmeta{\ensuremath{\mathsf{pdc}}}
\newcommand\downcharfn[1]{\downcharmeta(#1)}

\newcommand{\allocstrategy}{\ensuremath{\alpha}}

\newcommand{\stratapply}[2]{\ensuremath{#1.#2}}
\newcommand{\allocstate}{A}
\newcommand{\reservedmem}{R}

\newcommand{\stratawaredom}[1]{\ensuremath{\domain{Strat}_{#1}}}

\newcommand{\lval}{\ensuremath{\mathit{lval}}}
\newcommand{\lvalptr}[1]{\ensuremath{\code{*}\codeparens{#1}}}
\newcommand{\lvalcast}[2]{\ensuremath{#1\codeeq\cast{#2}}}
\newcommand{\lvalmalloc}[2]{\ensuremath{#1\codeeq\code{malloc(}#2\code{)}}}

\newcommand{\rulename}[1]{\DefTirName{#1}}
\newcommand{\play}{\ensuremath{\mathrel{\rhd}}}
\newcommand{\ff}{\ensuremath{\mathrel{\rhd\!\rhd}}}
\newcommand{\playrel}[5]{#1, #2 \mid #3 \play #4, #5}
\newcommand{\ffrel}[5]{#1, #2 \mid #3 \ff #4, #5}
\newcommand{\invseqrel}[1]{\ensuremath{#1 \vdash}}
\newcommand{\invmch}[1]{\ensuremath{#1 \vdash}}
\newcommand{\stackaddresses}[1]{\img{#1}}
\newcommand{\WF}[1]{\ensuremath{\mathsf{WF}(#1)}}

\definecolor{memsafeBlue}{RGB}{44, 123, 182}

\newcommand{\otherlang}{Memsafe}
\newcommand{\memsafe}[1]{\ensuremath{{\color{memsafeBlue}#1}}}
\newcommand{\notac}[1]{\ensuremath{{\color{black}#1}}}
\newcommand{\MemsafeToNotac}[1]{\ensuremath{{\color{black}\llparenthesis} \memsafe{#1} {\color{black}\rrparenthesis}}}
\newcommand{\MemsafeToNotacDef}[2]{\ensuremath{\MemsafeToNotac{#1}&\triangleq&\notac{#2}}}
\newcommand{\transvar}[1]{\texttt{\$#1}}
\newcommand{\oom}{\transvar{oom}} 
\newcommand{\alloci}{\transvar{i}}
\newcommand{\wguard}{\transvar{\transvar{g}}}
\newcommand{\memguard}[1]{{\color{black}\mathbf{G}\! \left ( \notac{#1} \color{black} \right )}}

\newcommand{\vars}[1]{\ensuremath{\mathsf{vars}(#1)}}
\newcommand{\addrmap}{\theta}
\newcommand{\statecompatsingle}[1]{\rightharpoonup_{#1}}
\newcommand{\statecompat}[3][\addrmap]{\ensuremath{#2 {\statecompatsingle{#1}} #3}}
\newcommand{\emptrace}{\epsilon}

\newcommand{\msskip}{\mathsf{skip}}
\newcommand{\msnil}{\mathsf{nil}}
\newcommand{\msptr}[3]{\left ( #1, #2, #3 \right )}
\newcommand{\msevalnotation}[1]{\llbracket #1 \rrbracket}
\newcommand{\mseval}[2][s]{\msevalnotation{#2} (#1)}
\newcommand{\msadd}[2]{#1 + #2}
\newcommand{\mssub}[2]{#1 - #2}
\newcommand{\msmul}[2]{#1 \times #2}
\newcommand{\mseq}[2]{#1 = #2}
\newcommand{\msleq}[2]{#1 \leq #2}

\newcommand{\mscase}[1]{\begin{array}{@{}ll} #1 \end{array}}
\newcommand{\mscases}[1]{\left \{ \mscase{#1} \right.}
\newcommand{\msbind}{\mathsf{bind}}
\newcommand{\msif}{\mathsf{if}}
\newcommand{\msthen}{\mathsf{then}}
\newcommand{\mselse}{\mathsf{else}}
\newcommand{\mswhile}{\mathsf{while}}
\newcommand{\msdo}{\mathsf{do}}
\newcommand{\msend}{\mathsf{end}}
\newcommand{\mserror}{\mathsf{error}}
\newcommand{\msevalc}[1]{\msevalnotation{#1}_{+}}
\newcommand{\msevalcs}[2][l,m]{\msevalc{#2} (#1)}
\newcommand{\msseq}[2]{#1 \code{;} #2}
\newcommand{\msalloc}{\mathsf{alloc}}
\DeclareMathAlphabet{\pazocal}{OMS}{zplm}{m}{n} 
\newcommand{\msstate}[1]{\pazocal{#1}}
\newcommand{\msfinmap}{\rightharpoonup_{\mathrm{fin}}}

\newcommand{\eagerF}[1]{\ensuremath{\mathsf{allocAddrs}(#1)}}
\newcommand\alloceager{\allocstrategy_{\mathit{eager}}}
\newcommand\eagernull{\stratapply{\alloceager}{\NULL}}
\newcommand\eagerinit[1]{\stratapply{\alloceager}{\INIT{#1}}}
\newcommand\eagermalloc[1]{\stratapply{\alloceager}{\MALLOC{#1}}}
\newcommand\eagerfree[1]{\stratapply{\alloceager}{\FREE{#1}}}
\newcommand\allocbump{\allocstrategy_{\mathit{bump}}}
\newcommand\bumpnull{\stratapply{\allocbump}{\NULL}}
\newcommand\bumpinit[1]{\stratapply{\allocbump}{\INIT{#1}}}
\newcommand\alloccurio{\allocstrategy_{\mathit{curio}}}

\newcommand\defn{\mathrel{\triangleq}}
\newcommand{\ruleref}[1]{\textsc{#1}}

\newcommand{\trchar}[1]{\ensuremath{\chi(#1)}}

\newif\ifnoappendix\noappendixfalse
\ifnoappendix
\newcommand\extorappendix{extended version of this paper}
\else 
\newcommand\extorappendix{appendix}
\fi

\ifnoappendix
\title{The Downgrading Semantics of Memory Safety}
\else
\title{The Downgrading Semantics of Memory Safety \\ (Extended Version)}
\fi
\begin{abstract}
Memory safety is traditionally characterized in terms of bad things that cannot happen. This approach is currently embraced in the literature on formal methods for memory safety. However, a general semantic principle for memory safety, that implies the negative items, remains elusive.

This paper focuses on the allocator-specific aspects of memory safety, such as null-pointer dereference, use after free, double free, and heap overflow. To that extent, we propose a notion of \emph{gradual allocator independence} that accurately captures the allocator-dependent aspects of memory safety. 
Our approach is inspired by the previously suggested connection between memory safety and noninterference, but extends that connection in a fundamentally important direction towards downgrading.

We consider a low-level language with access to an allocator that provides malloc and free primitives in a flat memory model. Pointers are just integers, and as such it is trivial to write memory-unsafe programs. The basic intuition of gradual allocator independence is that of noninterference, namely that allocators must not influence program execution. This intuition is refined in two important ways that account for the allocators running out-of-memory and for programs to have pointer-to-integer casts. The key insight of the definition is to treat these extensions as forms of downgrading and give them satisfactory technical treatment using the state-of-the-art information flow machinery.

\end{abstract}

\author{Ren\'{e} Rydhof Hansen}
\email{rrh@cs.aau.dk}
\orcid{0000-0002-5688-6432}
\affiliation{
	\institution{Aalborg University}
	\city{Aalborg}
	\country{Denmark}
}
\author{Andreas Stenb\ae{}k Larsen}
\email{astenbaek@cs.au.dk}
\orcid{0009-0001-2232-7720}
\affiliation{
	\institution{Aarhus University}
	\city{Aarhus}
	\country{Denmark}
}
\author{Aslan Askarov}
\email{aslan@cs.au.dk}
\orcid{0000-0002-9035-4034}
\affiliation{
	\institution{Aarhus University}
	\city{Aarhus}
	\country{Denmark}
}

\ccsdesc[500]{Security and privacy~Information flow control}
\ccsdesc[500]{Theory of computation~Operational semantics}
\ccsdesc[300]{Software and its engineering~Allocation / deallocation strategies}
\keywords{memory safety, allocators, declassification, cast}

\ifnoappendix
\acmDOI{10.1145/3808260}
\acmYear{2026}
\acmJournal{PACMPL}
\acmVolume{10}
\acmNumber{PLDI}
\acmArticle{182}
\acmMonth{6}
\acmSubmissionID{pldi26main-p49-p}
\received{2025-11-13}
\received[accepted]{2026-04-03}
\fi

\ifnoappendix
\setcopyright{cc}
\setcctype{by}
\fi

\ifnoappendix\else
\fi
\begin{document}
\maketitle
\ifnoappendix\else
\pagestyle{plain}
\thispagestyle{empty}
\renewcommand\thefootnote{$*$}%
\footnotetext{Extended version of a paper of the same title to appear in \emph{Proc.\ ACM Program.\ Lang.} 10, PLDI, Article 182 (June 2026). \url{https://doi.org/10.1145/3808260}}%
\renewcommand\thefootnote{\arabic{footnote}}%
\fi
\section{Introduction}

Memory safety is traditionally characterized in terms of bad things
that cannot
happen~\cite{hicks:blog2014:whatismemsafe,azevedo2018meaning}. This
characterization is used in the literature on memory safety, typically
for reasoning about optimizations in C
compilers~\cite{Krebbers:jar2016,beck2024twophase,KangHMGZV:pldi2015:formalc},
or basic security properties such as access control~\cite{watson2015cheri}.

The approach to defining memory safety in terms of negative behavior has recognized limitations~\cite{memarian2019:provenance,memarian2016into}.
For example, when programs exhibit complex behavior, such as 
casting pointers to integers and back, the associated reasoning principles become semantically fragile~\cite{beck2024twophase}.
It is unclear how to build upon approaches based on negative lists for 
applications where semantic consistency is necessary, such as end-to-end security.
These limitations align with the 2014 observation by Hicks~\cite{hicks:blog2014:whatismemsafe} that 
a satisfactory semantic definition of memory safety -- the one that would imply the traditionally accepted negative behavior -- remains elusive.

In their 2018 paper,
\citeauthor*{azevedo2018meaning}~\cite{azevedo2018meaning}  
study
memory safety for an abstract low-level language and derive a set of
reasoning principles that a memory safe language must guarantee. One
such principle is a form of noninterference. It stipulates
that ``code cannot affect or be affected by unreachable memory''.
That noninterference is intrinsic to memory safety is evident already
in the 2006 work of Berger et al.~\cite{BergerZ:pldi2006:diehard} on
the DieHard allocator that has a version with replicated execution and
voting; reading that paper in 2026 one sees similarities in the
DieHard's replication and voting to information flow approaches such
as self-composition~\cite{self-composition} and
secure-multi-execution~\cite{SecureMultiExecution}.

This paper continues the search for a principled
definition of memory safety.
We focus on the allocator-related violations, such as null-pointer dereference, use after free, double free, and heap overflow, that constitute a large fragment of memory errors~\cite{BergerZ:pldi2006:diehard,NovarkB:ccs2010:dieharder}. We propose a notion of \emph{gradual allocator independence} that accurately captures the allocator-related aspects of memory safety. 
We do not yet address 
stack and sub-object safety, leaving these issues for
future work.

We study heap memory errors in the setting of a low-level language with access to an allocator that provides malloc and free primitives in a flat memory model. Pointers are just integers, and as such, it is trivial to write
memory-\emph{unsafe} programs. This is a fundamentally different approach from \citeauthor*{azevedo2018meaning}~\cite{azevedo2018meaning}, where pointers are opaque and the language is memory-safe.

The basic intuition of gradual allocator independence is that of
noninterference, which is that allocators must not influence program
execution. This intuition is refined in two important ways. First,
allocators may run out of memory. If the program is judicious about
checking the return value of malloc, this kind of influence is
permitted. Second, pointer to integer casts expose allocator
internals, which the definition permits, for as long as casts are
involved. In other words, the slogan for our definition is that ``an
allocator must not influence program execution other than (i) running
out of memory in a checked manner and (ii) casts of allocated
pointers''. Note that using an integer as a pointer is not only
allowed, it does not even require a cast.

The two refinements above present a form of \emph{downgrading}. In the
literature on secure information flow control, downgrading~\cite{li2005downgrading} -- often
studied in its confidentiality dimension as declassification -- is a
well-known generalization of noninterference.
At a technical level, we utilize the recent advancements in
declassification literature~\cite{bpini} in order to sculpt the
definition.
We argue that it is the downgrading that has been the missing
component for formally establishing a connection between memory safety
and information flow.

The significance of our definition is that it decouples having a defined (in our case concrete)
behavior from having a good behavior, the latter conventionally associated with symbolic memory models where pointers are opaque.
This decoupling allows us to accept programs that
use idioms such as arbitrary pointer comparison that fall under the
realm of undefined behavior in~C~\cite{KR1988:c,iso-c}.
Furthermore, the definition %
clarifies how one should think about casts. A cast is a
semantic no-op, and must not change program semantics, but its use has
apparent implications for memory safety. In information flow,
downgrading is also a no-op, but it is widely accepted to have
importance for the semantic definitions of security. Treating casts as
downgrades therefore unveils a semantically well-grounded technical
device for understanding the meaning of casts.

The contributions of this paper are:
\begin{itemize}
\item It develops a novel framework for understanding allocator
  behavior and introduces a notion of allocator well-formedness
  (\Cref{sec:allocation:strategies}).
\item It introduces a definition of \emph{gradual allocator
    independence} that captures the downgrading semantics of the allocator-related aspects of memory safety
  (\Cref{sec:gai}). We incorporate the allocator model of
  \Cref{sec:allocation:strategies} in a simple low-level imperative
  language (\Cref{sec:notac}). The adequacy of our extensional
  definitions is illustrated through a set of examples traditionally
  associated with memory safety (\Cref{sec:common:examples}).
\item It establishes a transparency result between gradual allocator
  independence and a memory safe language in the style
  \citeauthor*{azevedo2018meaning}~\cite{azevedo2018meaning} with
  a CompCert based block-identifier memory
  model~\cite{compcert-mem-model}. The transparency further reinforces
  the adequacy of the definition, and its relation to standard
  enforcement mechanisms (\Cref{sec:memsafe}).
\end{itemize}

As far as we are aware, gradual allocator independence is the first
semantic definition that faithfully captures allocator-related aspects
of memory safety. 

Finally, while this work is obviously motivated by the problem of
memory safety in C, we refrain from postulating how our results apply
to real C (or unsafe Rust), because of the many complexities of the
standards~\cite{memarian2016into}. We do note however, that many
pointer operations that give rise to undefined behavior in C are
well-defined in our language and easily handled by our approach.
Despite C's prevalence, there is emergence of new low-level languages
that try to shed some of C's baggage: in addition to Rust, there is D
with a non-trivial user base, and Zig has a devoted community. Our
work is not tied to any of those efforts either, but we want to
emphasize that if there is no theoretical foundation for a clean
semantic definition of memory safety, we are depriving future efforts
to improve upon C.

\section{Information Flow Intuition for Memory Safety}%
\label{gai-by-example}
We start by setting up an informal intuition for our 
approach.

\newcommand\codeifc[1]{\mathtt{#1}}

\subsection{Allocators as Secrets}
\label{sec:gai:allocators-as-secrets}
In language-based information flow
control~\cite{sabelfeld:myers:jsac}, program security is defined in an
end-to-end way using a variant of
noninterference~\cite{goguen1982security}. No information about
secrets must be revealed to the attacker by the program. In simple
settings, confidential information is stored in initial variables. A
standard example of insecurity is the program 
\lstinline|output(secret)|
that outputs the secret directly.
%
%
Another standard example is the program 
\lstinline|if (secret) output(1) else output (2)|
that reveals the secret via an implicit flow.
\begin{figure}[b]
\centering
\begin{subfigure}[t]{0.3\textwidth}
\centering
\begin{lstlisting}[numbers=none,xleftmargin=1em,framexleftmargin=1em]
p = malloc (128); 
q = malloc (128);  
if (p > q) {
  print (1)
} else {
  print (2)
}    
\end{lstlisting}
\caption{Pointer comparison \label{fig:implicit:alloc:different:branches}}
\end{subfigure}
\hfill
\begin{subfigure}[t]{0.3\textwidth}
\centering
\begin{lstlisting}[numbers=none,xleftmargin=1em,framexleftmargin=1em]
p = malloc (128); 
q = malloc (128);  
if (p > q) {
  print (1)
} else {
  print (1)
}    
\end{lstlisting}
\caption{Same code in branches\label{fig:implicit:alloc:same:branches}}
\end{subfigure}
\hfill
\begin{subfigure}[t]{0.37\textwidth}
\centering
\begin{lstlisting}[numbers=none,xleftmargin=1em,framexleftmargin=1em]
p = malloc (LARGE);
if (p == NULL) {
  print ("Alloc failed");
} else {
  print ("Alloc OK");
}    
\end{lstlisting}
\caption{NULL comparison\label{fig:alloc:null:cmp}}
\end{subfigure}%
\label{fig:implicit:flows}
\caption{Standard implicit flow, pointer comparison, and NULL comparison}
\Description{A figure with three programs demonstrating standard implicit flow, pointer comparison, and NULL comparison}
\end{figure}

In complex programming models, with interactive I/O and concurrency,
modeling secrets in terms of initial variables is ill-fitting. An
often-used technical device is that of \emph{user
strategies}~\cite{strategies:csfw:2006} that encapsulates all
confidential information of a user. Strategies are invoked by the
underlying semantics at the times of inputs. For example, the first
example would now be rewritten to have an explicit input from a user
(Alice), and output to another user (Bob).
\begin{lstlisting}[numbers=none,xleftmargin=1em,framexleftmargin=1em]
secret = input (Alice) // invokes Alices's strategy
output (Bob, secret)
\end{lstlisting}
In such settings, noninterference means the program must not reveal
anything about the strategies.
    
We use this information flow intuition, but \emph{with allocators in
place of secrets}: a program must not depend on the internal aspects
of the allocator.  For example, the following program that allocates a
buffer of 128 bytes and prints the value of the returned pointer
address is rejected.
\begin{lstlisting}[numbers=none,xleftmargin=1em,framexleftmargin=1em]
p = malloc (128)    
print (p)
\end{lstlisting}
Another rejected example is the program in
\Cref{fig:implicit:alloc:different:branches}.
From the perspective of memory safety, the judgment to reject above
programs is sound. It is not a good idea to rely on the implementation
decisions of the allocator, as allocators may differ in how they
manage memory.
In fact, the pointer comparison in the last program is classified as
undefined behavior by the C standard, because the pointers originate
from two separate mallocs.

What about program in 
\Cref{fig:implicit:alloc:same:branches} that prints 1 in both
branches? Gradual allocator independence accepts such program --
clearly, despite the pointer comparison, that program does not reveal
anything about the allocator. Yet, the presence of the pointer
comparison would still be an undefined behavior in C.
As we say in the introduction, our goals are not to decree what C
standard should say.
Instead, we search for a semantic definition that would faithfully
capture known safe and unsafe behavior.

\subsection{The Case for Downgrading}%
\label{sec:informal:downgrading}
It turns out that the baseline approach of
noninterference-for-allocators is too strong. Consider program in
\Cref{fig:alloc:null:cmp}.
The NULL-check implements graceful handling of when the memory cannot
be allocated. This is good programming practice, but it contradicts
the intuition of the previous section.
Instead of abandoning our intuition outright, we propose to treat the
example the way information flow treats declassification (or
downgrading): a carefully-articulated weakening of the baseline
noninterference.

\paragraph{Declassification in Information Flow}
A standard example of declassification is the program 
revealing an average of two secrets: \lstinline|p = declassify((s1 + s2)/2); output(p)|.
%
%
Operationally,
declassification is a no-op, but its presence in the source program
has security significance.
Other standard examples of declassification are login, where only
partial information about the password is revealed, or sealed auctions.

While declassification policies are application-specific -- what is
considered appropriate information disclosure in one system may be
unacceptable in another -- there are general semantic principles for
declassification~\cite{declassification:dimensions}. For example,
declassification of one piece of information must not excuse other
leaks in the system. In other words, the weakening of noninterference
should not be its wreckening.

\paragraph{Progress-(In)sensitivity and Declassification}
Another important way in which noninterference is often weakened is
termination-insensitivity~\cite{tini} or, more generally,
\emph{progress-insensitivity}~\cite{askarov:sabelfeld:2009,hedin:sabelfeld:2012:perspective}. Consider
the program

\begin{lstlisting}
secret = input (Alice);
while (secret) { }       // potential infinite loop
output (Bob, "ping")    
\end{lstlisting}
Strictly speaking, this program is insecure. When Bob receives the
ping they learn that the infinite loop is not taken. Pragmatically,
there is a case to \emph{permit} such programs, typically justified by
the difficulty to implement a practical enforcement mechanism that is
both precise and sound: in place of the infinite loop there may be a
resource-exhausting computation, a blocking input, or some other way
of preventing the computation from progressing. Semantically such
noninterference definitions are dubbed progress-insensitive.

Traditionally, the weakening of progress-insensitivity and of
declassification have been studied orthogonally. Recently, the two
approaches have been reconciled~\cite{bpini} by recognizing that
progress-insensitivity is a special form of declassification. This
insight allows proper treatment of the following subtle example, for
progress-\emph{sensitive} (i.e., \emph{not} allowing leaks via loops)
security.

\begin{lstlisting}
secret = input (Alice);
secret2 = input (Charlie);
while (secret) { }        // potential infinite loop
x = declassify (secret2)  // declassification
output (Bob, "ping") 
\end{lstlisting}
The ping to Bob is insecure because its reachability depends on the
secret-guarded infinite loop. The declassification of Charlie's
secret2 is unrelated to Alice: it must not create a semantic loophole
that accepts this program. The solution is that reachability of
declassifying statements must not depend on secrets~\cite{bpini}.

\paragraph{Downgrading for Allocators}
Returning to memory safety, what dependencies on allocators should be
acceptable? We have identified two cases for downgrading:
\begin{inparaenum}
\item checking success of allocation. 
\item explicit cast of pointers to integers.    
\end{inparaenum}
Consider a variant of the earlier program with added explicit cast and
the execution outlined in the comments.
\begin{lstlisting}
p = malloc (128)       // allocation succeeds
x = cast (p)           // x = 1000
print (x)              // prints 1000
\end{lstlisting}
We want to accept this program. The cast despite being a no-op, just
as declassification, signifies the important intent to allow
dependence on the allocation implementation.

Our definition uses the methodology of the epistemic approach to
downgrading~\cite{gradual-release}. Given a concrete trace of a
program, we define a semantic gadget of \emph{allocator impact} that
is the set of allocators that --- when executed with the same program
--- produce a similar trace. Generally, the larger the allocator
impact is, the fewer allocator dependencies there are in the program.

For the above example run, there are four allocator impact sets.
\begin{itemize}
    \item $A_0$ --- the initial allocator impact:  includes all possible allocators 
    \item $A_1$ --- allocator impact after the malloc: all allocators able to allocate 128 bytes.
    \item $A_2$ --- allocator impact after the cast: all allocators that result in  $p = 1000$. 
    \item $A_3$ --- allocator impact after the print: all allocators that result in printing 1000. 
\end{itemize}
By construction, the allocator impact is monotonic. If $i-1$ and $i$
are two consecutive events in the trace, it holds that $A_{i}
\subseteq A_{i-1}$: the more observations there are the fewer
allocators are possible.

\label{sec:gai:intuition}
The idea behind gradual allocator independence, in line with the
established methodology in epistemic approaches to information
flow~\cite{banerjee2008expressive,askarov:sabelfeld:2009,ahmadian2022dynamic},
is to specify bounds on the allocator impact $A_i$ at each position
$i$ in the trace. We distinguish ordinary trace points, corresponding
to events like print, and downgrading events, corresponding to malloc
and cast.

At ordinary points $i$ in the trace, the allocator impact $A_i$ must
be also bounded by $A_{i-1}$, that is, $A_{i} \supseteq A_{i-1}$. In
our example, the only ordinary event is the print; and indeed $A_3 =
A_2$, because all allocators resulting in printing 1000 are also the
ones that result in x = 1000.  This supports our baseline intuition
for noninterference.

What should the requirements be for the downgrading points in the
race, like $A_1$ and $A_2$? The first instinct is to not require
anything at all, which means treating them as ``all-bets-are-off''
downgrades. Such treatment would match the declassification approach
of gradual release~\cite{gradual-release}. There is however a subtlety
here.
Consider the following program, and the terminating run
\begin{lstlisting}
p = malloc (128)               // allocation succeeds
while (p == CONSTANT_PTR) { }  // loop not taken
q = malloc (64)                // allocation succeeds  /*@\label{lst:malloc:64}@*/
print ("DONE");                // prints DONE
\end{lstlisting}
The while loop in this example is problematic, as it exposes more
information about the allocator than just success of the first
malloc. There are four allocator sets here.
\begin{itemize}
    \item $A_0$ --- the initial allocator impact: includes all possible allocators 
    \item $A_1$ --- the set of allocators that succeed in allocating 128 bytes
    \item $A_2$ --- the set of allocators that succeed in allocating 64 bytes on Line~\ref{lst:malloc:64}.
    \item $A_3$ --- the set of allocators that result in printing DONE. 
\end{itemize}
If nothing is required of $A_2$ (that is, it is unbounded), the only
requirement there is in this trace is $A_3 \supseteq A_2$. But this
misses the problem of the while-loop, because $A_2$ is already too
small: any allocator that succeeds in allocating the 64 bytes on
Line~\ref{lst:malloc:64} -- when executed with this program -- must
have chosen the address $p \neq \mathtt{CONSTANT\_PTR}$; otherwise the second malloc
would be unreachable. In the information flow literature, this
subtlety is recognized as the principle of
non-occlusion~\cite{declassification:dimensions}.

To tackle this subtlety we use the insight of reconciling
progress-insensitive security as downgrading: \emph{the reachability}
of the second malloc must not be allocator dependent! There must be a
bound at the downgrading events. To define this bound, we introduce
another semantic gadget: \emph{progress impact for malloc}. In the
context of this example, let $P_{1}^{+64}$ be the set of all
allocators that succeed in allocating 128 bytes (that is after the
first event) that also reach an allocation of 64 bytes, regardless of
whether the 64 byte allocation is successful or not. With that, we
require that the reachability of the second malloc does not leak:
$P_{1}^{+64} \supseteq A_{1}$, and in general when event $i$ is a
malloc of $n$ bytes, it must be that $P_{i-1}^{+n} \supseteq A_{i-1}$.
We define a similar gadget $P_{i}^{\mathit{cast}}$ for reachability of
a cast statement. Putting everything together, the shape of our
definition is as follows:

\paragraph{Gradual Allocator Independence, Informally}
Given a program $c$, that produces a trace of $i > 0$ events, the
following must hold that
\begin{inparaenum}[\itshape (1)]
    \item if event $i$ is not downgrading, then $A_i \supseteq A_{i-1}$.
    \item if event $i$ is a malloc of $n$ bytes, then $P_{i-1}^{+n} \supseteq A_{i-1}$
    \item if event $i$ is a cast, then $ P_{i-1}^{\mathit{cast}} \supseteq A_{i-1}$.
\end{inparaenum}
Note that gradual allocator independence corresponds to a progress-sensitive notion of information flow.

\subsection{Technical Perspective}
The definition of gradual allocator independence is about memory
safety of \emph{programs} linked with allocators (that provide malloc,
free). This inherently assumes that allocators do their job. But what
does it mean for an allocator to do its job?
\Cref{sec:allocation:strategies} develops a novel extensional framework for 
allocator well-formedness. This
framework is language-agnostic; it is specified only in terms of
allocator interactions and client updates. \Cref{sec:notac} defines a
low-level client language, and \Cref{sec:gai} pulls all these concepts
together to formally define gradual allocator independence (the
intuition of which we gave in this section).

\section{Memory Model and Allocation Strategies}%
\label{sec:allocation:strategies}

We use a flat memory model, with memory addresses $a\in \domain{Addr}
= \mathbb{N}_0$ as natural numbers, values $v\in \domain{Val} =
\mathbb{Z} $ as (unbounded) integers, and the memory $H\in
\domain{Mem} = \domain{Addr} \rightharpoonup_{\mathrm{Fin}}
\domain{Val}$ as a finite partial function from addresses to values.
The domain of allocation sizes $n\in \domain{Size} =
\mathbb{N}_0$ is also natural numbers.
The unbounded integers simplify the technical 
development without loss of the insights. 
\subsection{Allocation Strategies}
We model allocators as \emph{allocation strategies} (or simply
strategies) that encapsulate the decisions for malloc and free calls.
For technical convenience, we use strategies for the secondary role of
providing the null value and initializing the allocator.
\begin{definition}[Allocation strategy]%
  \label{def:allocation:stratey}
  An \emph{allocation strategy} $\allocstrategy$ is
  a structure exposing an abstract domain of allocator states $\domain{Alloc}$ and the following elements:
  \begin{displaymath}
    \begin{array}{rclrcl}
      \NULL & : & \domain{Addr}
&
      \MALLOC{\cdot}
            & : & \domain{Mem} \times \domain{Alloc} \times \domain{Size} \to
                \domain{Mem} \times \domain{Alloc} \times \domain{Addr}
      \\                      
      \INIT{\cdot}
            & : & \domain{Mem} \to \domain{Mem} \times \domain{Alloc}
&
      \FREE{\cdot}
            & : & \domain{Mem} \times \domain{Alloc} \times \domain{Addr} \to
                  \domain{Mem} \times \domain{Alloc}
      \\
    \end{array}
  \end{displaymath}
\end{definition}
We use dot notation, e.g., \stratapply{$\allocstrategy$}{\MALLOC{H, A, k}}
to refer to the individual components of the allocation strategy. 
The meaning of these components is as follows.  The address
$\stratapply{\allocstrategy}{\NULL}$ is the value used by the strategy
to represent the $\nullexpr$ value at runtime.
The function $\stratapply{\allocstrategy}{\INIT{H}}$ initializes an
allocator for heap $H$, returning a tuple $(H', A')$, containing the
updated heap $H'$, along with the allocator state $A'$.
Function $\stratapply{\allocstrategy}{\MALLOC{H, A, k}}$ takes the
heap $H$, allocator state $A$, and the number of bytes to allocate
$k$.  It returns the triple $(H', A', a)$ corresponding to the updated
heap, updated allocator state, and value $a$ such that if allocation
is successful then $a$ is the address, and otherwise $a$ is the
$\nullexpr$ value.  Function $\stratapply{\allocstrategy}{\FREE{H, A,
a}}$ takes the heap $H$, allocator state $A$, and the address to be
freed, returning the tuple $(H', A')$ of the updated heap and
allocator state.

The allocation strategy interface is sufficient to set up a semantics
of a low-level imperative language, which is what we do
in~\Cref{sec:notac}.  Before that, we define
the well-formedness of the allocation strategies.  There are a number
of properties we expect to hold.
\paragraph{Allocation Properties, Informally}%
\label{sec:alloc:informal}
We reinforce our intuition about allocator behavior by listing  
their expected properties in natural language.
\begin{inparaenum}
\item The allocator must not allocate non-free or inaccessible memory,
  including the ``error address'' \nullexpr.
\item Individual blocks of allocated memory are contiguous of at least
  the requested size.
\item The allocator does not modify (contents of) allocated memory.
\end{inparaenum}
As a consequence of the above, one may expect the following to hold:
\begin{inparaenum}
\item Calls to malloc may return \NULL\ even if memory is available
  (but never as a valid address);
\item The allocator may use either the heap or separate allocator
  state (or both) for bookkeeping;
\item The allocator may modify non-allocated memory;
\item The allocator may or may not mark freed memory as invalid;
\item The allocator may or may not initialize/scramble memory on
  allocation and/or freeing;
\item The allocator may or may not release freed memory
  (i.e., as in the bump allocator).
\end{inparaenum}
These informal properties guide the construction of the formal
machinery of allocator well-formedness.
\subsection{Symbolic Allocation Sequence}
We start by introducing the \emph{symbolic allocation events
and sequences} that abstractly capture the malloc/free
behavior of the program.
Symbolic allocation events are given by the grammar
\[
  \symbevent ::= \symbmalloc{k} \mid \symbfail{k} \mid \symbfree{z}
\]
where $\symbmalloc{k}$ corresponds to a successful allocation of $k
\geq 0$ many bytes using malloc, $\symbfail{k}$ corresponds to a
failed allocation of $k \geq 0$ many bytes using malloc, and
$\symbfree{z}$ corresponds to freeing memory allocated $z \geq 0$ many
successful mallocs earlier.  We denote a sequence of symbolic
allocation sequence using the notation~$\symbseq$.

\begin{figure}
\centering
\begin{subfigure}{0.45\textwidth}
\centering
\begin{lstlisting}[numbers=none,xleftmargin=1em,framexleftmargin=1em]
p1 = malloc(100); // succeeds
p2 = malloc(800); // fails
p3 = malloc(200); // succeeds
free (p3); free (p1);
\end{lstlisting}
\caption{Example program}%
\label{fig:symbolic:example}
\end{subfigure}
\hfill
\begin{subfigure}{0.45\textwidth}
\centering
\includegraphics[scale=0.7]{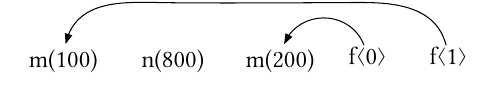}
\caption{Symbolic allocation sequence}%
\label{fig:symbolic:sequence}
\end{subfigure}
\caption{Example program and a corresponding symbolic allocation sequence}%
\label{fig:symbolic}
\Description{On the left, a four-line program performing three mallocs (of sizes 100, 800, and 200) and two frees; the 800-byte malloc is annotated as failing while the other two succeed. On the right, the corresponding symbolic allocation sequence as a horizontal list of events --- successful malloc(100), failed malloc(800), successful malloc(200), free matching the third malloc, free matching the first malloc --- with back-arches linking each free event to its matching successful malloc.}
\end{figure}

\Cref{fig:symbolic:sequence} presents an example of a simple sequence
of malloc and frees (\Cref{fig:symbolic:example}), and the
corresponding symbolic allocation sequence, with the back-arches
illustrating the relationship between the frees and mallocs
(\Cref{fig:symbolic:sequence}).
\paragraph{Well-Formedness of Symbolic Allocation}
A symbolic allocation sequence is well-formed, if there are no
double-frees and there is a malloc for each free, intuitively
capturing proper use of the malloc-free interface.  This definition of
symbolic allocation sequences is not sufficiently strong to enforce
well-formedness structurally, so we introduce auxiliary machinery.

First, we define the operator $\freeindex{\symbseq}{z}$ that returns
the index of the $z$-th malloc, counting backwards, in symbolic
sequence $\symbseq$.
\begin{definition}[Index of freed malloc]
Given a symbolic allocation sequence $\symbseq = \symbevent_1 \ldots
\symbevent_n$ of length $n > 0$, define $\freeindex{\symbseq}{z}$
inductively as 
$n$, if $\symbevent_n = \symbmalloc{k} \land z = 0$;
or 
$\freeindex{\symbevent_1 \ldots \symbevent_{n-1}}{z - 1}$, if $\symbevent_n = \symbmalloc{k} \land z > 0$;
or 
$\freeindex{\symbevent_1 \ldots \symbevent_{n-1}}{z}$, if $\symbevent_n \neq \symbmalloc{k}$.
\end{definition}
The above definition allows allows us to formalize the back-arch
intuition of~\Cref{fig:symbolic:sequence}:
\begin{definition}[Malloc-Free relation $i \mallocfreerel{\symbseq} j$ in allocation sequence]%
  \label{def:malloc-free:relation}
  Given a symbolic allocation sequence $\symbseq = \symbevent_1 \ldots
  \symbevent_n$ of length $n$, say that a free event in position $j$
  corresponds to a malloc event in position $i$, $i < j$, denoted $i
  \mallocfreerel{\symbseq} j$, if $\symbevent_i = \symbmalloc{k}$ for
  some $k$, and $\symbevent_j = \symbfree{z}$ such that
  $\freeindex{\symbevent_1 \ldots \symbevent_{j-1}}{z} = i$.
\end{definition}
The above allows us to define well-formedness of symbolic allocation sequence:
\begin{definition}[Well-formedness of symbolic allocation sequence]%
  \label{def:symseq:wf}
  Given a symbolic allocation sequence $\symbseq$, where events in
  positions $j_1, \ldots, j_p$ are free events, say that this sequence
  is \emph{well-formed}, if the list $ [i_1, \ldots, i_p \mid i_r
  \mallocfreerel{\symbseq} j_r, 1 \leq r \leq p ] $ of the indices of
  the freed malloc events from $\symbseq$ is defined and has no
  duplicates.
\end{definition}
\Cref{def:symseq:wf} requires that symbolic allocation sequences must
be self-contained w.r.t. free operations. For example, $\symbfree{0}$,
in isolation, is not a well-formed sequence, because it does not
contain the allocation that the free corresponds to. Note that
well-formedness permits unmatched mallocs.
\subsection{Allocator Well-Formedness}

This section develops the notion of \emph{allocator well-formedness}
that stipulates allocator behavior w.r.t. memory it has given to the
program.  There are two aspects to allocator well-formedness.  First,
rather straightforward, an allocator must not change the allocated
memory.  Second, the allocator must not behave in a way that exposes
it to corruption by (well-behaving clients), which is possible when
the allocator uses parts of the heap for internal bookkeeping. This
also means that allocator decisions about whether an allocation
succeeds or fails, must not depend on how the client code uses the
allocated memory.  The subtlety of the second aspect is also the cause
of the technical complexity of this section.

\begin{wrapfigure}{r}{0.5\textwidth}
\centering
\includegraphics[width=\linewidth]{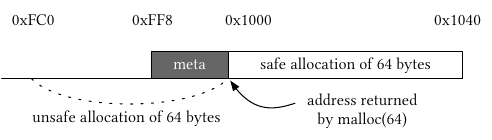}
\caption{Example unsafe allocation}
\label{fig:unsafe:allocation}
\Description{Schematic memory layout illustrating an allocator that stores per-allocation metadata immediately before each allocated region. A first 64-byte allocation occupies addresses 0x1000 to 0x1040, with an 8-byte metadata region at 0xFF8 shown as a shaded band to its left. A second 64-byte allocation proposed at 0xFC0 is shown overlapping the shaded metadata band, visually explaining why that placement is unsafe even though it does not overlap the client-visible first allocation.}
\end{wrapfigure}

We avoid specifically operationalizing aspects of allocator behavior
such as where on the heap and how the allocator stores
meta-information. Consider the example memory layout
in~\Cref{fig:unsafe:allocation}. Suppose we ask to allocate one block
of 64 bytes, and the allocator returns the address 0x1000; it
furthermore uses the memory to the left of the allocated address --
the gray region at address 0xFF8 -- for storing meta information about
the allocation, e.g., the allocated size to use upon freeing. Note
that the client does not know about the meta information, they can
only update the memory in the interval $\intvL{0x1000, 0x1040}$. Now,
consider another allocation of 64 bytes. From the client perspective,
it is possible that the second allocation returns 0xFC0 as it does not
overlap with what is allocated from their perspective. However, for
this allocator returning 0xFC0 is unsafe, because it allows client
updates into the region of memory containing the meta-information.

Another allocator may implement a different approach for meta
information, perhaps by storing it in a separate data structure
elsewhere, and therefore for them returning 0xFC0 is acceptable.  In
other words, well-formedness (that is
~\Cref{def:allocator:heap:well-formedness} below) only mandates that
whatever address the allocator returns the allocator's ability must
not depend on how the client uses the memory.
\paragraph{Client Update} The definition of allocator well-formedness
is language-agnostic.  We abstract over the writes into the allocated
memory via a notion of \emph{client updates}.
\begin{definition}[Client update]
For a heap $H$ and a set of addresses $D\subseteq \domain{Addr}$, a
\emph{client update (of $H$ over $D$)} is a function
$\clientupd{H}{D}$ that modifies the heap $H$ on the addresses
specified in $D$. Thus, if $H' = \clientupd{H}{D}$ then $\forall a \in
\domain{Addr}\colon a\not\in D \implies H'(a) = H (a)$.
\end{definition}
\paragraph{Reserved Memory}
Allocator well-formedness is parametrized to include reserved memory
$\reservedmem$ considered always allocated. Later  in \Cref{sec:notac}, the reserved memory
corresponds to the region that stores local variables (the stack
region in practice).
In the definition below the reserved memory often appears in a union
with the addresses of an allocation map $\addressesof{\allocsymmap}$
that tracks currently allocated memory, also formally defined
below. The intuition is that the union $\addressesof{\allocsymmap}
\cup \reservedmem$ captures all client-accessible memory associated
with the allocation map $\allocsymmap$ and reserved memory
$\reservedmem$.

\paragraph{Sequence Feasibility}
The key ingredient for well-formedness of an allocator is the notion
of \emph{sequence feasibility} that we formulate below. The intuition
for sequence feasibility is that it expresses whether a concrete
allocator can realize the malloc/free events of a symbolic allocation
sequence. The exact addresses the allocator uses for the malloc calls
are not important for feasibility. For example, given a symbolic
allocation sequence $\symbmalloc{64}\cdot \symbfree{0}\cdot
\symbmalloc{64}$ of two mallocs with a free in between, one allocation
strategy may return addresses {0x1000} for both mallocs, while another
may return addresses {0x1000} and {0x1040}.

\begin{definition}[Sequence Feasibility]%
  \label{def:sequence:feasibility}
  An allocation sequence $\symbseq$ is feasible for a given allocation
  strategy $\allocstrategy$, reserved memory $\reservedmem$, heap $H$,
  allocator state $A$, and update sequence $\clientupdseq$ if, for some
  $H', \allocstate', \allocsymmap'$, the sequence feasibility relation
  \(
    \invseqrel{\allocstrategy, \reservedmem}\ffrel{\emptyset}%
    {H,A}{\clientupdseq, \symbseq}{H', A'}{\allocsymmap'},
  \)
  defined in~\Cref{fig:sequence:feasibility:stepwise} holds.
\end{definition}
\begin{figure}
  \centering
  {%
\begin{mathpar}
  \inferrule[\play-malloc-ok]{%
    (H', A', a) = \stratapply{\allocstrategy}{\MALLOC{H, A, k}} \\
    a \neq \stratapply{\allocstrategy}{\NULL} \\
    i = \lenof{\symbseq} + 1 \\
    \allocsymmap' = \allocsymmap \cup \{ (a, k; i) \}
  }{%
    \invseqrel{\allocstrategy}
    \playrel{\allocsymmap}{H,A}{\symbseq\cdot \symbmalloc{k}}{H',A'}{\allocsymmap'}
  }
  \and
  \inferrule[\play-malloc-fail]{%
    (H', A', a) = \stratapply{\allocstrategy}{\MALLOC{H, A, k}} \\
    a = \stratapply{\allocstrategy}{\NULL}
  }{%
    \invseqrel{\allocstrategy}
    \playrel{\allocsymmap}{H,A}{\symbseq\cdot \symbfail{k}}{H',A'}{\allocsymmap}
  }
  \and
  \inferrule[\play-free]{%
    j = \lenof{\symbseq} + 1 \and 
    i \mallocfreerel{\symbseq \cdot \symbfree{z}} j \and
    \symbseqlookup{\symbseq}{i} = \symbmalloc{k} \and
    (H', A') = \stratapply{\allocstrategy}{\FREE{H,A,a}} \and 
    \allocsymmap' = \allocsymmap \setminus \{ (a,k;i) \}
  }{%
    \invseqrel{\allocstrategy}
    \playrel{\allocsymmap}{H,A}{\symbseq\cdot \symbfree{z}}{H',A'}{\allocsymmap'}
  }
  \\
  \inferrule[$\ff$-empty]{~}{%
    \invseqrel{\allocstrategy,R}
    \ffrel{\allocsymmap}{H,A}{\clientupdseq, \emptytr}{H,A}{\allocsymmap}
  }
  \and
  \inferrule[$\ff$-step]{%
    \invseqrel{\allocstrategy,R}
    \ffrel{\allocsymmap}{H,A}{\clientupdseq, \symbseq}{H',A'}{\allocsymmap'}
    \\%
    H'_{\clientupdmeta} = \clientupd{H'}{\addressesof{\allocsymmap'} \cup R}%
    \\%
    \invseqrel{\allocstrategy}
    \playrel{\allocsymmap'}{H'_{\clientupdmeta},A'}{\symbseq\cdot \symbevent}{H'',A''}{\allocsymmap''}
  }{%
    \invseqrel{\allocstrategy,R}
    \ffrel{\allocsymmap}{H,A}%
    {\clientupdseq \cdot \clientupdmeta, \symbseq\cdot \symbevent}%
    {H'',A''}{\allocsymmap''}
  }
\end{mathpar}}
 \caption{Sequence feasibility (fast-forward $\ff$ notation) and step feasibility (play $\play$ notation)}%
    \label{fig:sequence:feasibility:stepwise}
\Description{Five inference rules shown in a framed mathpar environment. The first three rules define the single-step play relation: successful malloc (updates heap and allocator state and records the allocation in the symbolic map), failed malloc (returns the allocator's null address and leaves the map unchanged), and free (locates the matching malloc via the malloc-free relation and removes that entry from the symbolic map). The last two rules define the multi-step fast-forward relation: an empty-sequence base case, and an inductive step that extends a fast-forward derivation by applying a client update to the live addresses and then one more play step.}
\end{figure}

The sequence feasibility relation that supports
\Cref{def:sequence:feasibility} has the form
$\invseqrel{\allocstrategy, \reservedmem}\ffrel{\allocsymmap}{H,
A}{\clientupdseq, \symbseq}{H', A'}{\allocsymmap'}$.  Here, the
components $\allocsymmap, \allocsymmap'$ are \emph{allocation
maps}. They are sets of triples of the form $(a,k; i)$, where $a$ is
the address chosen by the allocator for the malloc of size $k$ at
position $i$ in the sequence.
The intuition of sequence feasibility is that if a symbolic allocation
sequence is feasible for an allocator $\allocstrategy$ it means it can
be realized by $\allocstrategy$.
The idea is to iterate over the symbolic allocation sequence,
formalized by the rules (\rulename{$\ff$-empty}) and
(\rulename{$\ff$-step}), and ``simulate'' each symbolic memory
operation using the actual allocator, potentially interleaved by
client updates to allocated memory.  The individual steps, performed
in rule~(\rulename{$\ff$-step}), are formalized in the
\emph{step feasibility relation}, written
$\invseqrel{\allocstrategy}\playrel{\allocsymmap}{H, A}%
{\symbseq}{H',A'}{\allocsymmap'}$ that is also shown in
Figure~\ref{fig:sequence:feasibility:stepwise}. For client update
$\clientupd{H_1}{\addressesof{\allocsymmap_1}}$, the notation
$\addressesof{\allocsymmap_1}$ denotes the allocated addresses of
$\allocsymmap_1$, defined as:
\begin{math}
  \addressesof{\allocsymmap} \defn \bigcup_{ (a,k; i)
    \in \allocsymmap}\intvL{a,a+k}
\end{math}
where $[a, a + k)$ is the interval over the natural numbers. 

Rule (\textsc{$\play$-malloc-ok}) corresponds to an event
$\symbmalloc{k}$ of successful allocation of $k$ bytes.  The first
premise refers to the result of a malloc call with heap $H$ and
allocator state $\allocstate$, where $H'$ is the updated heap,
$\allocstate'$ is the updated allocator state, and $a$ the heap
address of the allocated memory.  The second premise ensures the
allocation is successful, by requiring the address $a$ of the
allocation is not $\NULL$.  The last two premises of
(\textsc{$\play$-malloc-ok}) relate to updating the set of allocations
$\allocsymmap$, by selecting an index $i$ for the allocation,
corresponding to the number of the symbolic event, and adding the
allocation information $(a,k;i)$ to the set of active allocations.

Rule (\textsc{$\play$-malloc-fail}) corresponds to an event
$\symbfail{k}$ of a failed allocation of $k$ bytes.  Like in the rule
for a successful allocation, $H', A'$, and $a$, refer to the updated
heap, allocator state, and allocated address respectively.  The key
difference is that $a$ is the $\NULL$ address of the allocator
strategy, indicating an allocation failure, therefore the set of
active allocations $\allocsymmap$ remains unchanged.

Rule (\textsc{$\play$-free}) corresponds to the symbolic free event
$\symbfree{z}$.  The relation $i \mallocfreerel{\symbseq \cdot
\symbfree{z}} j$ (\Cref{def:malloc-free:relation}) in the premises
links the free event to a previous successful malloc $\symbmalloc{k}$
at index $i$ in the symbolic event sequence $\symbseq$.  The fourth
premise ensures the allocator can free address $a$ in heap state $H$
and allocator state $\allocstate$, resulting in the updated heap state
$H'$ and updated allocator state $\allocstate'$.  Lastly, the rule
ensures the active allocation set $\allocsymmap'$ removes the region from the allocation map.
\paragraph{Heap Equivalence}
One last auxiliary piece is a notion of two heaps $H_1$ and $H_2$
agreeing on a set of addresses $S \subseteq \domain{Addr}$, which we
write $H_1 \heapeq{S} H_2$, defined as
\begin{math}
  H_1 \heapeq{S} H_2 \defn \forall a \in S .~ H_1(a) = H_2(a)
\end{math}.
\paragraph{Allocator Well-Formedness}
We are finally ready to define allocator well-formedness. 
The definition divides the well-formedness conditions into three
groups, of which the basic and the zero-allocation groups are
single-execution properties. The third group is about relational
properties that qualify allocator behavior w.r.t. alternative client
updates. The basic group is self-explanatory and includes many
expected properties (cf. \Cref{sec:alloc:informal}).
For zero-allocation, our definition is inspired by the C standard that
allows $\code{malloc(0)}$, where the typical use-case is
implementation of variable-size data-structures when the runtime
object size computes to zero. The allocated object must be
distinguished from NULL (the latter is about failed allocation) and
prior allocations. Here, our design choice aligns with the semantics
of~\cite{azevedo2018meaning}, which is relevant for
\Cref{sec:memsafe}, and is consistent with common implementations, e.g.,
returning a minimal ``chunk'' like
\texttt{dlmalloc}~\cite{lea:1996:dlmalloc}. Other approaches are possible~\cite{torvalds2007cpuset} but are orthogonal to our goals. 
The relational properties capture the subtlety that allocator
decisions must not depend on the client writes, as motivated earlier.

\begin{definition}[Allocator Well-Formedness]%
  \label{def:allocator:heap:well-formedness}
  An allocator strategy $\allocstrategy$ is
  \emph{well-formed for reserved memory $\reservedmem$}
  if given: 
    heap $H$ where $\reservedmem \subseteq \dom{H}$,
    initialized heap and allocator state $(H_0, A_0) =
    \stratapply{\allocstrategy}{\INIT{H}}$,
    client update sequence $\clientupdseq_1$,
   heap $H_1$, allocation state $\allocstate_1$, and allocation
    map $\allocsymmap_1$ that are related by the feasibility as 
    $
      \invseqrel{\allocstrategy, \reservedmem}
      \ffrel%
      {\emptyset}%
      {H_0,\allocstate_0}%
      {\clientupdseq_1,\symbseq}%
      {H_1, \allocstate_1}%
      {\allocsymmap_1}%
    $,
  then we require:
  \begin{itemize}
  \item the following basic allocator properties hold
    \begin{description}
    \item[Basic-1]%
      \label{wf:disjoint:alloc}
      allocated regions are disjoint:
      $\forall (a,k;i), (a',k';i') \in \Phi_1\colon i \neq i' \implies
      \intvL{a,a+k} \cap \intvL{a',a'+k'} = \emptyset$
    \item[Basic-2]%
      \label{wf:client:memory:final:heaps}
      client-accessible memory is in the final heaps:
      $\addressesof{\allocsymmap_1} \cup \reservedmem \subseteq \dom{H_1}$ 
    \item[Basic-3]%
      \label{wf:init:no:modify:client}
      allocator initialization does not update reserved memory: 
      $H  \heapeq{\reservedmem} H_0$
    \item[Basic-4]%
      \label{wf:alloc:no:modify:client}
      allocator does not modify client-accessible memory:
      if $\symbseq = \symbseq_{\text{pre}} \cdot \symbevent$
      then for any $H', \allocstate', \allocsymmap'$,
      such that
      \(
        \invseqrel{\allocstrategy,R}%
        \ffrel%
        {\emptyset}%
        {H_0,A_0}%
        {\clientupdseq, \symbseq_{\mathit{pre}}}%
        {H',A'}%
        {\allocsymmap'}
      \)
      and
      \[
        \invseqrel{\allocstrategy}%
        \playrel%
        {\allocsymmap'}%
        {H', \allocstate'}%
        {\symbseq_{\text{pre}} \cdot \symbevent}%
        {H_1, \allocstate_1}%
        {\allocsymmap_1}%
      \]
      then if $\symbevent = \symbfree{z}$
      it holds that
      $H' \heapeq{\addressesof{\allocsymmap_1}\cup \reservedmem} H_1$;
      otherwise
      $H' \heapeq{\addressesof{\allocsymmap'}\cup \reservedmem} H_1$
    \item[Basic-5]%
      \label{wf:alloc:no:overlap:reserved}
      allocated addresses do not overlap with reserved memory:
      $\addressesof{\allocsymmap_1} \cap \reservedmem = \emptyset$ 
    \item[Basic-6]%
      \label{wf:null:not:accessible}
      null is not client-accessible:
      $\stratapply{\allocstrategy}{\NULL} \not \in
      \addressesof{\allocsymmap_1}\cup \reservedmem$
    \end{description}
  \item the following zero-allocation--related properties hold
    \begin{description}
    \item[Zero-Alloc-1]%
      \label{wf:zero:no:overlap}
      allocated addresses are not reused: 
      $\forall (a,k;i), (a',k';i') \in \Phi_1\colon i \neq i' \implies a \neq a'$
    \item[Zero-Alloc-2]%
      \label{wf:zero:no:space}
      zero-sized allocations are disjoint from the client-updateable
      memory: $\forall (a,k;i) \in \Phi_1\colon k = 0 \implies a
      \notin\addressesof{\Phi_1} \cup \reservedmem$
    \end{description}
  \item for all $\clientupdseq_2$ where $\lenof{\clientupdseq_1} =
    \lenof{\clientupdseq_2}$, there exist $H_2, \allocstate_2,
    \allocsymmap_2$ that satisfy the following:
    \begin{description}
    \item[Rel-1]%
      \label{wf:update:influence}
      client updates to the heap do not influence allocator decision:
      \[
        \invseqrel{\allocstrategy, \reservedmem}
        \ffrel%
        {\emptyset}%
        {H_0, \allocstate_0}%
        {\clientupdseq_2, \symbseq}%
        {H_2, \allocstate_2}%
        {\allocsymmap_2}%
      \]
    \item[Rel-2]%
      \label{wf:alloc:map:equiv}
      final allocation maps match: $\forall (a,k;i)\in \Phi_1\colon
      \exists (a',k';i')\in \Phi_2\colon k = k' \land i = i'$
    \end{description}
  \end{itemize}
\end{definition}
We let $\WF{\reservedmem}$ denote the set of all well-formed allocator
strategies for reserved memory $\reservedmem$.

Note that well-formedness has no requirements on monotonicity of
allocation, that is if an allocator fails to allocate $k$ bytes it
should also fail to allocate $n > k$ bytes. We have seen no need for
such a restriction and therefore do not impose it. Furthermore, should
such a requirement be added, it would have to properly qualify the
case of zero-sized allocations: for example, an allocator that fails
on zero-sized allocations (by returning null) may succeed on non-zero
allocations.

\subsubsection{Basic Allocators}%
The \extorappendix{} includes formalization of simple
allocation strategies such as eager allocator, that allocates and frees memory in 
the straightforward way, a bump allocator, and a null allocator that always fails. We show their well-formedness.

\subsubsection{The Curious Allocator}
\label{ex:curious}\label{ex:curious-formal}
We also show formalization and well-formedness for the following allocation strategy, denoted \emph{curious allocator} that
illustrates the flexibility and end-to-end nature
of~\Cref{def:allocator:heap:well-formedness}. Informally,
the allocator works as follows:
\begin{inparaenum}[\itshape (1)]
\item Upon the first call to malloc, it allocates the first
  available address, and remembers it as $p$.
\item It partitions the rest of the heap in two equally-sized semispaces. 
\item Upon the second call to malloc it commits to using one of the
  semispaces for the rest of the program. If $p[0] > 0$ it picks the
  first semi-space. Otherwise, it picks the second one. The second and
  subsequent calls to malloc are served from the chosen semispace; the
  other semispace is never used.
\end{inparaenum}

The curious allocator is well-formed despite the fact that the address
it returns directly depends on the client memory. However, for as long
as the semi-spaces are of equal size, the ability of the allocator to
do its job is not affected by the client memory. 
\section{\thelanguage: a Memory Unsafe Language}%
\label{sec:notac}

This section presents \thelanguage\ --- a low-level imperative
language closely inspired by C, including C-style pointer operations,
but with fewer restrictions on the allowed
operations. \Cref{notac:syntax} presents the syntax of the language,
where $x \in \domain{Vars}$ denotes variables.

One notable feature of \thelanguage\ is the inclusion of an
\texttt{observe} instruction. The instruction is a placeholder for
non-memory related side effects, e.g., system calls; it makes the
result of evaluating an expression into an observable event (see event
traces below). This offers a convenient way to explicitly specify
parts of the program state that should not depend on the allocator.
\subsection{\thelanguage\ Semantics}
The semantics of \thelanguage\ includes the allocator strategy
function, $\allocstrategy$, as part of its operational configuration.
\Cref{notac:semantics:excerpt} presents a few selected rules for the
semantics of the language. Semantics of expressions has the big-step
form $\ejudg{e}{v}$, where $E$ is the variable environment, and $H$ is
the heap. Similarly, the semantics of lvals uses the judgement
$ \ljudg{\lval}{a}$. Environments keep track of (the addresses of)
local variables:
\begin{math}
  E\in \domain{Env}= \domain{Vars} \to \domain{Addr}
\end{math}.

The semantics is mostly standard; note the rule \ruleref{Exp-Null}
that invokes the allocator strategy for obtaining the NULL value. The
semantics for commands has the small-step form
$ \invmch{E,\allocstrategy} \cjudgtr[H,\allocstate]{c}{\event}{H',
  \allocstate'}$, where $\event$ is the event we explain below. We
write $\invmch{E, \allocstrategy}\conf{c;H,\allocstate} \to_{\tr}^{*}$
to indicate that executing a number of steps of the program $c$
produces the trace $\tr$; in particular the trace may be just a single
event $\tr = \event$. The local variable environment does not change
throughout the program execution (the set of declared variables is
fixed). The allocator state is threaded explicitly in the
configurations. Note that side effects
(\ruleref{C-Malloc},\ruleref{C-Cast}) do not expand the memory domain.
That happens only via the allocator.

\begin{figure}
  {\parbox{\linewidth}{\centering
$ 
  \begin{array}{rcl}
    e & ::= & n
              \mid x
              \mid \binop{e}{\mathit{bop}}{e}
              \mid \deref{e}
              \mid \addrof{x}
              \mid \nullexpr
    \\
    \lval & ::= & x \mid \lvalptr{e} 
    \\ 
    c & ::= & \assg{\lval}{e}
              \mid \lvalcast{\lval}{e} 
              \mid \lvalmalloc{\lval}{e}
              \mid \free{e}
              \\ &  & 
              \mid \cmdskip
              \mid \sequence{c}{c}
              \mid \cond{e}{c}{c}
              \mid \while{e}{c}
              \mid \obs{e}
  \end{array}
$}}
\caption{\thelanguage\ syntax\label{notac:syntax}}
\Description{A boxed BNF grammar for the \thelanguage\ language with three non-terminals. Expressions are integer constants, variables, binary operations, pointer dereference, address-of, or the null expression. Lvalues are either a variable or a pointer dereference. Commands are assignment, pointer-to-integer cast, malloc, free, skip, sequential composition, conditional, while loop, or an observe instruction.}
\end{figure}

\begin{figure}
  {
\begin{mathpar}
  \inferrule[Exp-Null]{~}{\ejudg{\nullexpr}{\stratapply{\allocstrategy}{\NULL}}}
  \and
  \inferrule[C-Cast]{%
    \ljudg{\lval}{a} \\
    \ejudg{e}{v} \\
    a \in \dom{H}
  }{
    \invmch{E,\allocstrategy}
    \cjudgtr[H,\allocstate]{\lvalcast{\lval}{e}}{\evcast{v}}{\upd{H}{a}{v},\allocstate}
  }
  \and
  \inferrule[C-Malloc]{%
    \ejudg{e}{n} \\
    n \in \domain{Size} \\
    (H',\allocstate',a) = \stratapply{\allocstrategy}{\MALLOC{H,\allocstate,n}} \\
    \ljudg{\lval}{a_{\mathit{lval}}} \\
    a_{\mathit{lval}} \in \dom{H'} \\
    \event = {\left\lbrace
        \begin{array}{@{}ll@{}}
          \evmalloc{n}{a} & a \neq \stratapply{\allocstrategy}{\NULL} \\
          \evmallocfail{n} & a = \stratapply{\allocstrategy}{\NULL}
        \end{array}
      \right.
    }
  }{
    \invmch{E, \allocstrategy}
    \cjudgtr[H,\allocstate]{\lvalmalloc{\lval}{e}}
    {\event}{\upd{H'}{a_{\mathit{lval}}}{a},\allocstate'}
  }
\end{mathpar}}
\caption{Example \thelanguage\ semantic rules}\label{notac:semantics:excerpt}
\Description{Three inference rules shown in a framed mathpar environment. Rule Exp-Null evaluates the null expression to whatever address the allocator strategy designates as null. Rule C-Cast defines a pointer-to-integer cast as a heap update of the lvalue that also emits a cast event recording the cast value. Rule C-Malloc invokes the allocator strategy, emits either a successful-malloc event carrying the returned address and size or a malloc-failure event with just the requested size depending on whether the returned address equals the allocator's null, and stores the returned address into the target lvalue.}
\end{figure}

\subsection{Event Traces}%
\label{sec:event:traces}
In addition to specifying how programs are evaluated, the semantics
also defines \emph{event traces} that capture all the ``memory
relevant'' actions of a program execution. In a sense, traces
represent what a program can observe and learn about an underlying
allocator and thus also what should be allocator-independent. The
possible events making up a trace cover all actions involving memory,
i.e., (successful and non-successful) memory allocation and
de-allocation, casts, and anything explicitly noted as relevant by the
\texttt{observe} instruction:
\begin{displaymath}
  \begin{array}{rcl}
    \event & ::= & \evobs{v}
                   \mid \evmalloc{v}{a}
                   \mid \evmallocfail{v}
                   \mid \evfree{a}
                   \mid \evcast{a}
  \end{array}
\end{displaymath}
An event trace $\tr$ is then simply a sequence of events: $\tr =
\event_0 \cdot \event_1 \cdots \event_n$ writing `$\cdot$' for
concatenation of events, e.g., for $\tr$ as above, we have $\tr \cdot
\event_{n+1} = \event_0 \cdots \event_n \cdot \event_{n+1}$.   
\subsection{Memory and Heap Compatibility}
To ensure that the program evaluation takes place in well-formed
configurations, we define the notion of heap and variable environment
compatibility. For local variables this means that memory is
accessible, that is $\img{E} \subseteq H$. Given a variable
environment $E : \domain{Var} \to \domain{Addr}$ and heap $H$, say
that $E$ and $H$ are \emph{compatible}, if $\img{E} \subseteq H$. To
express that strategies are well-formed w.r.t. $\img{E}$, we define
$\stratawaredom{E}$ as the set of allocator strategies that are
well-formed with respect to the memory reserved by $E$:
$\stratawaredom{E} \triangleq \WF{\stackaddresses{E}}$.

\subsection{A Note on Casts}%
\label{sec:notac:note:on:casts}
Note that by the rule \ruleref{C-Cast} the casts in \thelanguage\ are
no-ops, because of the underlying memory model. Casting pointers to
integers is sometimes necessary if the program's observable behavior
can depend on the allocator, which is the the focus of the next
section. Casting integers back to pointers is generally unnecessary.
For example, the following is a valid \thelanguage\ program.
\begin{lstlisting}[numbers=none,xleftmargin=1em,framexleftmargin=1em]
p = malloc (1024);
if (p != NULL && p == 10000000000) {
    *(10000000001) = 42      // no int to pointer cast required
}
\end{lstlisting}

\paragraph{XOR Linked List}
The \extorappendix{} includes a case study of how different fragments of
XOR-doubly linked list can be implemented in \thelanguage. A notable
feature of this study is that no casts are required in the
implementation.
\section{Gradual Allocator Independence}%
\label{sec:gai}
This section formalizes \emph{gradual allocator independence}.  We
start off by introducing the auxiliary notions of symbolic filtering,
trace similarity, and allocator impact.
\subsection{Symbolic Filter and Concrete Residue}
First, we define \emph{symbolic filters} that given a trace, extract
the events that do not correspond to ``well-behaved'' memory
operations. This includes casts, observes, and nonsensical frees. 
Symbolic filtering is the primary vehicle in the definition of trace
similarity below, where a common symbolic allocation sequence
acts as a kind of connecting tissue between traces. The inference
rules for symbolic filtering can be found in the \extorappendix{}.

\begin{definition}[Characteristic filter and residue]%
\label{def:trace:correspondence}%
\label{def:trace:filtering}
Given a trace $\tr$, let the \emph{characteristic filter},
denoted $\trchar{\tr}$, be the unique symbolic allocation sequence
such that there exists a residue $\tr_r$ satisfying the
symbolic filter relation
$\stepcorresponds{\emptyset}{\tr}{\emptytr}{\trchar{\tr}}{\tr_r}$.
Refer to $\tr_r$ as the \emph{characteristic residue},
and let $\symfilter{\tr}{\trchar{\tr}}$ be an operation that
returns $\tr_r$.
\end{definition}

The characteristic filter $\trchar{\tr}$ for a trace $\tr$ always exists and is unique (the proof is in the \extorappendix{}).
The auxiliary relation defining characteristic filters has the form
$\stepcorresponds{\allocsymmap}{\tr }{\symbseq_{\mathit{pre}} }{
\symbseq}{\tr_r}$ where $\allocsymmap$ is an allocation map, similarly
to how it appears in \Cref{sec:allocation:strategies}, $\tr$ is the
trace, and ${\symbseq_{\mathit{pre}}} \cdot {\symbseq}$ is a symbolic
allocation sequence. The slash is a cursor splitting the events into
the matched and yet-to-be-matched.  We explain the operating
principles of the symbolic filter with a few small examples.

Consider program \lstinline|p = malloc(8); free(p)|, where all memory
operations are well-behaved. Suppose this yields a concrete trace
$\evmalloc{8}{\hexlit{1000}} \cdot \evfree{\hexlit{1000}}$.  There are
no casts or observes. Consider symbolic allocation sequence
$\symbmalloc{8} \cdot \symbfree{0}$. All of the memory operations pass
through the filter, leaving no residue:
$\symfilter{\evmalloc{8}{\hexlit{1000}} \cdot
\evfree{\hexlit{1000}}}{\symbmalloc{8} \cdot \symbfree{0}} =
\emptytr$.

Consider another program \lstinline|p = malloc(8); free(p + 1)| and a
concrete trace $\evmalloc{8}{\hexlit{1000}} \cdot
\evfree{\hexlit{1001}}$. Here the free event is not filterable,
it is the part of the residue:
$\symfilter{ \evmalloc{8}{\hexlit{1000}} \cdot \evfree{\hexlit{1001}}
}{\symbmalloc{8}} = \evfree{\hexlit{1001}}$.
\subsection{Trace Similarity}
Second, we define what it means for two traces to be
\emph{similar}. Intuitively, this is when observable actions of the
two traces are the same.
\begin{definition}[Trace similarity]%
\label{def:trace:similarity}
Say that two traces $\tr_1$ and $\tr_2$ are \emph{similar}, written
$\tracesim{\tr_1}{\tr_2}$, if $\trchar{\tr_1} = \trchar{\tr_2}$
and $\symfilter{\tr_1}{\trchar{\tr_1}} = \symfilter{\tr_2}{\trchar{\tr_2}}$.
\end{definition}

\begin{figure}
    \centering
\begin{subfigure}{0.3\textwidth}
        \centering
\begin{lstlisting}[numbers=none,xleftmargin=1em,framexleftmargin=1em]
p = malloc(8); 
free (p);
observe (1);
observe (p);
\end{lstlisting}
\caption{Example program}%
\label{fig:trace:similarity:program}
\end{subfigure}
    \hfill
\begin{subfigure}{0.65\textwidth}
        \centering
    \includegraphics[scale=0.7]{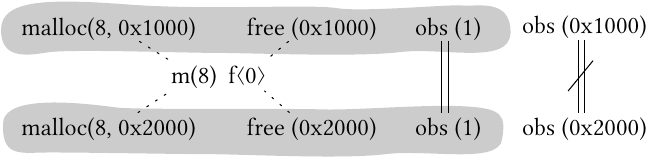}
    \caption{The highlighted traces are similar}%
    \label{fig:trace:similarity:diagram}
\end{subfigure}
    \caption{Trace similarity example}%
    \label{fig:trace:similarity:example}
\Description{On the left, a four-line program that allocates an 8-byte buffer, frees it, and then performs two observe instructions --- one of the constant 1 and one of the pointer p. On the right, two event traces produced by this program under different allocators, drawn side by side. The first three events of each trace --- a malloc, a free, and an observe of 1 --- are highlighted to indicate that the prefixes are similar; the fourth event (the observe of p) differs between the two traces because the observed pointer value depends on the allocator.}
\end{figure}

The characteristic filter and residue connects the two traces when
they originate from different allocators.
\Cref{fig:trace:similarity:example} illustrates a simple
program and two possible traces filtered by
the symbolic allocation sequence in the middle of the diagram,
leaving the observe events in the characteristic residues.
For this program, the full 4-event traces are not similar because
of the disagreement on the last observe.
However, the 3-event prefixes --- highlighted in the figure ---
are similar.

\subsection{Allocator Impact}
Third, we formalize the semantic gadgets of allocator impact, and the
notion of progress impact that we motivate
in~\Cref{sec:gai:intuition}.  Our definitions are based on the concept
of attacker knowledge~\cite{gradual-release} and attacker
impact~\cite{Askarov2011fb} that we adapt to allocators.

\begin{definition}[Allocator impact]
Given a program $c$, initially compatible $E$, $H$, and trace $\tr$,
define \emph{allocator impact} to be the set of allocators that can
produce traces that are similar to $\tr$.
\[
  \allocatorimpactmeta (c, E, H, \tr)  \defn 
  \{ \allocstrategy \in \stratawaredom{E}  \mid \,
  \exists H_0,\allocstate_0.~(H_0, \allocstate_0) = \stratapply{\allocstrategy}{\INIT{H}}
     \land\ \invmch{E, \allocstrategy}\conf{c;H_0,\allocstate_0} \to_{\tr'}^{*} 
     \land\ {\tracesim {\tr} {\tr'} } 
  \}
\]
\end{definition}

\noindent
Next, we define the notion of $\progclass$-progress impact that is a
generalization of the concept of progress
knowledge~\cite{tini,askarov:sabelfeld:2009}.

\begin{definition}[\progclass-progress impact]
Given a program $c$, initially compatible $E$, $H$, and trace $\tr$,
define \emph{\progclass-progress impact} to be the set of allocators
that produce traces that are similar to $\tr$, and where the execution
can further reach an event from the set of events $\progclass$.
\[
  \allocatorimpactproggen (c, E, H, \tr, \progclass) \defn 
  \bigcup_{\event \in \progclass}
  \allocatorimpactmeta (c, E, H, \tr \cdot \event)
\]
\end{definition}

Note that, by definition, singleton-progress impact, i.e., $\progclass
= \{ \event\}$, is exactly the allocator impact for $t \tracecons
\event$.  In addition, the following two set families
are relevant for $\progclass$-progress impact: the set
$\eventclassalloc{n}$ of all malloc or mfail events of size $n$, and
the set $\eventclasscast$ of all possible casts, defined as follows:
\begin{displaymath}
  \eventclassalloc{n} =  
  \{ \evmalloc{n}{a} \mid a \in \domain{Addr} \} \cup
  \{\evmallocfail{n}\}
  \quad\mbox{and}\quad
  \eventclasscast = \{ \evcast{a} \mid a \in \domain{Addr} \}
\end{displaymath}

\subsection{Gradual Allocator Independence}

Using the definitions of $\eventclassalloc{n}$ and $\eventclasscast$,
we can construct the progress gadgets for malloc and cast, the
intuition for which we give in \Cref{sec:informal:downgrading}.
We formally capture this using the notion of downgrading
characterization, defined as follows.

\begin{definition}[Progress downgrading characterization]
Given an event $\event$, define its \emph{progress downgrading characterization
function}, denoted $\downcharfn{\event}$, as follows
\[
  \downcharfn{\event} = 
  \begin{cases}
    \eventclassalloc{n}  & \text{if}\ \event = \evmalloc{n}{a} \text{\ for some}\ n \text{\ and}\ a, \text{or}\  \event= \evmallocfail{n} \text{\ for some}\ n  \\ 
    \eventclasscast & \text{if}\ \event = \evcast{a}\ \text{\ for some}\ $a$\\ 
    \{ \event \} & \text{otherwise}
  \end{cases}
\]
\end{definition}

The progress downgrading characterization specifies the bound on how
much information about allocator behavior may be exposed with each
event. The singleton case corresponds to the events that do not have
any downgrading semantics: frees and observes. The others are
associated with the corresponding event sets. Putting all this formal
machinery, we finally arrive at the main definition of gradual
allocator independence.

\begin{definition}[Gradual allocator independence]%
\label{def:gai:main:definition}
Given program $c$, initially compatible $E$, $H$,
say that $c$ satisfies 
\emph{gradual allocator independence} for $E, H$, written 
$\mathit{GAI}(c, E, H)$, 
if for all $\allocstrategy \in \stratawaredom{E}$, traces $\tr$, and events $\event$ such that
for $(H_0, \allocstate_0) = \stratapply{\allocstrategy}{\INIT{H}}$
where $\invmch{E, \allocstrategy}
  \conf{c; H_0,\allocstate_0} \to_{\tr}^{*}
  \conf{c'; H',\allocstate'} \
  \to_{\event}$
  it holds that
\begin{displaymath}
  \allocatorimpactproggen (c, E, H, \tr, \downcharfn{\event}) 
  \supseteq \allocatorimpactmeta (c, E, H, \tr) 
\end{displaymath}
\end{definition}

It helps to unfold this definition into three scenarios, corresponding
to the cases in $\downcharfn{\event}$.
\begin{enumerate}
\item When $\event$ is free or observe, the singleton characterization
  means that the actual requirement for this event is 
  $\allocatorimpactmeta (c, E, H, \tr \tracecons \event) \supseteq
  \allocatorimpactmeta (c, E, H, \tr)$. This is effectively a
  noninterference, because it requires that allocators possible before
  the event $\event$ also produce the event $\event$.
\item When $\event$ is a malloc or mfail of size $n$, the requirement
  is weaker than above; only reachability of an allocating event of size
  $n$ is required.
\item Last, when $\event$ is a cast, only reachability of a cast
(without committing to a specific address) is required.
\end{enumerate}

\section{Example Programs and Gradual Allocator Independence}
In the following, we show a number of \thelanguage\ programs to
illustrate how gradual allocator independence captures common idioms
and memory behavior, both safe and unsafe.
Recall that \emph{every malloc}, where
allocated memory is accessed, involves downgrading.

\needspace{6\baselineskip}
\subsection{Common Examples of Memory Safety Issues}%
\label{sec:common:examples}

\begin{figure}
\centering
\begin{subfigure}[t]{0.28\textwidth}
\centering
\begin{lstlisting}[mathescape=true,numbers=none,xleftmargin=1em,framexleftmargin=1em]]
// UNSAFE
p = malloc(87);
// NULL pointer 
// deref
*p = 42;         $\label{line:p0:asgn}$
observe(*p);
\end{lstlisting}
\caption{Null pointer dereference\label{fig:npe:bad}}
\end{subfigure}
\hfill
\begin{subfigure}[t]{0.4\textwidth}
\centering
\begin{lstlisting}[numbers=none,xleftmargin=1em,framexleftmargin=1em]
// UNSAFE
p = malloc(87);
if(p != 0) { // assumes  
  *p = 42;   // NULL is 0
  observe(*p);
}
\end{lstlisting}
\caption{Dependency on 0 as NULL\label{fig:npe:variation}}
\end{subfigure}
\hfill
\begin{subfigure}[t]{0.28\textwidth}
\centering
\begin{lstlisting}[numbers=none,xleftmargin=1em,framexleftmargin=1em]
// SAFE
p = malloc(87);
if(p != NULL) {
  *p = 42;
  observe(*p);
}
\end{lstlisting}
\caption{Satisfies GAI\label{fig:npd:safe}}
\end{subfigure}
\caption{Null pointer dereference examples}
\Description{Three short programs illustrating variations of a null-pointer dereference. The first program allocates an 87-byte buffer and immediately writes to the returned pointer with no null check, which is unsafe. The second adds a guard that compares the pointer against the literal integer 0 before dereferencing; this is still unsafe because an allocator may use a non-zero address as its null value. The third program guards the dereference with a comparison against the symbolic NULL expression and is safe.}
\end{figure}

\subsubsection{NULL Pointer Dereference.}
The program in \Cref{fig:npe:bad} contains a potential $\NULL$ pointer
dereference and is thus not memory safe.  To show that the program
does not satisfy gradual allocator independence,
consider two event traces: one generated by the program using an
allocator that makes the $\null$ address inaccessible and one
generated with an allocator that does not protect the $\null$
address. In the former case, the semantics will be stuck at the
assignment in line~\ref{line:p0:asgn} and the corresponding event
trace will \emph{not} include the event generated by the final
\texttt{observe} instruction. The latter trace, on the other hand,
\emph{will} include the observation event.

The variation of the first program in \Cref{fig:npe:variation} takes
care to check that the return value of the allocator is not `0' (the
zero address), to avoid the possible $\NULL$ pointer
dereference.  However, since an allocator may choose a different
address as the $\NULL$ address, this program may still potentially
perform such a dereference. Similarly to the previous program, it
suffices to consider the event traces generated by two allocators: one
that uses zero (0) as the $\NULL$ address and one that uses a
different address. Memory allocation failure would yield two
non-similar traces, one that includes the result of the
\texttt{observe} instruction and one that does not.

In the final variation of the program, in \Cref{fig:npd:safe}, the
error check is performed correctly, using the symbolic $\nullexpr$
value. This program can be shown to satisfy gradual allocator
independence.

\subsubsection{Use After Free}
Accessing memory after it has been released can result in numerous
errors, e.g., \NULL\ pointer dereference and buffer overwrite. In the
remainder of the examples, we assume that $\code{error()}$ stops
program execution.
\begin{figure}
\centering
\begin{subfigure}[t]{0.35\textwidth}
\centering
\begin{lstlisting}[mathescape=true]
// UNSAFE - use after 
//          free
p = malloc(42);
if(p == NULL) error();
free(p);
*p = 87; $\label{line:p1:asgn}$ // use after
          // free 
observe(*p);
\end{lstlisting}
\caption{Use after free\label{fig:ex:use:after:free}}
\end{subfigure}
\hfill
\begin{subfigure}[t]{0.6\textwidth}
\centering
\begin{lstlisting}[mathescape=true]
// UNSAFE - buffer overflow
p = malloc(4); q = malloc(4);
if(p == NULL || q == NULL) error();
*q = 42; i = 0;
while(i < 6) {
  *(p + i) = i;  // buffer overflow $\label{line:p1a:asgn}$
  i = i + 1;
}
observe(*q);
\end{lstlisting}
\caption{Buffer overflow\label{fig:ex:buffer:overflow}}
\end{subfigure}
\caption{Use after free and buffer overflow}
\Description{Two unsafe programs shown side by side. The left program allocates a 42-byte buffer, checks it against NULL, frees it, and then writes to the freed pointer and observes the result --- a use-after-free. The right program allocates two adjacent 4-byte buffers p and q, writes 42 into q, and then runs a while loop that writes bytes 0 through 5 into p[0] through p[5]; because p is only 4 bytes long, the last two writes overflow into the memory region holding q.}
\end{figure}

The program in \Cref{fig:ex:use:after:free} does not satisfy gradual
allocator independence, which can be shown by considering two event
traces generated by using an allocator that makes the $\null$ address
inaccessible and an allocator that does not protect the $\null$
address respectively. In the former case, the semantics will be stuck
at the assignment in line~\ref{line:p1:asgn}, whereas in the latter
case, it will not.
\subsubsection{Buffer Overflow}
Buffer overflows, possibly the most ``famous'' type of memory
management bug, are still the cause of serious security
vulnerabilities, e.g., the recent heap buffer overflow in the Rsync
application (assigned CVE-2024-12084). The program in
\Cref{fig:ex:buffer:overflow} is a simple illustration of such a heap
buffer overflow in which the program makes a ``+/-~1'' error and
writes beyond the bounds of the `\texttt{p}' buffer:

To show that this program does not satisfy gradual allocator
independence consider the event traces generated from runs
respectively with an allocator that places allocated memory blocks in
sequence and with any allocator that does not place allocated memory
regions in immediate sequence, e.g., an allocator that stores metadata
in a header preceding allocated memory or even an allocator that
separates allocated memory by ``memory guards'', regions of
inaccessible memory, to catch buffer overflows. In the latter case,
the above program will be stuck at line~\ref{line:p1a:asgn}.

\subsubsection{Double Free}
The \emph{double free} bug is another common error in memory
management, potentially leading to a security vulnerability. A double
free bug occurs (in C) when a memory area is released twice, i.e.,
free is called twice on the same pointer. This is often due to an
overlooked call to free, e.g., on an error path as illustrated in the
program in \Cref{fig:example:double:free}. Freeing a pointer twice (or
more) is not in itself a problem, but becomes a problem if the
(doubly) freed memory is allocated in between the two frees. In this
case the memory is erroneously released by the second call to free:
\begin{figure}
\centering
\begin{subfigure}[t]{0.45\textwidth}
\centering
\begin{lstlisting}[mathescape=true]
/* UNSAFE - double free */
p1 = malloc(17);
if(p1 == NULL) error();
if(some_other_err) 
   free(p1);
p2 = malloc(87);
if(p2 == NULL) error();
*p2 = 42;
free(p1); // potential double 
          // free
*p2 = 117;         $\label{line:p4a:asgn}$
observe(*p2);
free(p2);
\end{lstlisting}
\caption{Double free\label{fig:example:double:free}}
\end{subfigure}
\hfill
\begin{subfigure}[t]{0.52\textwidth}
\centering
\begin{lstlisting}[mathescape=true]
// SAFE - find ptr with least addr
p = arr; 
/* `arr' is an array of 
    5 pointers */
res = *p;
i = 1;
while(i < 5) {
  if(*(p + i) < res) $\label{line:p12a:cmp}$// pointer 
                     // comparison 
    res = *(p + i);
  i = i + 1;
}
observe(res <= *p);
\end{lstlisting}
\caption{Pointer comparison\label{fig:ex:pointer:comparison}}
\end{subfigure}
\caption{Double free and pointer comparison examples}
\Description{Two programs shown side by side. The left program allocates p1, optionally frees it on an error path, then allocates p2, writes to p2, frees p1 a second time, writes to p2 again, and observes p2 --- a potential double free whose second write is unsafe when the intervening allocation of p2 reuses the freed memory. The right program walks an array of five pointers and uses the less-than operator to find the numerically smallest pointer address, storing it in a variable res and then observing whether res is no greater than the first element. This program is safe despite performing relational pointer comparisons that are undefined behavior in standard C.}
\end{figure}

Using again an allocator the makes freed memory inaccessible and one
that does not, the latter would lead to a stuck semantics (and hence
event trace) at the assignment on line~\ref{line:p4a:asgn}, while the
former would result in an event trace including the result of the
\texttt{observe} instruction.

\subsubsection{Pointer Comparison}
Finally, we show how gradual allocator independence can be used to
reason about pointer comparison also in cases that are categorized as
``undefined behavior'' in standard C~\cite{KR1988:c,iso-c}, where only
pointers to the same memory object, e.g., the same array, may be
compared using relational operators such as $<$. This is illustrated
in \Cref{fig:ex:pointer:comparison}, that looks up the (numerically)
smallest pointer (address) in an array of five pointers called
`\texttt{arr}':

The pointer comparison on line~\ref{line:p12a:cmp} potentially
compares pointers to different memory objects, i.e., undefined
behavior\footnote{The GCC compiler also allows this comparison by
first casting the pointer to unsigned integers:
\url{https://www.gnu.org/software/c-intro-and-ref/manual/html_node/Pointer-Comparison.html}}. However,
the program satisfies gradual allocator independence.

\section{Translation From a Memory Safe Language to Notac}%
\label{sec:memsafe}%
\label{memsafe-to-notac-translation}
This section presents a translation from a low-level memory-safe
language \otherlang{} to \thelanguage{}. \otherlang{} is similar to
the language of~\cite{azevedo2018meaning} using a CompCert-style
memory model of identifiers and offsets with runtime enforcement of
bounds, automatic memory initialization, and infinite memory. Its
semantics enforces a form of fat pointers by design.
The translation serves the main purpose of showing that \otherlang{}
programs that terminate without error, i.e., programs that are memory
safe in the sense of~\cite{azevedo2018meaning}, also satisfy gradual
allocator independence. This result, a form of
\emph{transparency}~\cite{SecureMultiExecution}, shows that the two
notions coincide on memory safe programs.
The transparency result is also a useful vehicle for validating the
definition of the allocator well-formedness
(\Cref{def:allocator:heap:well-formedness}) as the translation
correctness relies on the allocator well-formedness.

We deviate from~\cite{azevedo2018meaning} in a few ways:
\begin{enumerate}
\item \otherlang{} does not include a free command since it is not
  strictly needed with infinite memory; furthermore, in a program like
  $\memsafe{x \leftarrow \mathsf{alloc}(8);}$
  $\memsafe{\mathsf{free}(x);}$
  $\memsafe{y \leftarrow \mathsf{alloc}(8)}$ the expression $x == y$
  will never evaluate to true due to provenance semantics (again
  enabled by the infinite memory). Thus our translation would be
  unsound for such programs;

\item \otherlang{} does not use $\msnil$ values to propagate errors
  during expression evaluation, because it leads to anomalies, such as
  $\msnil + 1 == \msnil$ evaluating to true. Instead, expression
  evaluation in \otherlang{} is a partial function;

\item it omits the \textsf{offset} operator, as its semantics is
  inherently related to the identifier and offset-based memory model.

\item for simplicity of the translation, it does not have booleans.
\end{enumerate}

The formalization of the translation is presented in the \extorappendix{}.
Informally, the translation embeds out-of-memory checks after each
allocation and uses a dedicated global variable $\oom$ to track
whether any of the allocations have failed. Since \thelanguage{} does
not have an \texttt{exit} command, a guard macro is introduced. Memory
is zero initialized using another global variable $\alloci$ and loop
execution is instrumented to be also guarded and ensures the actual
loop guard is not evaluated (as it may deviate) when $\oom$ is true.

For the translation theorem, note that in \otherlang{} a local store
is a finite map from variables to values, a heap is a finite map from
pointers to values, and a state is a pair of a local store and a
heap. Error-free evaluation of program $\memsafe{p}$ is
$\mseval[\memsafe{s}]{\memsafe{p}} = \memsafe{(l',m')}$.
The proof of \Cref{thm:memsafe:notac:gai} is in the \extorappendix{}.
\begin{theorem}[Gradual allocator independence of \otherlang{} to \thelanguage{} translation]%
  \label{thm:memsafe:notac:gai}
  Given a program $\memsafe{p}$ in \otherlang{} and initial state
  $\memsafe{s = (l, m_{\mathnormal{init}})}$, where
  \begin{inparaenum}[\itshape (1)]
  \item $\memsafe{m_\text{init}}$ is an empty map,
  \item $\mseval[\memsafe{s}]{\memsafe{p}} = \memsafe{(l',m')}$, and
  \item $\forall v \in \img{\memsafe{l}}. \, v \neq \memsafe{(i,b,n)} \land v \neq \memsafe{\msnil}$,
  \end{inparaenum}
  then for any \thelanguage{} program $\notac{c}$, environment $E$,
  and heap $H_0$, such that
  \begin{flushleft}
    \begin{minipage}[t]{0.28\linewidth}
      \begin{itemize}
      \item $\notac{c} = \MemsafeToNotac{p}$
      \item $\vars{\memsafe{p}} \subseteq \dom{E}$
      \end{itemize}
    \end{minipage}
    \begin{minipage}[t]{0.28\linewidth}
      \begin{itemize}
      \item $\oom, \transvar{i} \notin \vars{\memsafe{p}}$
      \item $H_0(E(\oom)) = 0$
      \end{itemize}
    \end{minipage}
    \begin{minipage}[t]{0.40\linewidth}
      \begin{itemize}
      \item $H_0(E(\alloci)) = 0$
      \item
        $\forall x \in \dom{\memsafe{l}}. \, H_0(E(x)) =
        \memsafe{l}(x)$
      \end{itemize}
    \end{minipage}
  \end{flushleft}
  it holds that $\mathsf{GAI}(\notac{c}, E, H_0)$.
  Moreover for any $\allocstrategy \in \stratawaredom{E}$
  where $(H,\allocstate) = \stratapply{\allocstrategy}{\INIT{H_0}}$
  it holds that $\invmch{E, \allocstrategy} \conf{\notac{c}; H, \allocstate} \to_{\tr}^{*}
  \conf{\notac{\code{stop}}; H', \allocstate'}$
  and if the execution did not run out of memory $H'(E(\oom)) = 0$ then
  for any $x \in \dom{\memsafe{l'}}$ it holds that:
  \[
    \memsafe{l'}(x) \neq (i,b,n) \land \memsafe{l'}(x) \neq \msnil
    \implies
    \memsafe{l'}(x) = H'(E(x))
  \]
\end{theorem}

\section{Related Work and Outlook}%
\label{sec:rw}
\paragraph{Memory Safety: Definitions and Allocator Dependence.}
Motivated by the lack of a clear and common definition,
Hicks~\cite{hicks:blog2014:whatismemsafe} defines memory safety, for a
single program execution, as 
not \emph{accessing undefined
memory}, where ``undefined'' means memory that has never been allocated
nor deallocated. This avoids 
null pointer
dereference and use after free, but does not cover certain kinds of
buffer overflows, nor illegal freeing or use of uninitialized
memory. Some of these can be recovered through various extensions and
assumptions. To deal with buffer overflows, Hicks introduces
the notion of \emph{pointers as capabilities} (aka.\ ``fat'' pointers)
which are pointers packaged with their corresponding bounds. This
enables bounds checking of pointer operations and prevents (some)
buffer overflows, resulting in a monitoring mechanism rather than a
definition of memory~safety.

Berger et al.~\cite{BergerZ:pldi2006:diehard} defines a program to be
\emph{fully memory safe} if ``it never reads uninitialized memory,
performs no illegal operations on the heap (no invalid/double frees),
and does not access freed memory (no dangling pointer errors)''.  It
further defines the notion of \emph{infinite heap semantics}, noting
that an infinite heap would allow C programs that are not ``fully
memory safe'' to run safely, since effects of memory faults (and
attacks?) would be contained. However, infinite heap semantics does
not prevent/mitigate reading uninitialized memory. The paper suggests
running several replicas of the program using differently randomized
allocators: if the output differs, then the program reads from
uninitialized memory. It is then shown that infinite heap semantics
can be approximated probabilistically. The \emph{DieHard} memory
allocator (and monitor) implementing these ideas. The described
replication is a \emph{monitoring} solution that \emph{masks}
incorrect (memory unsafe) programs.

Novark et al.~\cite{NovarkB:ccs2010:dieharder} presents the first
systematic analysis of memory allocator impact on (heap) memory
security including an overview and description of many common
heap-based attacks; also presents the \emph{DieHarder} allocator
designed to mitigate the described heap-based attacks with modest
performance overhead.

Amorim et al.~\cite{azevedo2018meaning} define memory safety of a
language in terms of \emph{reasoning principles}, as opposed to a list
of bad things to avoid, to allow for program proofs using separation
logic. They derive a noninterference property showing that (safe)
programs cannot be affected by unreachable memory.  The defined memory
model has many limitations: it does not support manipulating pointers
as integers, casting pointers to integers, usage of uninitialized
memory etc.\ While the paper argues that the formal model can be
relaxed to allow partial support of (some of) these constructs, the
general and full support seems infeasible in their model. From the
perspective of our work, we attribute those limitations to the simple
approach of defining noninterference on memories, and not having
principled means for relaxations. Our strategy-based approach,
together with the novel allocator model, and epistemic techniques for
information flow, lifts all these restrictions, and enables
reasoning about memory safety of individual programs in memory-unsafe
languages.

\paragraph{Provenance Semantics} Recent work on memory safety uses
provenance semantics~\cite{memarian2019:provenance,mswasm} 
grounded in the complexity of the C specification. Compared to
provenance-based approaches, GAI is an
end-to-end definition; it also operates at a lower level of
abstraction. We see provenance as an important reasoning and
enforcement mechanism for end-to-end notions like ours, not unlike
how compositional techniques are related to end-to-end semantic
security definitions~\cite{bastys2018:principles:ifc}.

Our informal intuition is that provenance semantics to GAI is what
taint tracking is to information flow. In security, taint tracking is
a useful, widely used, and easy to understand mechanism, but it is also
recognized to lack semantic finesse. IFC properties have semantic
precision but are difficult to enforce, and are in fact often
approximated with taint tracking. 
In the specific context of real C (with standards, etc),
provenance semantics are a valuable tool.
If we consider what memory safety should be
for a clean low-level untyped language in a flat memory model, the
case for a GAI-style definition is more appealing.
This is also why Notac is not a C, and  
the 
intuition behind GAI is ``Notac programs should not depend on the allocator
except OOM and pointer to integer casts'' is intuitive, easy to
understand, and is well-backed by the underlying formal machinery.

As an example of why provenance semantics is not an ideal basis for a
semantic definition, revisit the program in 
Figure~\ref{fig:implicit:alloc:same:branches}.
All approaches based on pointer provenance that we are aware of have
the limitation that they either reject this program or have to compromise and accept programs such as Figure~\ref{fig:implicit:alloc:different:branches} that exposes the program behavior to allocator decisions (cf. \Cref{sec:gai:allocators-as-secrets}).
Such approaches include the
MSWasm, Amorim et al paper (cf.\
Section~\ref{memsafe-to-notac-translation}), and the Cerberus project.
To some extent, those approaches inherit this semantic deficiency
because they are tied to C and require special treatment of integer-to-pointer casts.
The pointer comparison \texttt{(p < q)} is an undefined  behavior in
C~\cite{KR1988:c,iso-c}, but is valid 
in Notac. This gives us an opportunity to embrace a clean semantic
approach, and GAI accepts the above program.
With GAI and Notac, no special treatment is necessary. For
example, if a program dereferences a nonsensical pointer, GAI does not
discriminate between whether that value is obtained via a simple
pointer arithmetic or via a complex roundtrip to integers and back (or
even a cast of a literal).
Finally, GAI has a graceful handling of out-of-memory. In Cerberus'
provenance semantics, the OOM is treated as an edge case, e.g.,
evaluation of the program from \Cref{fig:alloc:null:cmp} is stuck at the attempt to
allocate a large buffer. The original MSWasm model assumes infinite memory. 

\paragraph{Sub-Object Memory Safety and Capability-Based Confinement}
In languages with structs, sub-object memory safety means that 
a pointer to one field of a struct should not be usable to access other fields.
Contemporary techniques tackle sub-object access by associating pointers with bounds and offsets that delimit the access. This is enforced  either through hardware  as in the line of work on CHERI~\cite{watson2015cheri,AmarCCFLLNMTWX:micro2023:cheriot} or abstractly as in~MSWasm~\cite{mswasm,iris-mswasm}.
In the case of MSWasm, \emph{handles} are capability-like unforgeable fat pointers. MSWasm defines memory safety as an absence of violations using an abstract execution monitor, for which it introduces a memory model where memory addresses are associated with colors and shades. Colors correspond to allocation provenance (as discussed above), and shades delimit ranges within an allocation, motivated by sub-object access. The abstract monitor deems an execution safe  when memory accesses through  handles agrees with the associated coloring. MSWasm's notion of memory safety is allocator-independent, but the language specification supports only a restricted set of pointer comparison operations and does not support casts. These restrictions are sufficient for running the PolyBench/C~\cite{PolyBenchC} -- the original Wasm benchmark suite  -- that, as far as we see, 
has a relatively clean code base.
\citeauthor{jangda2019:not-so-fast}~\cite{jangda2019:not-so-fast} 
 note already in 2019 that PolyBench/C benchmarks are small scientific kernels, designed for benchmarking polyhedral loop optimizations in compilers, that do not represent larger applications.

Iris-MSWasm~\cite{iris-mswasm} observes that because of their fine-grained capability nature, the handles in MSWasm enable reasoning about local state encapsulation when linking with adversarial code~\cite{devriese2016reasoning,swasey2017robust,georges2024cerise}, characterized as \emph{robust capability safety}.  However, the color and shade model cannot support the necessary reasoning, and Iris-MSWasm uses the ideas from separation logic instead.

As the name suggests, the focus of  GAI is only allocator-related aspects of memory safety, which the sub-object access is not, and Notac does not currently support structs. 
More importantly, Notac's semantics does not enforce access control (unlike e.g., capabilites).
We defer the support for sub-object memory safety to future work, but note that since the struct organization is compiler-dependent, GAI clarifies that memory-safety for the heap is a separate issue from sub-object access. We envision the complementary notion of \emph{struct-layout-independence} capturing that memory safe programs should be independent of the compiler's layout decisions (including stack layout). We will need to account for new downgraders leaking layout information, e.g., \texttt{sizeof}. 
In order to avoid using \texttt{sizeof} for most ordinary tasks, one can expose a primitive for allocating $N$ items of type $T$; the latter abstraction  is not surprising~\cite{ruef2019achieving}, but will be well-grounded in the underlying theory.
Alternatively, another concept from information flow control, namely information erasure~\cite{chong2005language}, for which an epistemic formulation is also available~\cite{askarov2015cryptographic}, can be applied here. 
With erasure, one can mandate that downgrading of the \texttt{sizeof} is localized~\cite{hunt:sands:2008just:forget:it} and should be forgotten after the malloc.

\paragraph{Low-Level Memory Models and Program Verification} 
There is a substantial amount of work on low-level memory models
useful for program verification. We survey a sample of particularly
relevant state-of-the-art related work.

Kell~\cite{kell:onward2017:c} argues for the usefulness of C's low
level memory model, mainly for ``systems programming'' where flexible,
explicit, and sometimes non-compliant memory access is often needed.
Kang et al.~\cite{KangHMGZV:pldi2015:formalc} and most recently Beck
et al.~\cite{beck2024twophase} formalize a C memory model with 
a refined semantics for undefined behavior that allows for
casts from integers to pointers, and apply it to show soundness of program
optimizations. 

Krebbers~\cite{Krebbers:jar2016} gives a Coq formalization of the
``(non-concurrent part of the) C11 standard'' (called the
\mbox{CH${}_2$O} memory model). A major feature of the model is the
focus on modeling both high-level and low-level aspects of memory in
order, among other things, to reason about their interaction, e.g.,
where the high-level, typed view of memory does not really correspond
to the low-level, untyped, view. Such reasoning is particularly useful
for proving the correctness of compiler optimizations and other
program transformations. The paper does not, as such, define memory
safety.

Tuch et al.~\cite{TuchKN:popl2007} present a verification framework
for C programs with a (more) realistic memory model, i.e., proving
functional correctness of C programs that may access memory in a
low-level non-standards compliant way. Undefined behavior is handled
by generating \emph{guards} explicitly checking against \emph{all}
possible undefined behavior(s). Similar to the noninterference
concepts underlying this paper, Tuch et al.\ note the challenge in
specifying that an allocation should not result in ``anything else''
changing on the heap: %
``[...] In the success case, we would ideally like to know that
nothing else in the heap changes. This `nothing else' is hard to nail
down formally[...]''.
Interestingly, the formal specification of the \texttt{alloc()}
function for the L4 kernel that is validated in the paper has elements
and reasoning similar to our symbolic allocator sequences.

Elliott et al.~\cite{ElliottRHT:secdev2018:checked-c} propose an
extension of C to allow for \emph{checked pointers}, i.e., pointers
that can be verified to access only memory that is within defined
bounds. The proposed extensions, cover many of the primary and typical uses of pointers in
non-trivial C programs. Bounds checking can either be statically proven or, if necessary, done by dynamic checks.
Allocator correctness is recognized as an important
problem~\cite{appel:naumann2020verified:malloc:free,reitz2024starmalloc}
usually approached using program logics~\cite{jung2018iris}.
Recent work on Bedrock2~\cite{erbsen2021:bedrock2} explores whole
system correctness in a flat memory model similar to the one
of \thelanguage. A related work on omnisemantics~\cite{omnisemantics}
uses hyperproperties~\cite{hyperproperties} based reasoning
that is similar in flavor to the epistemic approach~\cite{gradual-release} of our work.
\section{Future Work}
In addition to the future work on attaining sub-object access, covered in \Cref{sec:rw}, this section outlines other  future directions that we envision.

\paragraph{Information Disclosure Vulnerabilities}
A potential application of GAI is information disclosure vulnerabilities in C when allocators are under adversarial influence: they follow the basic contract (\Cref{sec:allocation:strategies}) and may not directly output secrets, but can indirectly disclose information by affecting the behavior of other functions. A well-known example of information disclosure vulnerability is Heartbleed, where an out-of-bounds read discloses whatever the allocator has placed in adjacent heap memory. We identify two future directions. First, one would expect that memory-safe functions should not be used as information conduits. GAI formalizes exactly this expectation. Second, allocators (and more generally, the underlying memory management abstractions) may exhibit different degrees of adversity. For example, hardening via zeroing secret buffers with \texttt{memset\_s} may scrub the original buffer~\cite{dsilva:correctness-security-gap,yang:usenix:2017:dse:harmful}, but the allocator may retain a copy, and can in principle even leak individual bits through the OOM behavior. Our work offers a foundation for reasoning about the degrees of allocator adversity and the corresponding effectiveness of such hardening.

\paragraph{Independent Extensions of the Allocator Model}
Our novel allocator model from cf. \Cref{sec:allocation:strategies} may
have independent future applications, such as relaxing it to support
parameterization in the style of OpenBSD and reasoning about
system techniques for memory management, such as deduplication.

\paragraph{Verification and Enforcement of GAI}
We anticipate that enforcement of GAI can leverage many of the
techniques for enforcing noninterference and declassification, most
notably type systems, e.g., as used to enforce \emph{robust
  declassification} and nonmalleable information
flow~\cite{Askarov2011fb,cecchetti2017nonmalleable,Cecchetti:NMIFC:progress}.
Similarly, the connection between provenance semantics and
taint-tracking~\cite{Schoepe2016bh} is worth pursuing and explore its
use as an enforcement mechanism. Additionally, GAI seems suitable as
foundation for incorrectness logics reasoning about/proving memory
errors, i.e., finding two allocators that
“disagree”~\cite{OHearn:popl2020:buglogic,LeRVBDO22:oopsla2022:bigbugs},
or possibly a more direct encoding of the set of acceptable allocators
in Outcome Logic~\cite{ZilbersteinDS:oopsla2023:ol}. Finally,
hyperproperty model checking and Hoare (hyper-)logics, based on event
traces and self-composition, could provide automated
enforcement~\cite{self-composition,hyperproperties,LamportS:csf2021:hypertla,AroraHLLP:spin2022:probhyper,DardinierLM:oopsla2024:hypra}.

\paragraph{Reasoning About Optimizations}
Presently, many optimizations that are sound in logical models (e.g., similar to the Memsafe language in \Cref{sec:memsafe}) are not sound in concrete models~\cite{besson2015concrete}, because the soundness criteria for optimizations are worded as ``preserving semantics for programs with defined behavior''; there are obviously more defined behaviors in concrete models.
GAI decouples the semantic definition of memory safety from the underlying memory model, which suggests that one can reword the soundness criteria from having-a-defined-behavior to satisfying-a-semantic-guarantee.
Because our transparency result indicates that intuition of the existing logical models indeed aligns with GAI, we see this as an opportunity to justify standard optimizations~\cite{beck2024twophase} in concrete models.
As a starting point, future work should investigate  soundness of intra-procedural optimizations,  assuming external functions satisfy a GAI-style property.
Furthermore,  the connection between memory safety and downgrading can link to quantitative information flow~\cite{alvim2020:qif:book}, opening questions, such as ``how much memory safety is broken by optimization X''. Such quantitative measures may be useful in comparing between optimizations.

\paragraph{Nonmalleable Casts}
Aiming beyond access control in capability systems, adversarial code that compares capabilities~\cite[§4.4\ Pointer Comparison]{cheri-c-guide} or casts them to integers can learn information about the trusted callers. 
An intuitive defense is to confine the adversary to only cast capabilities from their modules, and that bitwise representation of capabilities between the trust boundaries do not interfere. This intuition coincides with the idea of nonmalleable information flow control~\cite{zdancewic2001robust,Askarov2011fb,cecchetti2017nonmalleable,
ferraiuolo2018hyperflow,acay2021viaduct,zagieboylo2019using,Cecchetti:NMIFC:progress} that studies restrictions on attacker-controlled downgrades, and 
resource partitioning against side channels~\cite{zhang2012:pldi:language-based-mitigation,zhang2015hardware}.

\section{Conclusion}%
\label{sec:conclusion}
\newif\ifshortconclusion\shortconclusiontrue

The definition of gradual allocator independence establishes a formal
correspondence between two established decades-long research areas:
downgrading in information flow control and memory safety.
The downgrading approach faithfully models previously
difficult-to-capture aspects of memory safety, such as graceful
handling of out-of-memory errors and casts.
We achieve this by developing a novel allocator model and leveraging
the state-of-the-art information flow techniques.
\ifshortconclusion
This work creates an
opportunity for future cross-pollination between the areas. 
In the long term, the application of the rich formal machinery of downgrading, developed for  
controlled leaks, to a leaky abstraction~\cite{Spolsky:LeakyAbstractions}, such as memory management, 
suggests connections to other system abstractions and potentially a general 
theory of the downgrading semantics of leaky abstractions.

\else 

Upon reflection, we observe that memory management in malloc-free
systems is a kind of a leaky
abstraction~\cite{Spolsky:LeakyAbstractions}; it is not surprising
that the rich formal machinery developed for reasoning about intended
leaks appears suitable. 
The correspondence between downgrading and memory safety is both
encouraging and concerning. It is encouraging because it creates an
opportunity for future cross-pollination between the areas, e.g.,
connecting provenance semantics to
taint-tracking~\cite{Schoepe2016bh}. It is concerning because it links
\emph{one difficult problem -- declassification -- to another
difficult problem -- memory safety}. Yet, the concern has a positive
interpretation: it explains why both problems remained difficult for
such a long time in their respective areas, and why progress in either
of the directions demands full-system awareness.
\fi 

\section*{Acknowledgments}

This work was in part supported by the Danish National Defence Technology Centre (NFC) and Concordium Blockchain Research Center at Aarhus University. We thank the anonymous reviewers, and 
our colleagues in the section of Programming Languages, Logic, and Software Security at Aarhus University  for comments and suggestions on the earlier versions of this paper. We also thank Stephen Chong for the discussion on the generalization of our approach, and Emery Berger for helpful insights about the DieHard work. 

\section*{Data Availability Statement}
\ifnoappendix
\else
The artifact~\cite{notac:zenodo} that includes the Notac prototype implementation and code examples from the paper is publicly available.
\fi

\printbibliography{}

@inproceedings{azevedo2018meaning,
  title={The meaning of memory safety},
  author={Azevedo de Amorim, Arthur and Hri{\c{t}}cu, C{\u{a}}t{\u{a}}lin and Pierce, Benjamin C},
  booktitle={Principles of Security and Trust: 7th International Conference, POST 2018, Held as Part of the European Joint Conferences on Theory and Practice of Software, ETAPS 2018, Thessaloniki, Greece, April 14-20, 2018, Proceedings 7},
  pages={79--105},
  year={2018},
  doi = "10.1007/978-3-319-89722-6_4",
  organization={Springer}
}

@article{mswasm,
  title={{MSWasm}: Soundly enforcing memory-safe execution of unsafe code},
  author={Michael, Alexandra E and Gollamudi, Anitha and Bosamiya, Jay and Johnson, Evan and Denlinger, Aidan and Disselkoen, Craig and Watt, Conrad and Parno, Bryan and Patrignani, Marco and Vassena, Marco and others},
  journal={Proceedings of the ACM on Programming Languages},
  volume={7},
  number={POPL},
  pages={425--454},
  year={2023},
  publisher={ACM New York, NY, USA},
  doi =          "10.1145/3554344"
}

@InProceedings{kell:onward2017:c,
  author =       "Stephen Kell",
  title =        "Some were meant for {C:} the endurance of an
                  unmanageable language",
  year =         2017,
  booktitle =    "Proceedings of the 2017 {ACM} {SIGPLAN}
                  International Symposium on New Ideas, New Paradigms,
                  and Reflections on Programming and Software
                  (Onward!~2017)",
  pages =        "229--245",
  doi =          "10.1145/3133850.3133867"
}

@InProceedings{KangHMGZV:pldi2015:formalc,
  author =       "Jeehoon Kang and Chung{-}Kil Hur and William Mansky
                  and Dmitri Garbuzov and Steve Zdancewic and Viktor
                  Vafeiadis",
  title =        "A formal {C} memory model supporting integer-pointer
                  casts",
  year =         2015,
  booktitle =    "Proceedings of the 36th {ACM} {SIGPLAN} Conference
                  on Programming Language Design and Implementation
                  (PLDI~2015)",
  pages =        "326--335",
  doi =          "10.1145/2737924.2738005"
}

@InProceedings{NovarkB:ccs2010:dieharder,
  author       = "Gene Novark and Emery D. Berger",
  title        = "{DieHarder}: securing the heap",
  year         = 2010,
  booktitle    = "Proceedings of the 17th {ACM} Conference on Computer
                  and Communications Security (CCS~2010)",
  pages        = "573--584",
  doi          = "10.1145/1866307.1866371",
  publisher =    "ACM"
}

@InProceedings{BergerZ:pldi2006:diehard,
  author =       "Emery D. Berger and Benjamin G. Zorn",
  title =        "{DieHard}: probabilistic memory safety for unsafe
                  languages",
  year =         2006,
  booktitle =    "Proceedings of the ACM SIGPLAN Conference on
                  Programming Language Design and Implementation
                  (PLDI~2006)",
  pages =        "158--168",
  doi =          "10.1145/1133981.1134000"
}

@Online{hicks:blog2014:whatismemsafe,
  author =       "Michael Hicks",
  title =        "What is memory safety?",
  url =          "http://www.pl-enthusiast.net/2014/07/21/memory-safety/",
  note =         "Blog post. Last accessed: February 2026."
}

@InProceedings{TuchKN:popl2007,
  author =       "Harvey Tuch and Gerwin Klein and Michael Norrish",
  title =        "Types, bytes, and separation logic",
  year =         2007,
  booktitle =    "Proceedings of the 34th {ACM} {SIGPLAN-SIGACT}
                  Symposium on Principles of Programming Languages,
                  (POPL~2007)",
  pages =        "97--108",
  doi =          "10.1145/1190216.1190234"
}

@Article{Krebbers:jar2016,
  author =       "Robbert Krebbers",
  title =        "A Formal {C} Memory Model for Separation Logic",
  journal =      "J. Autom. Reason.",
  year =         2016,
  volume =       57,
  number =       4,
  pages =        "319--387",
  doi =          "10.1007/S10817-016-9369-1",
  url =          "https://link.springer.com/article/10.1007/S10817-016-9369-1"
}

@InProceedings{AmarCCFLLNMTWX:micro2023:cheriot,
  author =       "Saar Amar and David Chisnall and Tony Chen and
                  Nathaniel Wesley Filardo and Ben Laurie and Kunyan
                  Liu and Robert M. Norton and Simon W. Moore and
                  Yucong Tao and Robert N. M. Watson and Hongyan Xia",
  title =        "{CHERIoT}: Complete Memory Safety for Embedded
                  Devices",
  year =         2023,
  booktitle =    "Proceedings of the 56th International Symposium on
                  Microarchitecture (MICRO~2023)",
  pages =        "641--653",
  doi =          "10.1145/3613424.3614266"
}

@InProceedings{ElliottRHT:secdev2018:checked-c,
  author =       "Archibald Samuel Elliott and Andrew Ruef and Michael
                  Hicks and David Tarditi",
  title =        "Checked {C:} Making {C} Safe by Extension",
  year =         2018,
  booktitle =    "Proceedings of {IEEE} Cybersecurity Development
                  (SecDev~2018)",
  pages =        "53--60",
  doi =          "10.1109/SecDev.2018.00015",
  url = "https://www.microsoft.com/en-us/research/uploads/prod/2018/09/checkedc-secdev2018-preprint.pdf"
}

@INPROCEEDINGS{SecureMultiExecution,
  author={Devriese, Dominique and Piessens, Frank},
  booktitle={2010 IEEE Symposium on Security and Privacy}, 
  title={Noninterference through Secure Multi-execution}, 
  year={2010},
  volume={},
  number={},
  pages={109-124},
  doi={10.1109/SP.2010.15}}

@article{self-composition,
  title={Secure information flow by self-composition},
  author={Barthe, Gilles and D'argenio, Pedro R and Rezk, Tamara},
  journal={Mathematical Structures in Computer Science},
  volume={21},
  number={6},
  pages={1207--1252},
  year={2011},
  publisher={Cambridge University Press},
  doi =          "10.1017/S0960129511000193"
}

@inproceedings{bpini,
  title={Reconciling progress-insensitive noninterference and declassification},
  author={Bay, Johan and Askarov, Aslan},
  booktitle={2020 IEEE 33rd Computer Security Foundations Symposium (CSF)},
  pages={95--106},
  year={2020},
  organization={IEEE},
  doi =          "10.1109/CSF49147.2020.00015"
}

@inbook{compcert-mem-model,
  place =        {Cambridge},
  title =        {The CompCert memory model},
  booktitle =    {Program Logics for Certified Compilers},
  publisher =    {Cambridge University Press},
  author =       {Leroy, Xavier and Blazy, Sandrine and Appel, Andrew
                  W. and Stewart, Gordon},
  year =         {2014},
  pages =        {237–271},
  doi =          "10.1017/CBO9781107256552.037"
}

@inproceedings{gradual-release,
  title =        {Gradual release: Unifying declassification,
                  encryption and key release policies},
  author =       {Askarov, Aslan and Sabelfeld, Andrei},
  booktitle =    {2007 IEEE Symposium on Security and Privacy (SP'07)},
  pages =        {207--221},
  year =         {2007},
  organization = {IEEE},
  doi =          "10.1109/SP.2007.22"
}

@inproceedings{goguen1982security,
  title={Security policies and security models},
  author={Goguen, Joseph A and Meseguer, Jos{\'e}},
  booktitle={1982 IEEE Symposium on Security and Privacy},
  pages={11--11},
  year={1982},
  organization={IEEE},
  doi={10.1109/SP.1982.10014}
}

@inproceedings{strategies:csfw:2006,
  title =        {Information-flow security for interactive programs},
  author =       {O'Neill, Kevin R and Clarkson, Michael R and Chong,
                  Stephen},
  booktitle =    {19th IEEE Computer Security Foundations Workshop
                  (CSFW'06)},
  pages =        {12--pp},
  year =         {2006},
  organization = {IEEE},
  doi =          "10.1109/CSFW.2006.16"
}

@inproceedings{declassification:dimensions,
  title =        {Dimensions and principles of declassification},
  author =       {Sabelfeld, Andrei and Sands, David},
  booktitle =    {18th IEEE Computer Security Foundations Workshop
                  (CSFW'05)},
  pages =        {255--269},
  year =         {2005},
  organization = {IEEE},
  doi =          "10.1109/CSFW.2005.15"
}

@inproceedings{tini,
  title =        {Termination-insensitive noninterference leaks more
                  than just a bit},
  author =       {Askarov, Aslan and Hunt, Sebastian and Sabelfeld,
                  Andrei and Sands, David},
  booktitle =    {Computer Security-ESORICS 2008: 13th European
                  Symposium on Research in Computer Security,
                  M{\'a}laga, Spain, October 6-8, 2008. Proceedings
                  13},
  pages =        {333--348},
  year =         {2008},
  organization = {Springer},
  doi =          "10.1007/978-3-540-88313-5_22"
}

@inproceedings{askarov:sabelfeld:2009,
  title =        {Tight enforcement of information-release policies
                  for dynamic languages},
  author =       {Askarov, Aslan and Sabelfeld, Andrei},
  booktitle =    {22nd IEEE Computer Security Foundations
                  Symposium (CSF~2009)},
  pages =        {43--59},
  year =         {2009},
  organization = {IEEE},
  doi =          "10.1109/CSF.2009.22"
}

@incollection{hedin:sabelfeld:2012:perspective,
  title =        {A perspective on information-flow control},
  author =       {Hedin, Daniel and Sabelfeld, Andrei},
  booktitle =    {Software safety and security},
  pages =        {319--347},
  year =         {2012},
  publisher =    {IOS Press},
  doi =          "10.3233/978-1-61499-028-4-319"
}

@article{sabelfeld:myers:jsac,
  title={Language-based information-flow security},
  author={Sabelfeld, Andrei and Myers, Andrew C},
  journal={IEEE Journal on selected areas in communications},
  volume={21},
  number={1},
  pages={5--19},
  year={2003},
  publisher={IEEE},
  doi={10.1109/JSAC.2002.806121}
}

@article{memarian2019:provenance,
  title={Exploring C semantics and pointer provenance},
  author={Memarian, Kayvan and Gomes, Victor BF and Davis, Brooks and Kell, Stephen and Richardson, Alexander and Watson, Robert NM and Sewell, Peter},
  journal={Proceedings of the ACM on Programming Languages},
  volume={3},
  number={POPL},
  pages={1--32},
  year={2019},
  publisher={ACM New York, NY, USA},
  doi={10.1145/3290380}
}

@inproceedings{bastys2018:principles:ifc,
  title={Prudent design principles for information flow control},
  author={Bastys, Iulia and Piessens, Frank and Sabelfeld, Andrei},
  booktitle={Proceedings of the 13th Workshop on Programming Languages and Analysis for Security},
  pages={17--23},
  year={2018},
  doi={10.1145/3264820.3264824}
}

@inproceedings{erbsen2021:bedrock2,
  title={Integration verification across software and hardware for a simple embedded system},
  author={Erbsen, Andres and Gruetter, Samuel and Choi, Joonwon and Wood, Clark and Chlipala, Adam},
  booktitle={Proceedings of the 42nd ACM SIGPLAN International Conference on Programming Language Design and Implementation},
  pages={604--619},
  year={2021},
  doi={10.1145/3453483.3454065}
}

@article{omnisemantics,
  title={Omnisemantics: Smooth handling of nondeterminism},
  author={Chargu{\'e}raud, Arthur and Chlipala, Adam and Erbsen, Andres and Gruetter, Samuel},
  journal={ACM Transactions on Programming Languages and Systems},
  volume={45},
  number={1},
  pages={1--43},
  year={2023},
  publisher={ACM New York, NY},
  doi={10.1145/3579834}
}

@article{hyperproperties,
  title={Hyperproperties},
  author={Clarkson, Michael R and Schneider, Fred B},
  journal={Journal of Computer Security},
  volume={18},
  number={6},
  pages={1157--1210},
  year={2010},
  publisher={SAGE Publications Sage UK: London, England},
  doi={10.3233/JCS-2009-0393}
}

@inproceedings{appel:naumann2020verified:malloc:free,
  title={Verified sequential malloc/free},
  author={Appel, Andrew W and Naumann, David A},
  booktitle={Proceedings of the 2020 ACM SIGPLAN International Symposium on Memory Management},
  pages={48--59},
  year={2020},
  doi={10.1145/3381898.3397211}
}

@article{reitz2024starmalloc,
  title={StarMalloc: Verifying a Modern, Hardened Memory Allocator},
  author={Reitz, Antonin and Fromherz, Aymeric and Protzenko, Jonathan},
  journal={Proceedings of the ACM on Programming Languages},
  volume={8},
  number={OOPSLA2},
  pages={1757--1786},
  year={2024},
  publisher={ACM New York, NY, USA},
  doi={10.1145/3689773}
}

@article{jung2018iris,
  title={Iris from the ground up: A modular foundation for higher-order concurrent separation logic},
  author={Jung, Ralf and Krebbers, Robbert and Jourdan, Jacques-Henri and Bizjak, Ale{\v{s}} and Birkedal, Lars and Dreyer, Derek},
  journal={Journal of Functional Programming},
  volume={28},
  pages={e20},
  year={2018},
  publisher={Cambridge University Press},
  doi={10.1017/S0956796818000151}
}

@misc{rustGlobalAlloc,
  author =	 {{Rust Project Developers}},
  title =	 {GlobalAlloc Trait - Rust},
  year =	 2025,
  howpublished =
                  {\url{https://doc.rust-lang.org/stable/std/alloc/trait.GlobalAlloc.html}},
  note =	 {Accessed: 2025-05-30}
}

@Inbook{Spolsky:LeakyAbstractions,
  author =       "Spolsky, Joel",
  title =        "The Law of Leaky Abstractions",
  bookTitle =    "Joel on Software: And on Diverse and Occasionally
                  Related Matters That Will Prove of Interest to
                  Software Developers, Designers, and Managers, and to
                  Those Who, Whether by Good Fortune or Ill Luck, Work
                  with Them in Some Capacity",
  year =         "2004",
  publisher =    "Apress",
  address =      "Berkeley, CA",
  pages =        "197--202",
  isbn =         "978-1-4302-0753-5",
  doi =          "10.1007/978-1-4302-0753-5_26",
  url =          "https://doi.org/10.1007/978-1-4302-0753-5_26"
}

@misc{torvalds2007cpuset,
  author       = {Torvalds, Linus},
  title        = {{Re: [RFC] [PATCH] cpuset operations causes Badness at mm/slab.c:777 warning}},
  howpublished = {Linux Kernel Mailing List (LKML)},
  year         = {2007},
  month        = jun,
  day          = {1},
  note         = {Message ID and full thread available via LKML archive},
  url          = {https://lkml.org/lkml/2007/6/1/440}
}

@article{Askarov2011fb, 
 author    = {Aslan Askarov and Andrew C. Myers},
  title     = {Attacker Control and Impact for Confidentiality and Integrity},
  journal   = {Logical Methods in Computer Science},
  volume    = {7},
  number    = {3},
  year      = {2011},
  doi       = {10.2168/LMCS-7(3:17)2011},
  url       = {https://lmcs.episciences.org/987},
}

@inproceedings{banerjee2008expressive,
  title={Expressive declassification policies and modular static enforcement},
  author={Banerjee, Anindya and Naumann, David A and Rosenberg, Stan},
  booktitle={2008 IEEE Symposium on Security and Privacy (sp 2008)},
  pages={339--353},
  year={2008},
  organization={IEEE},
  doi={10.1109/SP.2008.20}
}

@inproceedings{ahmadian2022dynamic,
  title={Dynamic policies revisited},
  author={Ahmadian, Amir M and Balliu, Musard},
  booktitle={2022 IEEE 7th European Symposium on Security and Privacy (EuroS\&P)},
  pages={448--466},
  year={2022},
  organization={IEEE},
  doi={10.1109/EuroSP53844.2022.00035}
}

@article{memarian2016into,
  title={Into the depths of C: elaborating the de facto standards},
  author={Memarian, Kayvan and Matthiesen, Justus and Lingard, James and Nienhuis, Kyndylan and Chisnall, David and Watson, Robert NM and Sewell, Peter},
  journal={ACM SIGPLAN Notices},
  volume={51},
  number={6},
  pages={1--15},
  year={2016},
  publisher={ACM New York, NY, USA},
  doi={10.1145/2908080.2908081}
}

@article{beck2024twophase,
  title={A two-phase infinite/finite low-level memory model: Reconciling integer--pointer casts, finite space, and undef at the {LLVM IR} level of abstraction},
  author={Beck, Calvin and Yoon, Irene and Chen, Hanxi and Zakowski, Yannick and Zdancewic, Steve},
  journal={Proceedings of the ACM on Programming Languages},
  volume={8},
  number={ICFP},
  pages={789--817},
  year={2024},
  publisher={ACM New York, NY, USA},
  doi={10.1145/3674652}
}

@book{Schoepe2016bh, 
  year      = {2016}, 
  title     = {Explicit Secrecy: A Policy for Taint Tracking}, 
  author    = {Schoepe, Daniel and Balliu, Musard and Pierce, Benjamin C and Sabelfeld, Andrei}, 
  isbn      = {978-1-5090-1751-5}, 
  series    = {2016 {IEEE} European Symposium on Security and Privacy ({EuroS}\&P)}, 
  publisher = {{IEEE}}, 
  language  = {English}, 
  doi       = {10.1109/eurosp.2016.14}
}

@Book{KR1988:c,
  author =       "Brian W. Kernighan and Dennis M. Ritchie",
  title =        "The {C} Programming Language",
  publisher =    "Prentice Hall",
  year =         1988,
  edition =      "Second"
}

@TechReport{iso-c,
  author =       "ISO/IEC",
  type =         "Standard",
  title =        "{ISO/IEC 9899:202y N3685: Information technology ---
                  Programming languages --- C}",
  institution =  "International Organization for Standardization,
                  Geneva, Switzerland",
  shorttitle =   "ISO/IEC 9899:202y --- N3685 (working draft)",
  year =         2025,
  note =         "Working draft",
  url = "https://www.open-std.org/jtc1/sc22/wg14/www/docs/n3685.pdf"
}

@Misc{lea:1996:dlmalloc,
  author =       "Doug Lea",
  year =         2012,
  note =         "Source code for \texttt{dlmalloc()} version~2.8.6.",
  url = "https://gee.cs.oswego.edu/pub/misc/malloc.c"
}

@inproceedings{zdancewic2001robust,
  title={Robust Declassification.},
  author={Zdancewic, Steve and Myers, Andrew C},
  booktitle={14th {IEEE} Computer Security Foundations Workshop (CSFW~2001)},
  volume={1},
  pages={15--23},
  year={2001},
  doi={10.1109/CSFW.2001.930133}
}

@inproceedings{cecchetti2017nonmalleable,
  title={Nonmalleable information flow control},
  author={Cecchetti, Ethan and Myers, Andrew C and Arden, Owen},
  booktitle={Proceedings of the 2017 ACM SIGSAC Conference on Computer and Communications Security},
  pages={1875--1891},
  year={2017},
  doi={10.1145/3133956.3134054}
}

@inproceedings{li2005downgrading,
  title={Downgrading policies and relaxed noninterference},
  author={Li, Peng and Zdancewic, Steve},
  booktitle={Proceedings of the 32nd ACM SIGPLAN-SIGACT symposium on Principles of programming languages},
  pages={158--170},
  year={2005},
  doi={10.1145/1040305.1040319}
}

@inproceedings{watson2015cheri,
  title={CHERI: A hybrid capability-system architecture for scalable software compartmentalization},
  author={Watson, Robert NM and Woodruff, Jonathan and Neumann, Peter G and Moore, Simon W and Anderson, Jonathan and Chisnall, David and Dave, Nirav and Davis, Brooks and Gudka, Khilan and Laurie, Ben and others},
  booktitle={2015 IEEE Symposium on Security and Privacy},
  pages={20--37},
  year={2015},
  organization={IEEE},
  doi={10.1109/SP.2015.9}
}

@inproceedings {yang:usenix:2017:dse:harmful,
author = {Zhaomo Yang and Brian Johannesmeyer and Anders Trier Olesen and Sorin Lerner and Kirill Levchenko},
title = {Dead Store Elimination (Still) Considered Harmful},
booktitle = {26th USENIX Security Symposium (USENIX Security 17)},
year = {2017},
isbn = {978-1-931971-40-9},
address = {Vancouver, BC},
pages = {1025--1040},
OPTurl = {https://www.usenix.org/conference/usenixsecurity17/technical-sessions/presentation/yang},
publisher = {USENIX Association},
month = aug
}

@INPROCEEDINGS{dsilva:correctness-security-gap,
  author={D'Silva, Vijay and Payer, Mathias and Song, Dawn},
  booktitle={2015 IEEE Security and Privacy Workshops}, 
  title={The Correctness-Security Gap in Compiler Optimization}, 
  year={2015},
  volume={},
  number={},
  pages={73-87},
  doi={10.1109/SPW.2015.33}}

@INPROCEEDINGS{Cecchetti:NMIFC:progress,
  author={Cecchetti, Ethan},
  booktitle={2025 IEEE 38th Computer Security Foundations Symposium (CSF)}, 
  title={Nonmalleable Progress Leakage}, 
  year={2025},
  volume={},
  number={},
  pages={537-552},
  doi={10.1109/CSF64896.2025.00029}}

@inproceedings{chong2005language,
  title={Language-based information erasure},
  author={Chong, Stephen and Myers, Andrew C},
  booktitle={18th IEEE Computer Security Foundations Workshop (CSFW'05)},
  pages={241--254},
  year={2005},
  organization={IEEE},
  doi={10.1109/CSFW.2005.19}
}

@inproceedings{askarov2015cryptographic,
  title={Cryptographic enforcement of language-based information erasure},
  author={Askarov, Aslan and Moore, Scott and Dimoulas, Christos and Chong, Stephen},
  booktitle={2015 IEEE 28th Computer Security Foundations Symposium},
  pages={334--348},
  year={2015},
  organization={IEEE},
  doi={10.1109/CSF.2015.30}
}

@inproceedings{hunt:sands:2008just:forget:it,
  title={Just forget it--the semantics and enforcement of information erasure},
  author={Hunt, Sebastian and Sands, David},
  booktitle={European Symposium on Programming},
  pages={239--253},
  year={2008},
  organization={Springer},
  doi={10.1007/978-3-540-78739-6_19}
}

@inproceedings{ruef2019achieving,
  title={Achieving safety incrementally with Checked C},
  author={Ruef, Andrew and Lampropoulos, Leonidas and Sweet, Ian and Tarditi, David and Hicks, Michael},
  booktitle={International Conference on Principles of Security and Trust},
  pages={76--98},
  year={2019},
  organization={Springer},
  doi={10.1007/978-3-030-17138-4_4}
}

@article{iris-mswasm,
  title={Iris-MSWasm: Elucidating and mechanising the security invariants of memory-safe WebAssembly},
  author={Legoupil, Maxime and Rousseau, June and Georges, A{\"\i}na Linn and Pichon-Pharabod, Jean and Birkedal, Lars},
  journal={Proceedings of the ACM on Programming Languages},
  volume={8},
  number={OOPSLA2},
  pages={304--332},
  year={2024},
  publisher={ACM New York, NY, USA},
  doi={10.1145/3689722}
}

@inproceedings{jangda2019:not-so-fast,
  title={Not so fast: Analyzing the performance of $\{$WebAssembly$\}$ vs. native code},
  author={Jangda, Abhinav and Powers, Bobby and Berger, Emery D and Guha, Arjun},
  booktitle={2019 USENIX Annual Technical Conference (USENIX ATC 19)},
  pages={107--120},
  year={2019}
}

@misc{PolyBenchC,
  author = {Pouchet, Louis-No{\"e}l and Yuki, Tomofumi},
  title = {{PolyBench/C} 4.2.1},
  howpublished = {\url{http://polybench.sourceforge.net}},
  year = {2016},
  note={Accessed: March 2026}
}

@inproceedings{acay2021viaduct,
  title={Viaduct: an extensible, optimizing compiler for secure distributed programs},
  author={Acay, Co{\c{s}}ku and Recto, Rolph and Gancher, Joshua and Myers, Andrew C and Shi, Elaine},
  booktitle={Proceedings of the 42nd ACM SIGPLAN International Conference on Programming Language Design and Implementation},
  pages={740--755},
  year={2021},
  doi={10.1145/3453483.3454074}
}

@inproceedings{ferraiuolo2018hyperflow,
  title={HyperFlow: A processor architecture for nonmalleable, timing-safe information flow security},
  author={Ferraiuolo, Andrew and Zhao, Mark and Myers, Andrew C and Suh, G Edward},
  booktitle={Proceedings of the 2018 ACM SIGSAC Conference on Computer and Communications Security},
  pages={1583--1600},
  year={2018},
  doi={10.1145/3243734.3243743}
}

@inproceedings{zagieboylo2019using,
  title={Using information flow to design an ISA that controls timing channels},
  author={Zagieboylo, Drew and Suh, G Edward and Myers, Andrew C},
  booktitle={2019 IEEE 32nd Computer Security Foundations Symposium (CSF)},
  pages={272--27215},
  year={2019},
  organization={IEEE},
  doi={10.1109/CSF.2019.00026}
}

@TechReport{cheri-c-guide,
  author = {Watson, Robert N. M. and Richardson, Alexander and Davis, Brooks
            and Baldwin, John and Chisnall, David and Clarke, Jessica
            and Filardo, Nathaniel and Moore, Simon W.
            and Napierala, Edward and Sewell, Peter and Neumann, Peter G.},
  title = {{CHERI C/C++ Programming Guide}},
  year = 2020,
  month = jun,
  institution = {University of Cambridge, Computer Laboratory},
  number = {UCAM-CL-TR-947},
  doi = {10.48456/tr-947}
}

@article{swasey2017robust,
  title={Robust and compositional verification of object capability patterns},
  author={Swasey, David and Garg, Deepak and Dreyer, Derek},
  journal={Proceedings of the ACM on Programming Languages},
  volume={1},
  number={OOPSLA},
  pages={1--26},
  year={2017},
  publisher={ACM New York, NY, USA},
  doi={10.1145/3133913}
}

@inproceedings{devriese2016reasoning,
  title={Reasoning about object capabilities with logical relations and effect parametricity},
  author={Devriese, Dominique and Birkedal, Lars and Piessens, Frank},
  booktitle={2016 IEEE European Symposium on Security and Privacy (EuroS\&P)},
  pages={147--162},
  year={2016},
  organization={IEEE},
  doi={10.1109/EuroSP.2016.22}
}

@article{georges2024cerise,
  title={Cerise: Program verification on a capability machine in the presence of untrusted code},
  author={Georges*, A{\"\i}na Linn and Gu{\'e}neau*, Arma{\"e}l and Van Strydonck, Thomas and Timany, Amin and Trieu*, Alix and Devriese, Dominique and Birkedal, Lars},
  journal={Journal of the ACM},
  volume={71},
  number={1},
  pages={1--59},
  year={2024},
  publisher={ACM New York, NY},
  doi={10.1145/3623510}
}

@inproceedings{zhang2012:pldi:language-based-mitigation,
  title={Language-based control and mitigation of timing channels},
  author={Zhang, Danfeng and Askarov, Aslan and Myers, Andrew C},
  booktitle={Proceedings of the 33rd ACM SIGPLAN conference on Programming Language Design and Implementation},
  pages={99--110},
  year={2012},
  doi={10.1145/2254064.2254078}
}

@article{zhang2015hardware,
  title={A hardware design language for timing-sensitive information-flow security},
  author={Zhang, Danfeng and Wang, Yao and Suh, G Edward and Myers, Andrew C},
  journal={Acm Sigplan Notices},
  volume={50},
  number={4},
  pages={503--516},
  year={2015},
  publisher={ACM New York, NY, USA},
  doi={10.1145/2694344.2694372}
}

@inproceedings{besson2015concrete,
  title={A concrete memory model for CompCert},
  author={Besson, Fr{\'e}d{\'e}ric and Blazy, Sandrine and Wilke, Pierre},
  booktitle={International Conference on Interactive Theorem Proving},
  pages={67--83},
  year={2015},
  organization={Springer},
  doi={10.1007/978-3-319-22102-1_5}
}

@book{alvim2020:qif:book,
  title={The science of quantitative information flow},
  author={Alvim, M{\'a}rio S and Chatzikokolakis, Konstantinos and McIver, Annabelle and Morgan, Carroll and Palamidessi, Catuscia and Smith, Geoffrey},
  year={2020},
  publisher={Springer},
  doi={10.1007/978-3-319-96131-6}
}

@Article{OHearn:popl2020:buglogic,
  author =       "Peter W. O'Hearn",
  title =        "Incorrectness logic",
  journal =      "Proc. {ACM} Program. Lang.",
  year =         2020,
  volume =       4,
  number =       "POPL",
  pages =        "10:1--10:32",
  doi =          "10.1145/3371078"
}

@Article{LeRVBDO22:oopsla2022:bigbugs,
  author       = "Quang Loc Le and Azalea Raad and Jules Villard and
                  Josh Berdine and Derek Dreyer and Peter W. O'Hearn",
  title        = "Finding real bugs in big programs with incorrectness
                  logic",
  journal      = "Proc. {ACM} Program. Lang.",
  year         = 2022,
  volume       = 6,
  number       = "{OOPSLA1}",
  pages        = "1--27",
  doi          = "10.1145/3527325"
}

@Article{ZilbersteinDS:oopsla2023:ol,
  author       = "Noam Zilberstein and Derek Dreyer and Alexandra
                  Silva",
  title        = "Outcome Logic: {A} Unifying Foundation for
                  Correctness and Incorrectness Reasoning",
  journal      = "Proc. {ACM} Program. Lang.",
  year         = 2023,
  volume       = 7,
  number       = "{OOPSLA1}",
  pages        = "522--550",
  doi          = "10.1145/3586045"
}

@InProceedings{LamportS:csf2021:hypertla,
  author =       "Leslie Lamport and Fred B. Schneider",
  title =        "Verifying Hyperproperties With {TLA}",
  year =         2021,
  booktitle =    "Proceedings of the 34th {IEEE} Computer Security
                  Foundations Symposium (CSF~2021)",
  pages =        "1--16",
  doi =          "10.1109/CSF51468.2021.00012"
}

@inproceedings{AroraHLLP:spin2022:probhyper,
  title =        "Statistical Model Checking for Probabilistic
                  Hyperproperties of Real-Valued Signals",
  author =       "Shiraj Arora and Ren{\'e} Rydhof Hansen and Kim
                  Guldstrand Larsen and Axel Legay and Danny
                  B{\o}gsted Poulsen",
  year =         "2022",
  month =        aug,
  series =       "Lecture Notes in Computer Science",
  publisher =    "Springer",
  pages =        "61--78",
  booktitle =    "Proceedings of the 28th International Symposium on
                  Model Checking of Software (SPIN~2022)",
  doi =          "10.1007/978-3-031-15077-7_4",
}

@Article{DardinierLM:oopsla2024:hypra,
  author       = "Thibault Dardinier and Anqi Li and Peter
                  M{\"{u}}ller",
  title        = "Hypra: {A} Deductive Program Verifier for Hyper
                  Hoare Logic",
  journal      = "Proc. {ACM} Program. Lang.",
  year         = 2024,
  volume       = 8,
  number       = "{OOPSLA2}",
  pages        = "1279--1308",
  doi          = "10.1145/3689756"
}

@software{notac:zenodo,
  author       = {Hansen, René Rydhof and
                  Stenbaek Larsen, Andreas and
                  Askarov, Aslan},
  title        = {Notac interpreter and examples},
  month        = mar,
  year         = 2026,
  publisher    = {Zenodo},
  doi          = {10.5281/zenodo.19076610},
  url          = {https://doi.org/10.5281/zenodo.19076610},
}
\ifnoappendix\else
\appendix
\onecolumn
The appendix contains extended semantic rules, a more detailed review
of the curious allocator, of the translation from \otherlang{}\ to
\thelanguage, and the translation theorem.

\section{Semantic Rules}
\Cref{notac:semantics:expressions}, \Cref{notac:semantics:commands},
and \Cref{notac:semantics:commands:ext} present the semantic rules
for \thelanguage.

\begin{figure}
\framebox{
\begin{mathpar}
  \inferrule[Exp-Const]{\ }{\ejudg{n}{n}}
  \and
  \inferrule[Exp-Var]{%
    v = H(E(x)) 
  }{\ejudg{x}{v}}
  \and
  \inferrule[Exp-Null]{~}{\ejudg{\nullexpr}{\stratapply{\allocstrategy}{\NULL}}}
  \and
  \inferrule[Exp-Addressof]{\ }{\ejudg{\addrof{x}}{E(x)}}
  \and
  \inferrule[Exp-Deref]{%
    \ejudg{e}{a} \and
    v = H(a) 
  }{\ejudg{\deref{e}}{v}}
  \and
  \inferrule[Exp-Binop]{\ejudg{e_i}{n_i}, \quad i = 1,2}{
    \ejudg{\binop{e_1}{\mathit{bop}}{e_2}}{n_1 \mathop{\mathit{bop}} n_2}}
  \\
  \inferrule[Lval-Var]{
    E(x) = a
  }{
    \ljudg{x}{a}
  }
  \and 
  \inferrule[Lval-Deref]{
    \ejudg{e}{a}
  }{
    \ljudg{\lvalptr{e}}{a}
  }
\end{mathpar}}
\caption{\thelanguage\ semantics of expressions and lvals}\label{notac:semantics:expressions}
\end{figure}

\begin{figure}
\framebox{
\begin{mathpar}
  \inferrule[C-Obs]{
    \ejudg{e}{v}
  }{
    \invmch{E,\allocstrategy}\cjudgtr[H,\allocstate]{\obs{e}}{\evobs{v}}{H,\allocstate}
  }
  \and
  \inferrule[C-Assign]{%
    \ljudg{\lval}{a} \\
    \ejudg{e}{v} \\ 
    a \in \dom{H}
   }{%
    \invmch{E,\allocstrategy}\cjudg[H,\allocstate]{\assg{\lval}{e}}{\upd{H}{a}{v}, \allocstate}}
  \and
  \inferrule[C-Cast]{%
    \ljudg{\lval}{a} \\
    \ejudg{e}{v} \\
    a \in \dom{H}
  }{
    \invmch{E,\allocstrategy}
    \cjudgtr[H,\allocstate]{\lvalcast{\lval}{e}}{\evcast{v}}{\upd{H}{a}{v},\allocstate}
  }
  \and
   \inferrule[C-Malloc]{%
     \ejudg{e}{n} \\
     n \in \domain{Size} \\
     (H',\allocstate',a) = \stratapply{\allocstrategy}{\MALLOC{H,\allocstate,n}} \\
     \ljudg{\lval}{a_{\mathit{lval}}} \\
     a_{\mathit{lval}} \in \dom{H'} \\
     \event = {\left\lbrace
         \begin{array}{@{}ll@{}}
           \evmalloc{n}{a} & a \neq \stratapply{\allocstrategy}{\NULL} \\
           \evmallocfail{n} & a = \stratapply{\allocstrategy}{\NULL}
         \end{array}
       \right.
  }
}{
  \invmch{E, \allocstrategy}
  \cjudgtr[H,\allocstate]{\lvalmalloc{\lval}{e}}
                         {\event}{\upd{H'}{a_{\mathit{lval}}}{a},\allocstate'}
}
  \and
  \inferrule[C-Free]{%
    \ejudg{e}{a} \\
    (H',\allocstate') = \stratapply{\allocstrategy}{\FREE{H,\allocstate, a}}\\
  }{
    \invmch{E, \allocstrategy}
    \cjudgtr[H, \allocstate]{\free{e}}{\evfree{a}}{H',\allocstate'}
  }
\end{mathpar}
}
\caption{\thelanguage\ semantics of commands, selected rules}\label{notac:semantics:commands}
\end{figure}

\begin{figure}
\framebox{
\begin{mathpar}
\inferrule[C-Skip]{~}{
  \invmch{E, \allocstrategy}
  \cjudg[H, \allocstate]{\cmdskip}{H, \allocstate}
}\and 
\inferrule[C-Seq-1]{
  \invmch{E, \allocstrategy}
  \cjudgtr[H,\allocstate]{c_1}{\event}{H', \allocstate'}
}{
  \invmch{E, \allocstrategy}
  \cjudgtrext[H, \allocstate]{\sequence{c_1}{c_2}}{\event}{c_2}{H', \allocstate'}
}
\and 
\inferrule[C-Seq-2]{
  \invmch{E,\allocstrategy}
  \cjudgtrext[H, \allocstate]{c_1}{\event}{c'_1}{H', \allocstate'}
}{
  \invmch{E, \allocstrategy}
  \cjudgtrext[H, \allocstate]{\sequence{c_1}{c_2}}{\event}{\sequence{c'_1}{c_2}}
                             {H', \allocstate'}
}
\and 
\inferrule[C-If]{
  \ejudg{e}{v} \and v \neq 0 \implies i = 1 \and v = 0 \implies i = 2
}{
  \invmch{E, \allocstrategy}
  \cjudgtrext[H, \allocstate]{\cond{e}{c_1}{c_2}}{}{c_i}{H, \allocstate}
}
\and 
\inferrule[C-While]{~}{
  \invmch{E, \allocstrategy}
  \cjudgtrext[H, \allocstate]{\while{e}{c}}{}{
    \cond{e}{ \sequence{c}{\while{e}{c}} }{\cmdskip}
  }{H, \allocstate}
}
\end{mathpar}
}
\caption{\thelanguage\ semantics of commands, extended rules}%
\label{notac:semantics:commands:ext}
\end{figure}

\section{Details of the translation}\label{app:memsafe-translation}
Figures~\ref{fig:memsafe:to:notac:translation:expr}
and~\ref{fig:memsafe:to:notac:translation:cmd} present the details of
the translation. It embeds out-of-memory checks after each allocation
and uses a dedicated global variable $\oom$ to track whether any of
the allocations have failed. A guard macro $\memguard{c}$ executes $c$
only if $\oom$ is not set; this is needed because \thelanguage{} does
not have an \texttt{exit} command. The allocation also initializes the
memory to 0 with another dedicated global variable $\alloci$. Finally,
the while loops must also consult the guard; their execution is also
guarded; the translation ensures that the actual loop guard is not
evaluated (as it may deviate) when $\oom$ is true, using a fresh (for
each syntactically occurring while-statement) variable $\wguard$.

\begin{figure}
  \centering
  \framebox{{
    \begin{minipage}{0.48\linewidth}
      \centering
      $\begin{array}{rcl}
        \MemsafeToNotacDef{x\in \mathsf{var}}{x}\\
        \MemsafeToNotacDef{n\in \mathbb{Z}}{n \in \mathbb{Z}}\\
      \end{array}$
    \end{minipage}
    \begin{minipage}{0.48\linewidth}
      \centering
      $\begin{array}{rcl}
      \MemsafeToNotacDef{e_1 \oplus e_2}{
          \MemsafeToNotac{e_1}\, bop\, \MemsafeToNotac{e_2}
        }\\
        \MemsafeToNotacDef{\mathsf{nil}}{\nullexpr}
      \end{array}$
    \end{minipage}
  }}
  \caption{\otherlang{} expressions to \thelanguage{} translation.%
  \label{fig:memsafe:to:notac:translation:expr}}
\end{figure}

\begin{figure}
  \framebox{
    \begin{minipage}[t]{0.55\linewidth}
      $\begin{array}{rcl}
        \MemsafeToNotacDef{\mathsf{skip}}{\cmdskip}\\[0.7em]
        \MemsafeToNotacDef{c_1\, ; \, c_2}{
          \sequence{\MemsafeToNotac{c_1}}{\MemsafeToNotac{c_2}}
        }\\[0.7em]
        \MemsafeToNotacDef{
          \mathsf{if} \, e \, \mathsf{then} \, c_1 \, \mathsf{else} \, c_2
        }{
          \memguard{\cond{\MemsafeToNotac{e}}{\MemsafeToNotac{c_1}}{\MemsafeToNotac{c_2}}}
        }\\[0.9em]
        \MemsafeToNotacDef{
          \mathsf{while}\, e \, \mathsf{do} \, c \, \mathsf{end}
        }{
          \begin{array}{@{}l@{}}
            \assg{\wguard}{\codeparens{\oom \, \texttt{==} \, 0}};\\
            \code{while}\codeparens{\wguard} \,\{\\
            \quad \cond{\MemsafeToNotac{e}}{\MemsafeToNotac{c}}{\assg{\wguard}{0}};\\
            \quad \assg{\wguard}{\codeparens{\oom \, \texttt{==} \, 0} * \wguard}\\
            \}, \wguard\ \textrm{is fresh}
          \end{array}
        }\\\\[-0.5em]
        \MemsafeToNotacDef{
          x \leftarrow e
        }{
          \memguard{\assg{\MemsafeToNotac{x}}{\MemsafeToNotac{e}}}
        }\\[0.7em]
        \MemsafeToNotacDef{
          x \leftarrow [e]
        }{
          \memguard{\assg{\MemsafeToNotac{x}}{\deref{\MemsafeToNotac{e}}}}
        }\\[0.7em]
        \MemsafeToNotacDef{
          [e_1] \leftarrow e_2
        }{
          \memguard{\ptrassg{\MemsafeToNotac{e_1}}{\MemsafeToNotac{e_2}}}
        }
      \end{array}$
    \end{minipage}\quad 
    \begin{minipage}[t]{0.40\linewidth}
      $\begin{array}{rcl}
        \multicolumn{3}{l}{\MemsafeToNotac{x \leftarrow \mathsf{alloc}(e)} \defn}\\
        \multicolumn{3}{r}{
                   \textbf{G}\left(
          \begin{array}{@{}l@{}}
            \assg{\alloci}{\MemsafeToNotac{e}};\\
            \malloc{\MemsafeToNotac{x}}{\alloci};\\
            \cnd{\MemsafeToNotac{x}\,\texttt{==}\,\nullexpr}{\assg{\oom}{1}}\\
            \code{else}\;\{\\
            \quad \code{while}\codeparens{\alloci \geq 0}\;\{\\
            \qquad \ptrassg{x+\alloci}{0};\\
            \qquad \assg{\alloci}{\alloci - 1}\\
            \quad\}\\
            \}
          \end{array}\right)
        }\\\\[3em]
        \memguard{\notac{c}} &\triangleq& \notac{\cond{\oom}{\cmdskip}{c}}
      \end{array}$
    \end{minipage}
  }
  \caption{\otherlang{} commands to \thelanguage{} translation.}%
  \label{fig:memsafe:to:notac:translation:cmd}
\end{figure}

\begin{figure}%
  \label{fig:memsafe:syntax}
  \centering
\framebox{\parbox{\linewidth}{
    \centering
    \begin{align*}
      \oplus ::=\; &+ \mid \times \mid - \mid {=} \mid {\leq} &&\text{(operators)}\\
      e ::=\; &x \in \mathsf{var} \mid n \in \mathbb{Z} \mid \msnil \mid e_1 \oplus e_2 &&\text{(expressions)}\\
      c ::=\; &\msskip \mid \msseq{c_1}{c_2} \mid {\msif\, e\, \msthen \, c_1 \, \mselse\, c_2} \mid
          {\mswhile \, e \, \msdo \, c \, \msend} &&\text{(commands)}\\
      \mid\; &x \leftarrow e \mid x \leftarrow [e] \mid [e_1] \leftarrow e_2 
        \mid x \leftarrow \msalloc(e)
    \end{align*}
    \begin{align*}
      s \in \msstate{S} &\triangleq \msstate{L} \times \msstate{M} &&\text{(states)}\\
      l \in \msstate{L} &\triangleq {\textsf{var}} \msfinmap \msstate{V} &&\text{(local stores)}\\
      m \in \msstate{M} &\triangleq \mathbb{I} \times \mathbb{Z} \times \mathbb{Z} \msfinmap \msstate{V} &&\text{(heaps)}\\
      v \in \msstate{V} &\triangleq \mathbb{Z} \uplus \{\msnil\} \uplus \mathbb{I} \times \mathbb{Z} \times \mathbb{Z} &&\text{(values)}\\
      \msstate{O} &\triangleq \msstate{S} \uplus \{\mserror\} &&\text{(outcomes)}
    \end{align*}
    \begin{align*}
      \mathbb{I} &\triangleq \text{some countably infinite set}\\
      X \msfinmap Y &\triangleq \text{partial functions }X \text{ with finite domain}
    \end{align*}
  }
}
\caption{\otherlang{} syntax and program states}
\end{figure}
\begin{figure}%
  \label{fig:memsafe:eval:expression}
  \framebox{\parbox{\linewidth}{
      \centering
      \begin{align*}
        \mseval[l,m]{x} &\triangleq l(x)\\
        \mseval{n} &\triangleq n\\
        \mseval{\msnil} &\triangleq \msnil\\
        \mseval{\msadd{e_1}{e_2}} &\triangleq \mscases{
                                    \msadd{n_1}{n_2}, & \text{if } \mseval{e_1}= n_1 \text{ and } \mseval{e_2} = n_2\\
        \msptr{i}{b}{\msadd{n_1}{n_2}}, & \text{if } \mseval{e_1} = \msptr{i}{b}{n_1} \text{ and } \mseval{e_2} = n_2\\
                        &\text{or } \mseval{e_1} = n_1 \text{ and } \mseval{e_2} = \msptr{i}{b}{n_2}}\\
        \mseval{\mssub{e_1}{e_2}} &\triangleq \mscases{
                                    \mssub{n_1}{n_2}, & \text{if } \mseval{e_1}= n_1 \text{ and } \mseval{e_2} = n_2\\
        \msptr{i}{b}{\mssub{n_1}{n_2}}, & \text{if } \mseval{e_1} = \msptr{i}{b}{n_1} \text{ and } \mseval{e_2} = n_2}\\
        \mseval{\msmul{e_1}{e_2}} &\triangleq \mscase{\msmul{n_1}{n2}, & \text{if } \mseval{e_1} = n_1 \text{ and } \mseval{e_2} = n_2}\\
        \mseval{\mseq{e_1}{e_2}} &\triangleq \mscases{
          \mseq{n_1}{n_2}, &\text{if } \mseval{e_1} = n_1 \text{ and } \mseval{e_2} = n_2\\
          \mseq{p_1}{p_2}, &\text{if } \mseval{e_1} = p_1 \text{ and } \mseval{e_2} = p_2, 
                             \text{ where } p_1,p_2 \in \{\msnil{}\} \uplus \mathbb{I} \times \mathbb{Z} \times \mathbb{Z}\\
                             &\text{and } p_j = (i,b,n) \Rightarrow 0 \leq n < b, \text{ for } j = 1,2
        }\\
        \mseval{\msleq{e_1}{e_2}} &\triangleq \mscase{\msleq{n_1}{n2}, & \text{if } \mseval{e_1} = n_1 \text{ and } \mseval{e_2} = n_2}
      \end{align*}
    }
  }
  \caption{\otherlang{} expression evaluation}
\end{figure}
\begin{figure}%
  \label{fig:memsafe:eval:aux}
  \framebox{\parbox{\linewidth}{
      \centering
      \begin{align*}
        \msbind(f,\bot) &\triangleq \bot\\
        \msbind(f,\mserror) &\triangleq \mserror\\
        \msbind(f,(I,l,m)) &\triangleq \mscases{
          (I \cup I', l', m'), & \text{if } f(l,m) = (I',l',m')\\
          \mserror, & \text{if } f(l,m) = \mserror\\
          \bot, & \text{otherwise}
        }\\
        \msif(b,x,y) &\triangleq \mscases{
          x, & \text{if } b \neq 0 \text{ and } b \in \mathbb{Z}\\
          y, & \text{if } b = 0 \text{ and } b \in \mathbb{Z}\\
          \mserror, &\text{otherwise}
        } 
      \end{align*}
    }}
  \caption{\otherlang{} auxiliary operators}
\end{figure}
\begin{figure}%
  \label{fig:memsafe:eval:command}
  \framebox{\parbox{\linewidth}{
      \centering
      \begin{align*}
        \msevalcs{\msskip} &\triangleq (\emptyset, l, m)\\
        \msevalcs{\msseq{c_1}{c_2}} &\triangleq \msbind(\msevalc{c_2}, \msevalcs{c_1})\\
        \msevalcs{\msif\, e \, \mathsf{then} \, c_1 \, \mathsf{else} \, c_2} &\triangleq \msif(\mseval[l,m]{e}, \msevalcs{c_1}, \msevalcs{c_2})\\
        \msevalc{\mswhile \, e \, \msdo \, c \, \msend} &\triangleq 
           \mathsf{fix}(\lambda f (l,m).\, \msif(\mseval[l,m]{e},\msbind(\msevalc{c}, f(l,m)), (\emptyset,l,m)))\\
        \msevalcs{x \leftarrow e} &\triangleq (\emptyset, l[x \mapsto \mseval[l,m]{e}], m)\\
        \msevalcs{x \leftarrow [e]} &\triangleq \mscases{
          (\emptyset, l[x \mapsto v], m), &\text{if } \mseval[l,m]{e} = (i,b,n) \text{ and } m(i,b,n) = v\\
          \mserror, &\text{otherwise}
        }\\
        \msevalcs{[e_1] \leftarrow e_2} &\triangleq \mscases{
                                          (\emptyset, l, m[(i,b,n) \mapsto \mseval[l,m]{e_2}]), &\text{if } \mseval[l,m]{e_1} = (i,b,n)\\
                           &\text{and } m(i,b,n) \neq \bot\\
          \mserror, &\text{otherwise}
        }\\
        \msevalcs{x \leftarrow \msalloc(e)} &\triangleq \mscases{
                                              (\{i\}, l[x \mapsto (i,0)], m[(i,k) \mapsto 0 \mid 0 \leq k < n]), &\text{if } \mseval{e_1} = n\\
                           &\text{and } i = \mathsf{fresh}(\mathsf{ids}(i,m))\\
          \mserror, &\text{otherwise}
        }
      \end{align*}
    }}
  \caption{\otherlang{} command evaluation}
\end{figure}

\begin{definition}[\otherlang{} to \thelanguage{} state compatibility]%
  For a mapping $\addrmap$ from \otherlang{} pointers to \thelanguage{} pointers,
  a \otherlang{} state $(l, m)$ and \thelanguage{}
  variable environment $E$, heap $H$, and allocator strategy $\allocstrategy$,
  are compatible, written $\statecompat{(l,m)}{(E,H,\allocstrategy)}$,
  when the following properties are satisfied:
  \begin{itemize}
  \item $\dom{l} \subseteq \dom{E}$
  \item $\dom{m} \subseteq \dom{\addrmap}$
  \item $\forall \memsafe{(i,b,n)} \in \dom{m}, j \in \mathbb{Z}. \,
    \memsafe{(i,b,n+j)} \in \dom{m} \Rightarrow \addrmap(\memsafe{i,b,n + j}) =
    \addrmap(\memsafe{i,b,n}) + j$
  \item $\forall x \in \dom{l}. \, l(x) = \memsafe{n} \Leftrightarrow H(E(x)) = n$
  \item $\forall x \in \dom{l}. \, l(x) = \memsafe{\msnil} \Leftrightarrow H(E(x)) =
    \stratapply{\allocstrategy}{\NULL}$
  \item $\forall x \in \dom{l}. \, l(x) = \memsafe{(i,b,n)} \Leftrightarrow H(E(x)) =
    \addrmap(\memsafe{i,b,n})$
  \item $\forall \memsafe{(i,b,n)} \in \dom{m}. \, m(\memsafe{i,b,n}) = \memsafe{n}
    \Leftrightarrow H(\addrmap(\memsafe{i,b,n})) = n$
  \item $\forall \memsafe{(i,b,n)} \in \dom{m}. \, m(\memsafe{i,b,n}) = \memsafe{\msnil}
    \Leftrightarrow H(\addrmap(\memsafe{i,b,n})) = \stratapply{\allocstrategy}{\NULL}$
  \item $\forall \memsafe{(i,b,n)} \in \dom{m}. \, m(\memsafe{i,b,n}) = \memsafe{(i',b',n')}
    \Leftrightarrow H(\addrmap(\memsafe{i,b,n})) = \addrmap(\memsafe{i',b',n'})$
  \end{itemize}
\end{definition}
With this, we can now show that translating a \otherlang{} program,
which is memory safe by design, to \thelanguage{}, the resulting
program will satisfy gradual allocator independence.
\begin{lemma}[Semantic correctness of expression translation]%
  \label{lem:correct:expr:trans}
  Let $\memsafe{e}$ be an expression and $(l,m)$ a state in \otherlang{}.
  Then for map $\addrmap$, \thelanguage{} environment $E$,
  heap $H$, and allocator strategy $\allocstrategy$,
  where $\statecompat{(l,m)}{(E,H,\allocstrategy)}$,
  the following holds:
  \begin{itemize}
  \item If $\mseval[l,m]{\memsafe{e}} = \memsafe{n}$
    then $\ejudg{\MemsafeToNotac{e}}{n}$
  \item If $\mseval[l,m]{\memsafe{e}} = \memsafe{\msnil}$
    then $\ejudg{\MemsafeToNotac{e}}{\stratapply{\allocstrategy}{\NULL}}$
  \item If $\mseval[l,m]{\memsafe{e}} = \memsafe{(i,b,n)}$
    then $\ejudg{\MemsafeToNotac{e}}{\addrmap(\memsafe{i,b,0}) + n}$
  \end{itemize}
\end{lemma}
\begin{pfproof}\pf\
  By induction in $\memsafe{e}$.
  \STEP[case-var]{\pfcase{$\memsafe{x \in \texttt{var}}$}}
  \begin{pfproof}\pf\
    By the assumption $\statecompat{s}{(E,H,\allocstrategy)}$.
  \end{pfproof}
  \STEP[case-int]{\pfcase{$\memsafe{n \in \mathbb{Z}}$}}
  \begin{pfproof}\pf\
    By \textsc{e-const} since $\MemsafeToNotac{n} = n$.
  \end{pfproof}
  \STEP[case-bop]{\pfcase{$\memsafe{e_1 \oplus e_2}$}}
  \begin{pfproof}\pf\
    Consider the valid evaluations for $\memsafe{e_1}$ and $\memsafe{e_2}$
    If both evaluate to an integer, $\memsafe{n_1}$ and $\memsafe{n_2}$
    respectively, the statement follows directly from the
    induction hypothesis.
    The argument is mostly analogous when both expressions evaluate to
    either a pointer or $\memsafe{\msnil}$,
    except one important difference. 
    Due to \otherlang{} restricting equality comparison of pointers,
    to pointers that are within their allocation bounds, 
    a well-founded allocator strategy will not 
    influence the outcome of the evaluation, since the null address
    of the strategy will not be overlapping with allocated memory.
    \\[1mm]
    This leaves the case where one of $\memsafe{e_1}$ or $\memsafe{e_2}$
    evaluates to a pointer $\memsafe{(i,b,n)}$,
    and the other expression to some integer $\memsafe{n}$.
    In this case,
    the evaluation could only be valid if the operator was either
    an addition or subtraction.
    Since addition is commutative, it is safe to assume
    $\mseval[\memsafe{l,m}]{\memsafe{e_1}} = \memsafe{(i,b,n)}$
    and $\mseval[\memsafe{l,m}]{\memsafe{e_2}} = m$.
    The statement then follows by the assumption $\statecompat{(l,m)}{(E,H,\allocstrategy)}$
    and the fact that $(i,0) \in \dom{\addrmap}$ for all $i$.
  \end{pfproof}
  \STEP[case-nil]{\pfcase{$\memsafe{\msnil}$}}
  \begin{pfproof}\pf\
    By \textsc{e-null} since $\MemsafeToNotac{\msnil} = \nullexpr$.
  \end{pfproof}
\end{pfproof}
\begin{lemma}[Semantic correctness of command translation]%
  \label{lem:correct:cmd:trans}
  Given a \otherlang{} program $\memsafe{p}$ and states $(l, m), (l',m')$ such that
  \[
    \mseval[l,m]{\memsafe{p}} = (l',m')
  \]
  and for \thelanguage{} heaps $H_0, H, H'$, variable environment $E$,
  and allocator strategy $\allocstrategy \in \stratawaredom{E}$,
  such that
  \[
    (H, A) = \stratapply{\allocstrategy}{\INIT{H_0}}
    \quad\text{and}\quad
    \invmch{E, \allocstrategy}\conf{\MemsafeToNotac{p}; H, A}
    \to_t^{*}
    \conf{H', A'}
  \]
  Where the execution does not run out of memory.
  Then for any mapping $\addrmap$ satisfying $\statecompat{(l,m)}{(E,H,\allocstrategy)}$,
  there exists a mapping $\addrmap'$ such that
  \[
    \statecompat[\addrmap']{(l',m')}{(E,H',\allocstrategy)}
  \]
  In addition,
  if $q$ is the number of indices in $m$,
  then the following holds for trace $\tr$:
  \[
    \tr(i) = \evmalloc{k}{a} \iff |\{(j,b,n) \in \dom{m'} \mid j = i + q \}| = k
  \]
\end{lemma}
\begin{pfproof}\pf\
  Induction on $\memsafe{p}$.
  \STEP[case-skip]{$\memsafe{\msskip}$}
  \begin{pfproof}
    Trivial since no state change and no trace events produced.
  \end{pfproof}
  \STEP[case-seq]{$\memsafe{\msseq{c_1}{c_2}}$}
  \begin{pfproof}
    Follows from induction hypothesises.
  \end{pfproof}
  \STEP[case-ite]{$\memsafe{\msif \, e \, \msthen \, c_1 \, \mselse \, c_2}$}
  \begin{pfproof}
    Since the \otherlang{} evaluation is successful $\memsafe{e}$ must
    evaluate to an integer.
    Per assumption that $\statecompat{(l,m)}{(E,H,\allocstrategy)}$,
    Lemma~\ref{lem:correct:expr:trans} implies that if $\memsafe{e}$ evaluates to
    $n$, so too will $\MemsafeToNotac{e}$.
    This in turn also means if the \thelanguage{} computation will
    enter the same branch as the \otherlang{} evaluation.
    The statement then follows by case analysis on the selected branch
    using the respective induction hypothesis.
  \end{pfproof}
  \STEP[case-while]{$\memsafe{\mswhile \, e \, \msdo \, c \, \msend}$}
  \begin{pfproof}
    Induction in number of loops.
    Base case is analogous to $\memsafe{\msskip}$.
    Inductive step is analogous to the argument for if-then-else.
  \end{pfproof}
  \STEP[case-assg]{$\memsafe{x \leftarrow e}$}
  \begin{pfproof}
    Select $\addrmap$ as the assignment does not change how
    the states relate.
    Then per assumption $\statecompat{(l,m)}{(E,H,\allocstrategy)}$
    and Lemma~\ref{lem:correct:expr:trans} it holds
    \(
      \statecompat{(l',m)}{(E,H',\allocstrategy)}
      \).
  \end{pfproof}
  \STEP[case-load]{$\memsafe{x \leftarrow [e]}$}
  \begin{pfproof}
    Select $\addrmap$ as the command does not change
    how pointers relate.
    Then per assumption $\statecompat{(l,m)}{(E,H,\allocstrategy)}$
    and Lemma~\ref{lem:correct:expr:trans} it holds that
    \(
    \statecompat{(l',m)}{(E,H',\allocstrategy)}
    \).
  \end{pfproof}
  \STEP[case-write]{$\memsafe{[e_1] \leftarrow e_2}$}
  \begin{pfproof}
    Select $\addrmap$ as the command does not change
    how pointers relate.
    Then per assumption $\statecompat{(l,m)}{(E,H,\allocstrategy)}$
    and Lemma~\ref{lem:correct:expr:trans} it holds that
    \(
    \statecompat{(l,m')}{(E,H',\allocstrategy)}
    \).
  \end{pfproof}
  \STEP[case-alloc]{$\memsafe{x \leftarrow \msalloc(e)}$}
  \begin{pfproof}
    The execution of $\mseval[l,m]{\memsafe{x \leftarrow \msalloc(e)}} = (l',m')$,
    must add an additional block index to $m$ by definition.
    Note that the new index contains the same number of entries as the result
    of evaluating $\memsafe{e}$.
    Per our assumption the following run succeeds, without running
    out of memory:
    \[
      \invmch{E, \allocstrategy}\conf{\MemsafeToNotac{x \leftarrow \msalloc(e)}; H, A}
      \to_t^{*}
      \conf{H', A'}
    \]
    This will produce a single $\evmalloc{k}{a}$ event, where $k$ is the result of
    computing $\memsafe{e}$, per Lemma~\ref{lem:correct:expr:trans}.
    From this we get the relation between the trace and $m'$.
    Left to show is that some $\addrmap'$ exists such that
    $\statecompat[\addrmap']{(l',m)}{(E,H',\allocstrategy)}$.
    To create $\addrmap'$, take $\addrmap$ as a basis.
    The trace event reveals the address $a$.
    Say the latest index in $m'$ is $i$.
    Then for each entry with this index $(i, n) \in \dom{m'}$
    extend $\addrmap$ such that $\addrmap'(i,b,n) = a + n$.
    Since every memory slot in \otherlang{} is initialized to 0,
    we need to show that $H(a+n) = 0$ for every entry added.
    This holds if the while-loop initialization in the target program succeeds,
    which will only be the case if the allocated memory is
    contiguous and does not overlap with the null pointer or
    previous allocations, as this would make it impossible to
    satisfy the $\statecompatsingle{\addrmap}$ relation.
    But this follows from the assumption that the allocator strategy is well formed.
    Moreover, the well-formedness of the allocator also guarantees
    memory governed by the variable environment $E$ remains unmodified by
    the memory allocation.
    Because of this, we can conclude that $\addrmap'$ is a mapping
 that satisfies $\statecompat[\addrmap']{(l',m)}{(E,H',\allocstrategy)}$.
  \end{pfproof}
\end{pfproof}
\begin{theorem*}[(Restatement of \Cref{thm:memsafe:notac:gai}) Gradual allocator independence of \otherlang{} to \thelanguage{} translation]%
  Given a program $\memsafe{p}$ in \otherlang{} and initial state
  $\memsafe{s = (l, m_{\mathnormal{init}})}$, where
  \begin{itemize}
  \item $\memsafe{m_\text{init}}$ is an empty map.
  \item $\mseval[\memsafe{s}]{\memsafe{p}} = \memsafe{(l',m')}$.
  \item $\forall v \in \img{\memsafe{l}}. \, v \neq \memsafe{(i,b,n)} \land v \neq \memsafe{\msnil}$.
  \end{itemize}
  then for any \thelanguage{} program $\notac{c}$, environment $E$,
  and heap $H_0$, such that
  \begin{itemize}
  \item $\notac{c} = \MemsafeToNotac{p}$
  \item $\vars{\memsafe{p}} \subseteq \dom{E}$
  \item $\oom, \transvar{i} \notin \vars{\memsafe{p}}$
  \item $H_0(E(\oom)) = 0$
  \item $H_0(E(\alloci)) = 0$
  \item $\forall x \in \dom{\memsafe{l}}. \, H_0(E(x)) = \memsafe{l}(x)$
  \end{itemize}
  it holds that $\mathsf{GAI}(\notac{c}, E, H_0)$.
  Moreover for any $\allocstrategy \in \stratawaredom{E}$
  where $(H,\allocstate) = \stratapply{\allocstrategy}{\INIT{H_0}}$
  it holds that $\invmch{E, \allocstrategy} \conf{\notac{c}; H, \allocstate} \to_{\tr}^{*}
  \conf{\notac{\code{stop}}; H', \allocstate'}$
  and if the execution did not run out of memory $H'(E(\oom)) = 0$ then
  for any $x \in \dom{\memsafe{l'}}$ it holds that:
  \[
    \memsafe{l'}(x) \neq (i,b,n) \land \memsafe{l'}(x) \neq \msnil
    \implies
    \memsafe{l'}(x) = H'(E(x))
  \]
\end{theorem*}
\begin{pfproof}\pf\
  Start by selecting the empty map as $\addrmap$.
  Given $\memsafe{m_\text{init}}$ is empty, together with the assumptions of the theorem,
  it holds that:
  \[
    \statecompat{(\memsafe{l},\memsafe{m_\text{init}})}{(E,H,\allocstrategy)}
  \]
  If the \thelanguage{} program does not run out of memory,
  the above statement together with Lemma~\ref{lem:correct:cmd:trans} implies that
  there exists some $\addrmap'$ such that:
  \[
    \statecompat[\addrmap']{(\memsafe{l'},\memsafe{m'})}{(E,H',\allocstrategy)}
  \]
  By definition of $\statecompatsingle{\addrmap'}$, it follows that
  \[
    \memsafe{l'}(x) \neq (i,b,n) \land \memsafe{l'}(x) \neq \msnil
    \implies
    \memsafe{l'}(x) = H'(E(x))
  \]
  This leaves to show $\mathsf{GAI}(c,E,H_0)$.
  Given the only two events producible by the translation
  is either $\evmalloc{k}{a}$ or $\evmallocfail{k}$,
  it suffices to consider the case of
  \[
    \allocatorimpactproggen(\MemsafeToNotac{p}, E, H_0, t, \eventclassalloc{n})
    \supseteq
    \allocatorimpactmeta(\MemsafeToNotac{p}, E, H_0, \tr)
  \]
  Let some $\allocstrategy' \in \allocatorimpactmeta(\MemsafeToNotac{p}, E, H_0, \tr)$
  be given. Since $\allocstrategy' \in \stratawaredom{E}$, any trace produced
  by the strategy must respect the structure of $m'$, per Lemma~\ref{lem:correct:cmd:trans}.
  In addition, if the evaluation in \thelanguage{} under this strategy terminates,
  the trace will match the full structure of $m'$, again per Lemma~\ref{lem:correct:cmd:trans}.
  Therefore, $\allocstrategy'$ \emph{must} produce at least one event,
  which can only be either
  $\evmalloc{n}{a'}$ or $\evmallocfail{n}$.
  This in turn means
  $\allocstrategy' \in \allocatorimpactproggen(\MemsafeToNotac{p}, E, H_0, t, \eventclassalloc{n})$
  and thus it must hold that $\mathsf{GAI}(c,E,H_0)$.
\end{pfproof}

\section{Allocation strategies}
This section contains a collection of well-formed allocator strategies
along with proof of well-formedness.
\subsection{Eager allocator}
This allocator, denoted $\alloceager$, works on a memory segment
$[N_1, N_3)$ in which the initial segment $[N_1,N_2)$ (with
$N_2 \leq N_3$) is \emph{reserved}, e.g., for the stack. The
allocator follows the strategy of eagerly picking the first
contiguous block of unallocated memory in $[N_2, N_3)$ of at least
the requested size. Unallocated memory is made inaccessible to the
client (denoted $\bot$) and the allocated memory is tracked as a set
of pairs, each consisting of the starting address (as returned by
$\MALLOC{}$) and the number of bytes allocated:
\[
  \stratapply{\alloceager}{\domain{Alloc}} = 2^{\domain{Addr} \times \domain{Size}}
\]
As the null address, the strategy reserves $N_2$:
\[
  \stratapply{\alloceager}{\NULL} = N_2
\]
The eager allocator initializes the heap by preserving the values of
the reserved memory $[N_1,N_2)$ and marking the rest of the interval
$[N2,N_3)$ as unallocated (and thus inaccessible):
\[
  \stratapply{\alloceager}{\INIT{H}} = (H', \emptyset),
  \text{ where }
  H'(i) = \begin{cases}
    H(i), &\text{if }i \in [N_1, N_2)\\
    \bot, &\text{if }i \in [N_2, N_3)
  \end{cases}
\]
Let $\eagerF{H,\allocstate,s}$ be the set of addresses in $H$
that can be allocated for size $s$:
\[
  \eagerF{H,\allocstate,s} \defn \{ N_2 < a < N_3 \land
  a \notin \{ a' \mid (a',s') \in \allocstate \} \land
  \forall i \in [a, a+s).~ H(i) = \bot 
  \}
\]
To allocate memory, the eager allocator picks the first address
that can be used to allocate the requested size, makes the memory available,
and records the allocation in its internal state:
\[
  \stratapply{\alloceager}{\MALLOC{H, \allocstate, s}} =
  \begin{cases}
    (H, \allocstate, \stratapply{\alloceager}{\NULL}),
    &\text{if } \eagerF{H,\allocstate,s} = \emptyset\\
    (H[a \ldots a+s-1 \mapsto 0], \allocstate \cup \{(a, s)\}, a),
    &\text{if } \eagerF{H,\allocstate,s} \neq \emptyset \land 0 < s\\
    (H, \allocstate \cup \{(a,s)\},a),
    &\text{if } \eagerF{H,\allocstate,s} \neq \emptyset \land 0 = s
  \end{cases}
\]
where $a = \min \eagerF{H,\allocstate,s}$.

Freeing makes the freed memory inaccessible and updates the
allocator state to reflect this:
\[
  \stratapply{\alloceager}{\FREE{H,\allocstate,a}} =
  \begin{cases}
    (H, \allocstate), &\text{if } a \notin \{ a' \mid (a', s) \in \allocstate \}\\
    (H[a \ldots a+s-1\mapsto \bot], \allocstate \setminus \{(a,s)\}),
                      &\text{if } (a,s) \in \allocstate
  \end{cases}
\]
Note that calls to free memory that has not been registered as
allocated are ignored.
\subsubsection{Well-formedness of the eager allocator strategy}
We show the well-formedness of the eager allocation strategy $\alloceager$.
\begin{lemma}[Eager allocator client memory heap match]%
  \label[lemma]{lem:eager:allocator:heap:match}
  Let $\reservedmem$ be reserved memory as per the definition
  of $\alloceager$, meaning $\reservedmem = [N_1, N_2)$
  for some $N_1, N_2$.
  Then for all heaps $H,H_0,H_1$,
  allocator states $\allocstate_0, \allocstate_1$,
  allocation map $\allocsymmap_1$,
  symbolic sequence $\symbseq$,
  and update sequence $\clientupdseq_1$,
  if $\reservedmem \subseteq \dom{H}$,
  $(H_0,\allocstate_0) = \stratapply{\alloceager}{\INIT{H}}$,
  and $\invseqrel{\alloceager, \reservedmem}%
  \ffrel%
  {\emptyset}%
  {H_0,\allocstate_0}%
  {\clientupdseq_1,\symbseq}%
  {H_1, \allocstate_1}%
  {\allocsymmap_1}$
  then it holds that
  \[
    \dom{H_1} = \addressesof{\allocsymmap_1} \cup \reservedmem
    \quad\text{and}\quad
    \bigcup_{(a,s) \in \allocstate_1}[a,a+s) = \addressesof{\allocsymmap_1}
    \quad\text{and}\quad
    \{ a \mid (a,s) \in \allocstate_1 \} = \{ a \mid (a,s;i) \in \allocsymmap_1 \}
  \]
\end{lemma}
\begin{pfproof}\pf\
  We want to show that for the eager allocator strategy $\alloceager$,
  the domain of the heap matches client-accessible memory.
  The proof follows by induction in the $\ff$ relation on
  $\invseqrel{\alloceager, \reservedmem}%
  \ffrel%
  {\emptyset}%
  {H_0,\allocstate_0}%
  {\clientupdseq_1,\symbseq}%
  {H_1, \allocstate_1}%
  {\allocsymmap_1}$
  \STEP[base-case]{(\textsc{$\ff$-empty})
    $\invseqrel{\alloceager, \reservedmem}%
    \ffrel%
    {\emptyset}%
    {H_0,\allocstate_0}%
    {\clientupdseq_1,\emptytr}%
    {H_0,\allocstate_0}%
    {\emptyset}$:}
  \begin{pfproof}
    Since $\allocsymmap_1 = \emptyset$,
    it suffices to show that $\dom{H_0} = \reservedmem$,
    which is equivalent to showing $\dom{H_0} = [N_1,N_2)$.
    That this holds follows directly from the definition of $\eagerinit$
    giving us that: $\dom{H_0} = [N_1,N_2)$
    The second and third statement follows from the fact that $\allocstate_0 = \emptyset$
    by definition of $\eagerinit$.
  \end{pfproof}
  \STEP[step-case]{(\textsc{$\ff$-step}) $\invseqrel{\alloceager, \reservedmem}%
    \ffrel%
    {\emptyset}%
    {H_0,\allocstate_0}%
    {\clientupdseq_1' \cdot \clientupdmeta,\symbseq_{\text{pre}} \cdot \symbevent}%
    {H_1, \allocstate_1}%
    {\allocsymmap_1}$:}
  \begin{pfproof}
    We need to show \[
      \dom{H_1} = \addressesof{\allocsymmap_1} \cup \reservedmem
    \]
    and \[
      \bigcup_{(a,s) \in \allocstate_1}[a,a+s) = \addressesof{\allocsymmap_1}
    \]
    We assume some $H_1',H_1'',\allocstate_1'$, and $\allocsymmap_1'$
    such that
    \[
      \invseqrel{\alloceager, \reservedmem}%
      \ffrel%
      {\emptyset}%
      {H_0,\allocstate_0}%
      {\clientupdseq_1',\symbseq_{\text{pre}}}%
      {H_1', \allocstate_1'}%
      {\allocsymmap_1'}
    \]
    and \[
      \invseqrel{\alloceager}%
      \playrel%
      {\allocsymmap_1'}%
      {H_1'', \allocstate_1'}%
      {\symbseq_{\text{pre}} \cdot \symbevent}%
      {H_1, \allocstate_1}%
      {\allocsymmap_1}%
    \]
    where $H_1'' = \clientupd{H_1'}{\addressesof{\allocsymmap_1' \cup \reservedmem}}$.
    Per our induction hypothesis we have \[
      \dom{H_1'} = \addressesof{\allocsymmap_1'} \cup \reservedmem
    \]
    and \[
      \bigcup_{(a,s) \in \allocstate_1'}[a,a+s) = \addressesof{\allocsymmap_1'}
    \]
    Since client updates cannot change the domain of the heap,
    the induction hypothesis gives us that \[
      \dom{H_1''} = \addressesof{\allocsymmap_1'} \cup \reservedmem
    \]
    The proof now follows by case analysis in the $\play$ relation
    on \(
    \invseqrel{\alloceager}%
    \playrel%
    {\allocsymmap_1'}%
    {H_1'', \allocstate_1'}%
    {\symbseq_{\text{pre}} \cdot \symbevent}%
    {H_1, \allocstate_1}%
    {\allocsymmap_1}%
    \)
    \STEP[case-step-malloc-ok]{(\textsc{$\play$-malloc-ok}):}
    \begin{pfproof}
      For a successful allocation of $k$ addresses,
      we have $H_1 = H_1''[a\ldots a+k-1 \mapsto 0]$,
      $\allocsymmap_1 = \allocsymmap_1 \cup \{(a,k;i)\}$,
      and $\allocstate_1 = \allocstate_1' \cup \{(a,k)\}$,
      where per definition of $\eagermalloc$ we know that
      $[a,a+k) \cap \dom{H_1''} = \emptyset$.
      As such we have
      $\dom{H_1} = \dom{H_1''[a\ldots a+k-1 \mapsto 0]} = \dom{H_1''} \cup [a,a+k)$.
      \begin{align*}
        \dom{H_1} &= \dom{H_1''[a\ldots a+k-1 \mapsto 0]}\\
                  &= \dom{H_1''} \cup [a,a+k)\\
                  &= \addressesof{\allocsymmap_1'} \cup \reservedmem \cup [a,a+k)
                  &&\text{induction hypothesis}\\
                  &= \addressesof{\allocsymmap_1' \cup \{(a,k;i)\}} \cup reservedmem
                  &&\text{definition of allocation map}\\
                  &= \addressesof{\allocsymmap_1} \cup \reservedmem
                  &&\allocsymmap_1' \cup \{(a,k;i)\} = \allocsymmap_1
      \end{align*}
      Giving us the first part of the statement.
      We now show the other part of the statement:
      \begin{align*}
        \bigcup_{(a',s) \in \allocstate_1} [a',a'+s)
        &= \bigcup_{(a',s) \in (\allocstate_1' \cup \{(a,k)\})} [a',a'+s)
        &&\allocstate_1 = \allocstate_1' \cup \{(a,k)\}\\
        &= [a,a+k) \cup \bigcup_{(a',s) \in \allocstate_1'} [a',a'+s)\\
        &= [a,a+k) \cup \addressesof{\allocsymmap_1'}
        &&\text{induction hypothesis}\\
        &= \addressesof{\allocsymmap_1' \cup \{(a,k;i\}}
        &&\text{definition of allocation map}\\
        &= \addressesof{\allocsymmap_1}
        &&\allocsymmap_1' \cup \{(a,k;i)\} = \allocsymmap_1
      \end{align*}
      Lastly, we need to show that \[
        \{ a' \mid (a',s') \allocstate_1 \} = \{ a' \mid (a',s';i') \in \allocsymmap_1 \}
      \]
      Per induction hypothesis, we already know that this holds for $\allocstate_1'$
      and $\allocsymmap_1'$, so we just need to show the
      additions resulting from the successful allocation agrees.
      Per definition of $\eagermalloc$ we see that $\allocstate_1'$
      is extended with $(a,k)$, but that is example the same $a$ and $k$
      that extends $\allocsymmap_1'$, which concludes the case.
    \end{pfproof}
    \STEP[case-step-malloc-fail]{(\textsc{$\play$-malloc-fail}):}
    \begin{pfproof}
      In the case of a failed allocation
      we have that $H_1 = H_1''$, $\allocsymmap_1 = \allocsymmap_1'$,
      and $\allocstate_1' = \allocstate_1$.
      But in that case the statement follows directly from
      the induction hypothesis.
    \end{pfproof}
    \STEP[case-step-free]{(\textsc{$\play$-free}):}
    \begin{pfproof}
      Lastly, we consider the case of a free.
      Let us first assume that the freed address $a$
      and the corresponding size $k$
      is not in $\allocstate_1'$.
      In this case we get $\allocstate_1 = \allocstate_1'$
      and $H_1 = H_1''$.
      Since $(a,k)$ is not in $\allocstate_1'$ our induction hypothesis
      tells us that
      $[a,a+k) \cap \addressesof{\allocsymmap_1'} = \emptyset$,
      which would mean $\allocsymmap_1' \cap \{(a,k;i)\} = \emptyset$.
      But in that case we also have
      \[
        \allocsymmap_1 = \allocsymmap_1' \setminus \{(a,k;i)\}
        = \allocsymmap_1'
      \]
      Therefore, showing $\dom{H_1} = \addressesof{\allocsymmap_1} \cup \reservedmem$
      is equivalent to showing $\dom{H_1'} = \addressesof{\allocsymmap_1'} \cup \reservedmem$.
      But this is exactly what our induction hypothesis states.
      Similarly, showing $\bigcup_{(a',s) \in \allocstate_1}[a',a'+s) = \addressesof{\allocsymmap_1}$
      is equivalent to showing
      $\bigcup_{(a',s) \in \allocstate_1'}[a',a'+s) = \addressesof{\allocsymmap_1'}$,
      which we also know from our induction hypothesis.

      Now let us instead consider the case where $(a,s) \in \allocstate_1'$.
      In this case we instead have:
      \[
        \dom{H_1} = \dom{H_1'} \setminus [a,a+s)
        \quad\text{and}\quad
        \allocstate_1 = \allocstate_1' \setminus \{(a,s)\}
        \quad\text{and}\quad
        \addressesof{\allocsymmap_1} = \addressesof{\allocsymmap_1'} \setminus [a,a+s)
      \]
      The first part of the statement then follows from:
      \begin{align*}
        \dom{H_1}
        &= \dom{H_1'} \setminus [a,a+s)\\
        &= \addressesof{\allocsymmap_1'} \cup \reservedmem \setminus [a,a+s)
        &&\text{induction hypothesis}\\
        &= \left ( \addressesof{\allocsymmap_1'} \setminus [a,a+s)
          \right ) \cup \reservedmem
        &&\text{since }\reservedmem \cap [a,a+s) = \emptyset\\
        &= \addressesof{\allocsymmap_1} \cup \reservedmem
        &&\addressesof{\allocsymmap_1} = \addressesof{\allocsymmap_1'} \setminus [a,a+s)
      \end{align*}
      The second part of the statement follows from:
      \begin{align*}
        \bigcup_{(a',s') \in \allocstate_1} [a',a'+s')
        &= \left ( \bigcup_{(a',s') \in \allocstate_1'}[a',a'+s') \right ) \setminus [a,a+s)
        && \allocstate_1 = \allocstate_1' \setminus \{(a,s)\}\\
        &= \addressesof{\allocsymmap_1'} \setminus [a,a+s)
        &&\text{induction hypothesis}\\
        &= \addressesof{\allocsymmap_1}
        &&\addressesof{\allocsymmap_1} = \addressesof{\allocsymmap_1'} \setminus [a,a+s)
      \end{align*}
      The third part of the statement follows by realizing
      the $(a,s)$ removed with $\allocstate_1'$
      are the same $a$ and $k$ removed from $\allocsymmap_1'$.
    \end{pfproof}
  \end{pfproof}
  \qed
\end{pfproof}
\begin{theorem}[Eager allocator strategy $\alloceager$ is well-formed]
  The eager allocator strategy $\alloceager$ is well-formed as per
  \Cref{def:allocator:heap:well-formedness}.
\end{theorem}
\begin{pfproof}\pf\
  \STEP[case-basic-1]{\hyperref[wf:disjoint:alloc]{
      \emph{allocated regions are disjoint:}}}
  \begin{pfproof}
    Induction in the $\ff$ relation.
    \STEP[case-basic-1-base]{(\textsc{$\ff$-empty})
      $\invseqrel{\alloceager, \reservedmem}%
      \ffrel%
      {\emptyset}%
      {H_0,\allocstate_0}%
      {\clientupdseq_1,\emptytr}%
      {H_0,\allocstate_0}%
      {\emptyset}$:}
    \begin{pfproof}
      Trivial since $\allocsymmap_1 = \emptyset$.
    \end{pfproof}
    \STEP[case-basic-1-step]{(\textsc{$\ff$-step}) $\invseqrel{\alloceager, \reservedmem}%
      \ffrel%
      {\emptyset}%
      {H_0,\allocstate_0}%
      {\clientupdseq_1' \cdot \clientupdmeta,\symbseq_{\text{pre}} \cdot \symbevent}%
      {H_1, \allocstate_1}%
      {\allocsymmap_1}$:}
    \begin{pfproof}
      We assume some $H_1',H_1'',\allocstate_1'$, and $\allocsymmap_1'$
      such that
      \[
        \invseqrel{\alloceager, \reservedmem}%
        \ffrel%
        {\emptyset}%
        {H_0,\allocstate_0}%
        {\clientupdseq_1',\symbseq_{\text{pre}}}%
        {H_1', \allocstate_1'}%
        {\allocsymmap_1'}
      \]
      and \[
        \invseqrel{\alloceager}%
        \playrel%
        {\allocsymmap_1'}%
        {H_1'', \allocstate_1'}%
        {\symbseq_{\text{pre}} \cdot \symbevent}%
        {H_1, \allocstate_1}%
        {\allocsymmap_1}%
      \]
      where $H_1'' = \clientupd{H_1'}{\addressesof{\allocsymmap_1' \cup \reservedmem}}$.
      Proof proceeds by case analysis on the $\play$ relation.
      In the cases of \textsc{$\play$-malloc-fail} and \textsc{$\play$-free},
      we have
      $\addressesof{\allocsymmap_1} \subseteq \addressesof{\allocsymmap_1'}$
      and the statement in this case follows directly from the induction
      hypothesis.
      This leaves the case of \textsc{$\play$-malloc-ok}.
      Now let $a,k,i$ be given such that $\allocsymmap_1 = \allocsymmap_1' \cup \{(a,k;i)\}$.
      We only need to show that the new allocation
      does not overlap any previous allocations,
      since we know per our induction hypothesis,
      that the allocations in $\allocsymmap_1'$ are disjoint.
      As such, it suffices to show that
      $\addressesof{\allocsymmap_1'} \cap [a,a+k) = \emptyset$.
      From the definition of $\eagermalloc$ we can see
      that the new allocation is selected such that
      the addresses are not in the domain of $H_1'$.
      This allows us to deduce the following:
      \begin{align*}
        \emptyset
        &= \dom{H_1'} \cap [a,a+k)\\
        &= \left (\addressesof{\allocsymmap_1'} \cup \reservedmem \right )
          \cap [a,a+k)
        &&\text{by \Cref{lem:eager:allocator:heap:match}}\\
        &= \left (\addressesof{\allocsymmap_1'} \cap [a,a+k) \right )
          \cup \left( \reservedmem \cap [a,a+k) \right )\\
        &= \addressesof{\allocsymmap_1'} \cap [a,a+k)
        &&\text{since }\reservedmem \cap [a,a+k) = \emptyset
      \end{align*}
      This is exactly what we needed to show, concluding the case.
    \end{pfproof}
  \end{pfproof}
  \STEP[case-basic-2]{\hyperref[wf:client:memory:final:heaps]{
      \emph{client-accessible memory is in the final heaps:}}}
  \begin{pfproof}
    Follows directly from \Cref{lem:eager:allocator:heap:match}.
  \end{pfproof}
  \STEP[case-basic-3]{\hyperref[wf:init:no:modify:client]{
      \emph{allocator initialization does not update reserved memory:}}}
  \begin{pfproof}
    We know per our assumptions that $\reservedmem = [N_1,N_2)$
    and that $\reservedmem \subseteq \dom{H}$.
    By construction, $\eagerinit$ keeps exactly the contents
    at the addresses $[N_1, N_2)$ the same as in before initialization
    meaning $H \heapeq{\reservedmem} H_0$ must hold.
  \end{pfproof}
  \STEP[case-basic-4]{\hyperref[wf:alloc:no:modify:client]{
      \emph{allocator does not modify client-accessible memory:}}}
  \begin{pfproof}
    Let $H', \allocstate'$, and $\allocsymmap'$ be given,
    the proof then follow by case analysis in the $\play$
    relation on:
    \[
      \invseqrel{\alloceager}%
      \playrel%
      {\allocsymmap'}%
      {H', \allocstate'}%
      {\symbseq_{\text{pre}} \cdot \symbevent}%
      {H_1, \allocstate_1}%
      {\allocsymmap_1}%
    \]
    \STEP[case-basic-4-malloc-ok]{(\textsc{$\play$-malloc-ok}):}
    \begin{pfproof}
      Per definition of $\eagermalloc$ we know that
      the allocated memory is not in the domain of $H'$.
      By \Cref{lem:eager:allocator:heap:match} we also know
      that $\dom{H'} = \addressesof{\allocsymmap'} \cup \reservedmem$,
      and thus the allocation does not interact with any
      client-accessible memory,
      and therefore we have that $H' \heapeq{\addressesof{\allocsymmap'} \cup \reservedmem} H_1$.
    \end{pfproof}
    \STEP[case-basic-4-malloc-fail]{(\textsc{$\play$-malloc-fail}):}
    \begin{pfproof}
      Trivial as $\alloceager$ does not modify the heap
      on failed allocations,
      meaning $H_1 = H'$,
      giving us $H' \heapeq{\addressesof{\allocsymmap'} \cup \reservedmem} H'$.
    \end{pfproof}
    \STEP[case-basic-4-free]{(\textsc{$\play$-free}):}
    \begin{pfproof}
      Consider the call to free on address $a$ that was
      allocated with size $s$.
      If $(a,s) \notin \allocstate'$, then the heap is unmodified,
      and we are done.
      If instead $(a,s) \in \allocstate'$,
      we will have to show that client-accessible memory,
      except the freed memory, is untouched.
      Looking at the definition of $\eagerfree{}$,
      we see that the only difference between $H'$
      and $H_1$ is that the freed memory is undefined
      in $H_1$, which means the property holds.
    \end{pfproof}
  \end{pfproof}
  \STEP[case-basic-5]{\hyperref[wf:alloc:no:overlap:reserved]{
      \emph{allocated addresses do not overlap with reserved memory:}}}
  \begin{pfproof}
    Induction in the $\ff$ relation.
    Base case is trivial since \[
      \addressesof{\emptyset} \cap \reservedmem
      = \emptyset \cap \reservedmem
      = \emptyset
    \]
    The inductive step is then by case analysis on
    the $\play$ relation.
    Both \textsc{$\play$-malloc-fail} and \textsc{$\play$-free}
    follow directly from the induction hypothesis.
    In the case of \textsc{$\play$-malloc-ok},
    we just need to check the newly allocated memory
    does not overlap with $\reservedmem$
    since we know per the induction hypothesis,
    that any previously allocated memory is already disjoint
    from $\reservedmem$.
    By the definition of $\eagermalloc$ we know that
    any allocated memory will be in a address space
    greater than $N_2$,
    and since $\reservedmem = [N_1, N_2)$,
    then the newly allocated memory must necessarily
    also be disjoint from $\reservedmem$.
  \end{pfproof}
  \STEP[case-basic-6]{\hyperref[wf:null:not:accessible]{
      \emph{null is not client-accessible:}}}
  \begin{pfproof}
    Induction in the $\ff$ relation.
    \STEP[case-basic-6-base]{(\textsc{$\ff$-empty})
      $\invseqrel{\alloceager, \reservedmem}%
      \ffrel%
      {\emptyset}%
      {H_0,\allocstate_0}%
      {\clientupdseq_1,\emptytr}%
      {H_0,\allocstate_0}%
      {\emptyset}$:}
    \begin{pfproof}
      Since $\allocsymmap_1 = \emptyset$,
      it suffices to show $\eagernull \notin \reservedmem$,
      but this is true by definition since $\eagernull = N_2$
      and $\reservedmem = [N_1,N_2)$.
    \end{pfproof}
    \STEP[case-basic-6-step]{(\textsc{$\ff$-step}) $\invseqrel{\alloceager, \reservedmem}%
      \ffrel%
      {\emptyset}%
      {H_0,\allocstate_0}%
      {\clientupdseq_1' \cdot \clientupdmeta,\symbseq_{\text{pre}} \cdot \symbevent}%
      {H_1, \allocstate_1}%
      {\allocsymmap_1}$:}
    \begin{pfproof}
      By case analysis on the $\play$ relation.
      The \textsc{$\play$-malloc-fail} and \textsc{$\play$-free}
      do not grow the client-accessible memory,
      and these cases therefore follow directly from the
      induction hypothesis.
      This leaves us to show it also holds for \textsc{$\play$-malloc-ok}.
      Using our induction hypothesis, the statement reduces to
      showing that the newly allocated memory
      does not overlap with $\eagernull$.
      This holds per definition of $\eagermalloc$,
      since any allocated memory will only be in
      an address-space strictly greater than $N_2$.
    \end{pfproof}
  \end{pfproof}
  \STEP[case-zero-1]{\hyperref[wf:zero:no:overlap]{
      \emph{allocated addresses are not reused:}}}
  \begin{pfproof}
    Proof by induction in the $\ff$ relation.
    \STEP[case-rel-1-base]{(\textsc{$\ff$-empty})
      $\invseqrel{\alloceager, \reservedmem}%
      \ffrel%
      {\emptyset}%
      {H_0,\allocstate_0}%
      {\clientupdseq_1,\emptytr}%
      {H_0,\allocstate_0}%
      {\emptyset}$:}
    \begin{pfproof}
      Trivially true since $\allocsymmap_1 = \emptyset$.
    \end{pfproof}
    \STEP[case-basic-1-step]{(\textsc{$\ff$-step}) $\invseqrel{\alloceager, \reservedmem}%
      \ffrel%
      {\emptyset}%
      {H_0,\allocstate_0}%
      {\clientupdseq_1' \cdot \clientupdmeta,\symbseq_{\text{pre}} \cdot \symbevent}%
      {H_1, \allocstate_1}%
      {\allocsymmap_1}$:}
    \begin{pfproof}
      We assume some $H_1',H_1'',\allocstate_1'$, and $\allocsymmap_1'$
      such that
      \[
        \invseqrel{\alloceager, \reservedmem}%
        \ffrel%
        {\emptyset}%
        {H_0,\allocstate_0}%
        {\clientupdseq_1',\symbseq_{\text{pre}}}%
        {H_1', \allocstate_1'}%
        {\allocsymmap_1'}
      \]
      and \[
        \invseqrel{\alloceager}%
        \playrel%
        {\allocsymmap_1'}%
        {H_1'', \allocstate_1'}%
        {\symbseq_{\text{pre}} \cdot \symbevent}%
        {H_1, \allocstate_1}%
        {\allocsymmap_1}%
      \]
      where $H_1'' = \clientupd{H_1'}{\addressesof{\allocsymmap_1' \cup \reservedmem}}$.
      Per induction hypothesis we know that the statement holds for
      $\allocsymmap_1'$, and we need to show it also holds for $\allocsymmap_1$.
      Doing case analysis on the $\play$ relation, the
      statement follows directly from the induction hypothesis
      for \textsc{$\play$-malloc-fail} and \textsc{$\play$-free}
      since in these cases $\allocsymmap_1 \subseteq \allocsymmap_1'$.
      For the \textsc{$\play$-malloc-ok} case we start
      by noting that whichever $a$ is allocated cannot
      be in $\allocstate_1'$, by the definition $\eagermalloc$.
      But in that case \Cref{lem:eager:allocator:heap:match}
      tells us that this $a$ is also not present in $\allocsymmap_1'$.
      This gives us, together with the induction hypothesis,
      that no address therefore appears twice in $\allocsymmap_1$,
      which concludes the proof.
    \end{pfproof}
  \end{pfproof}
  \STEP[case-zero-2]{\hyperref[wf:zero:no:space]{
      \emph{zero-sized allocations are disjoint from the client-updateable memory:}}}
  \begin{pfproof}
    Induction in the $\ff$ relation.
    Base case is trivial since here $\allocsymmap_1 = \emptyset$.
    The inductive case is then by case analysis
    on the resulting $\play$ relation.
    Failed malloc and free follow directly by the
    induction hypothesis, since $\allocsymmap_1 \subseteq \allocsymmap_1'$
    in these cases.
    This leaves us to consider the case of a successful allocation.
    Concretely, we need to show that the address of the new allocation
    $a$ is not in $\addressesof{\allocsymmap_1} \cup \reservedmem$,
    if it has size $0$.
    By the definition of $\eagermalloc$, we that $H_1' = H_1$,
    and that $a$ is not in $\dom{H_1}$.
    But per \Cref{lem:eager:allocator:heap:match}
    we know that $\dom{H_1} = \addressesof{\allocsymmap_1} \cup \reservedmem$,
    and thus it holds that $a \notin \addressesof{\allocsymmap_1} \cup \reservedmem$.
  \end{pfproof}
  \STEP[case-relational]{Relational properties:}
  We want to show that there exists $H_2,\allocstate_2,\allocsymmap_2$
  such that the relational properties of
  \Cref{def:allocator:heap:well-formedness} hold.
  Let some $\clientupdseq_2$ be given such that
  $\lenof{\clientupdseq_1} = \lenof{\clientupdseq_2}$.
  Select $\allocstate_2$ to be $\allocstate_1$,
  and $\allocsymmap_2$ to be $\allocsymmap_1$.
  \begin{pfproof}
    \STEP[case-rel-1]{\hyperref[wf:update:influence]{
        \emph{client updates to the heap do not influence allocator decision:}}}
    \begin{pfproof}
      By induction in the $\ff$ relation.
      \STEP[case-rel-1-base]{(\textsc{$\ff$-empty})
        $\invseqrel{\alloceager, \reservedmem}%
        \ffrel%
        {\emptyset}%
        {H_0,\allocstate_0}%
        {\clientupdseq_1,\emptytr}%
        {H_0,\allocstate_0}%
        {\emptyset}$:}
      \begin{pfproof}
        Select $H_2$ to be $H_0$,
        then the statement follows by applying \textsc{$\ff$-empty}.
      \end{pfproof}
      \STEP[case-rel-1-step]{(\textsc{$\ff$-step}) $\invseqrel{\alloceager, \reservedmem}%
        \ffrel%
        {\emptyset}%
        {H_0,\allocstate_0}%
        {\clientupdseq_1' \cdot \clientupdmeta,\symbseq_{\text{pre}} \cdot \symbevent}%
        {H_1, \allocstate_1}%
        {\allocsymmap_1}$:}
      \begin{pfproof}
        Let $H_1', H_1'', \allocstate_1', \allocsymmap_1'$
        be given such that \[%
          \invseqrel{\alloceager, \reservedmem}%
          \ffrel%
          {\emptyset}%
          {H_0,\allocstate_0}%
          {\clientupdseq_1',\symbseq_{\text{pre}}}%
          {H_1', \allocstate_1'}%
          {\allocsymmap_1'}%
        \]
        and \[%
          \invseqrel{\alloceager}%
          \playrel%
          {\allocsymmap_1'}%
          {H_1'', \allocstate_1'}%
          {\symbseq_{\text{pre}} \cdot \symbevent}%
          {H_1, \allocstate_1}%
          {\allocsymmap_1}%
        \]
        where \(H_1'' = \clientupdmeta_1(H_1',\addressesof{\allocsymmap_1'} \cup \reservedmem)\).
        Per our assumptions we know that
        $\lenof{\clientupdseq_2} = \lenof{\clientupdseq_1' \cdot \clientupdmeta} = \lenof{\clientupdseq_1'} + 1$.
        Therefore, there must exist some $\clientupdseq_2'$ and $\clientupdmeta_2$ such that $\clientupdseq_2 = \clientupdseq_2' \cdot \clientupdmeta_2$.
        Per our induction hypothesis, we also know that there exists
        some $H_2'$ such that \[
          \invseqrel{\alloceager, \reservedmem}%
          \ffrel%
          {\emptyset}%
          {H_0,\allocstate_0}%
          {\clientupdseq_2',\symbseq_{\text{pre}}}%
          {H_2', \allocstate_1'}%
          {\allocsymmap_1'}%
        \]
        Let $H_2'' = \clientupdmeta_2(H_2',\addressesof{\allocsymmap_1'}) \cup \reservedmem)$.
        The proof proceeds by case analysis on the $\play$ relation
        in $\invseqrel{\alloceager}%
          \playrel%
          {\allocsymmap_1'}%
          {H_1'', \allocstate_1'}%
          {\symbseq_{\text{pre}} \cdot \symbevent}%
          {H_1, \allocstate_1}%
          {\allocsymmap_1}$.
        \STEP[case-rel-1-malloc-ok]{(\textsc{$\play$-malloc-ok}):}
        \begin{pfproof}
          Per the case analysis we know that a malloc call
          has succeeded for the ``1'' execution.
          Let $(a,k;i)$ be the element added to the allocation map
          such that: $\allocsymmap_1 = \allocsymmap_1' \cup \{(a,k;i)\}$.
          Select $H_2$ to be $H_2''$ if $k = 0$ and
          $H_2''[a\ldots a+k-1 \mapsto 0]$ otherwise.
          Per the \textsc{$\ff$-step} rule and our induction hypothesis
          it we just need to show that
          \[
            \invseqrel{\alloceager}%
            \playrel%
            {\allocsymmap_1'}%
            {H_2'', \allocstate_1'}%
            {\symbseq_{\text{pre}} \cdot \symbmalloc{k}}%
            {H_2, \allocstate_1}%
            {\allocsymmap_1}
          \]
          By \Cref{lem:eager:allocator:heap:match} we see that:
          \[
            \dom{H_2''} = \addressesof{\allocsymmap_1'} \cup \reservedmem = \dom{H_1''}
          \]
          By definition of $\eagermalloc$, the allocation will select some
          address $a'$, such that $a'$ the smallest
          address in the interval $[N_2,N_3)$ where
          $[a',a'+k) \cap \dom{H_2''} = \emptyset$.
          But since $\dom{H_2''} = \dom{H_1''}$, then
          $a'$ would also be the smallest valid allocation
          address for $H_1''$, which we know already know exists
          since $a$ was picked for the ``1'' execution.
          Therefore, it must be that $a' = a$.
          But in that case, we indeed have that:
          \[
            \invseqrel{\alloceager}%
            \playrel%
            {\allocsymmap_1'}%
            {H_2'', \allocstate_1'}%
            {\symbseq_{\text{pre}} \cdot \symbmalloc{k}}%
            {H_2, \allocstate_1}%
            {\allocsymmap_1}
          \]
          Since $\allocsymmap_1 = \allocsymmap_1' \cup \{(a,k;i)\}$
          and $\allocstate_1 = \allocstate_1' \cup \{(a,k)\}$.
        \end{pfproof}
        \STEP[case-rel-1-malloc-fail]{(\textsc{$\play$-malloc-fail}):}
        \begin{pfproof}
          Select $H_2$ to be $H_2''$.
          We know that allocation failed for heap $H_1''$,
          and if we can show that it also fails for $H_2''$ we are done.
          Assume on the contrary that allocation will not fail for $H_2''$.
          Since both execution start with the same allocator state $\allocstate_1'$,
          the allocation cannot have been of size zero, as otherwise
          it would have succeeded for $H_1''$.
          Then there must exist some address $a$ and size $k > 0$, such that \[
            [a,a+k) \cap \dom{H_2''} = \emptyset
          \]
          But by \Cref{lem:eager:allocator:heap:match} we have that
          \[
            \dom{H_2''} = \addressesof{\allocsymmap_1'} \cup \reservedmem = \dom{H_1''}
          \]
          which means $a$ would also have been a valid allocation address for
          $H_1''$ and it would not have failed, which is a contradiction.
          Thus we can conclude that if the $H_1''$ execution fails to allocate
          so too will the $H_2''$ execution.
        \end{pfproof}
        \STEP[case-rel-1-free]{(\textsc{$\play$-free}):}
        \begin{pfproof}
          We assume a successful free for the $H_1$ execution,
          with the symbolic free event $\symbfree{z}$.
          Let the free be of some address $a$ that was allocated with size $k$.
          Select $H_2$ to be $H_2''[a\ldots a+k-1 \mapsto \bot]$ if
          $(a,k) \in \allocstate_1'$ and $k > 0$,
          and $H_2''$ otherwise.
          We need to show that \[
            \invseqrel{\alloceager}%
            \playrel%
            {\allocsymmap_1'}%
            {H_2'', \allocstate_1'}%
            {\symbseq_{\text{pre}} \cdot \symbfree{z}}%
            {H_2, \allocstate_1}%
            {\allocsymmap_1}
          \]
          This must however be the case, since the result
          of $\eagerfree$ only depends on $\allocstate_1'$.
        \end{pfproof}
      \end{pfproof}
    \end{pfproof}
    \STEP[case-rel-2]{\hyperref[wf:alloc:map:equiv]{
        \emph{final allocation maps match:}}}
    \begin{pfproof}
      Trivially true since $\allocsymmap_2 = \allocsymmap_1$.
    \end{pfproof}
  \end{pfproof}
  \qed
\end{pfproof}

\subsection{Bump allocator}
A bump allocator governs a contiguous area of memory and allocates
memory by moving (bumping) a pointer to track what memory is being
used.  This ``bump pointer'' effectively splits the memory into used
memory and free memory.  The algorithm can be seen in the following
pseudocode:
\begin{lstlisting}
bump(size) {
  if (our capacity < size) {
    return null
  } else {
    bump_ptr = move bump_ptr by size bytes
    return pointer to freshly allocated space
  }
}
\end{lstlisting}
For a bump allocator the \texttt{free} operation is a no-op.
\subsubsection{Bump allocator strategy}
Let us define the bump allocator as an allocation strategy.  The bump
allocator will track the start and end point of the memory segment it
governs along with a ``bump pointer''.
\[
  \stratapply{\allocbump}{\domain{Alloc}} = \domain{Addr}
\]
For its null address, the bump allocator uses the first address in
governed memory segment:
\[
  \stratapply{\allocbump}{\NULL} = N_2
\]
The bump allocator strategy is initialized with a bump pointer right
to the null address:
\[
  \stratapply{\allocbump}{\INIT{H}} = (H', N_2+1), \text{ where }H'(i) = \begin{cases}
    H(i), &\text{if } i \in \dom{H}\\
    0, &\text{if }i \notin \dom{H} \land i \in [N_2+1,N_3)\\
    \bot &\text{if }i = N_2
  \end{cases}
\]
To allocate memory, the bump allocator moves the bump pointer by the
number of bytes requested and returns an address to the segment passed
by the bump pointer. If the requested memory was of size zero, the
allocator will act as if it was of size one. If the requested memory
would make the bump pointer move beyond the interval used by the bump
allocator, the allocation fails and returns the null address.
\[
  \stratapply{\allocbump}{\MALLOC{H,\allocstate,s}} = \begin{cases}
    (H, \allocstate + s, \allocstate), &\text{if }\allocstate + s \leq N_3\\
    (H, \allocstate + 1, \allocstate), &\text{if } s = 0 \land \allocstate + 1 \leq N_3\\
    (H, \allocstate, \stratapply{\allocbump}{\NULL}), &\text{otherwise}
  \end{cases}
\]
The bump allocator does not free memory, so call to free are no-ops.
\[
  \stratapply{\allocbump}{\FREE{H,\allocstate,a}} = (H,\allocstate)
\]
\subsubsection{Well-formedness of the bump allocator strategy}
We show the well-formedness of the bump allocation strategy.  To
begin, we prove the following lemma, that shows the bump pointer is
always larger than client-accessible memory.
\begin{lemma}[Bump allocator strategy $\ff$ invariant]%
  \label[lemma]{lem:bump:alloc:invariant}
  Let $\reservedmem$ be reserved memory as per the definition
  of $\allocbump$, meaning $\reservedmem = [N_1, N_2)$
  for some $N_1, N_2$.
  Then for all heaps $H,H_0,H_1$,
  allocator states $\allocstate_0, \allocstate_1$,
  allocation map $\allocsymmap_1$,
  symbolic sequence $\symbseq$,
  and update sequence $\clientupdseq_1$,
  if $\reservedmem \subseteq \dom{H}$,
  $(H_0,\allocstate_0) = \stratapply{\allocbump}{\INIT{H}}$,
  and $\invseqrel{\allocbump, \reservedmem}%
  \ffrel%
  {\emptyset}%
  {H_0,\allocstate_0}%
  {\clientupdseq_1,\symbseq}%
  {H_1, \allocstate_1}%
  {\allocsymmap_1}$
  then
  \[
    \forall a \in \addressesof{\allocsymmap_1}
    \cup \reservedmem
    \cup \{\stratapply{\allocbump}{\NULL}\}.~ a < \allocstate_1
  \]
\end{lemma}
\begin{pfproof}\pf\
  By definition of $\allocbump$ we know $\bumpnull = N_2$,
  and thus for $\reservedmem = [N_1,N_2)$ we have that
  $\forall a' \in \reservedmem.~ a' < \bumpnull$.
  Therefore, it suffices to show the statement:
  \[
    \forall a \in \addressesof{\allocsymmap_1} \cup
    \{N_2\}.~ a < \allocstate_1
  \]
  since $\allocstate_1$ greater than $\stratapply{\allocbump}{\NULL}$
  must transitively also be greater than any address in $\reservedmem$.
  The proof then follows by induction in the $\ff$ relation
  on $\invseqrel{\allocbump, \reservedmem}%
  \ffrel%
  {\emptyset}%
  {H_0,\allocstate_0}%
  {\clientupdseq_1,\symbseq}%
  {H_1, \allocstate_1}%
  {\allocsymmap_1}$.
  \STEP[base-case]{(\textsc{$\ff$-empty})
    $\invseqrel{\allocbump, \reservedmem}%
    \ffrel%
    {\emptyset}%
    {H_0,\allocstate_0}%
    {\clientupdseq_1,\emptytr}%
    {H_0, \allocstate_0}%
    {\emptyset}$:}
  \begin{pfproof}
    Since $\allocsymmap_1 = \emptyset$ we only need to show that
    $N_2 < \allocstate_0$.
    By definition of $\bumpinit$ we know that $\allocstate_0 = N_2 + 1$
    meaning the statement reduces to
    $N_2 < N_2 + 1$, which holds.
  \end{pfproof}
  \STEP[step-case]{(\textsc{$\ff$-step}) $\invseqrel{\allocbump, \reservedmem}%
    \ffrel%
    {\emptyset}%
    {H_0,\allocstate_0}%
    {\clientupdseq_1' \cdot \clientupdmeta,\symbseq_{\text{pre}} \cdot \symbevent}%
    {H_1, \allocstate_1}%
    {\allocsymmap_1}$:}
  \begin{pfproof}
    We assume some $H_1',H_1'',\allocstate_1'$, and $\allocsymmap_1'$
    such that
    \[
      \invseqrel{\allocbump, \reservedmem}%
      \ffrel%
      {\emptyset}%
      {H_0,\allocstate_0}%
      {\clientupdseq_1',\symbseq_{\text{pre}}}%
      {H_1', \allocstate_1'}%
      {\allocsymmap_1'}
    \]
    and \[
      \invseqrel{\allocbump}%
      \playrel%
      {\allocsymmap_1'}%
      {H_1'', \allocstate_1'}%
      {\symbseq_{\text{pre}} \cdot \symbevent}%
      {H_1, \allocstate_1}%
      {\allocsymmap_1}%
    \]
    where $H_1'' = \clientupd{H_1'}{\addressesof{\allocsymmap_1' \cup \reservedmem}}$.
    Per our induction hypothesis we know that:
    \[
      \forall a \in \addressesof{\allocsymmap_1'}
      \cup \{N_2\}.~ a < \allocstate_1'
    \]
    The rest of the proof goes by case analysis on the $\play$ relation:
    \STEP[case-step-malloc-ok]{(\textsc{$\play$-malloc-ok}):}
    \begin{pfproof}
      Let some $k$ such that $\symbevent = \symbmalloc{k}$, and let $i
      = \lenof{\symbseq} + 1$ be given.  We know that $\allocsymmap_1 =
      \allocsymmap_1' \cup \{(\allocstate_1', k; i)\}$ per the
      \textsc{$\play$-malloc-ok} rule.  Per our induction hypothesis we also
      know that $\allocstate_1'$ is greater than any address in
      $\addressesof{\allocsymmap_1'} \cup \{N_2\}$.  As such, the greatest
      element in $\addressesof{\allocsymmap_1'} \cup \{N_2\}$ must
      necessarily be $\allocstate_1'$, and therefore it suffices to show
      that $\allocstate_1' < \allocstate_1$.  If we look at the definition
      of malloc for $\allocbump$, we see that $\allocstate_1' + 1 \leq
      \allocstate_1$ (since $k=0$ behaves like $k=1$).  Because of this it
      must therefore hold that $\allocstate_1' < \allocstate_1$.
    \end{pfproof}
    \STEP[case-step-malloc-fail]{(\textsc{$\play$-malloc-fail}):}
    \begin{pfproof}
      Follows directly from induction hypothesis since
      $\allocsymmap_1 = \allocsymmap_1'$
      and that $\allocstate_1' = \allocstate_1$ since
      failed allocations do not change the allocator state.
    \end{pfproof}
    \STEP[case-step-free]{(\textsc{$\play$-free}):}
    \begin{pfproof}
      We have $\allocsymmap_1 = \allocsymmap_1' \setminus \{(a,k;i)\}$
      for some $a,k,i$.
      As such, we have that $\addressesof{\allocsymmap_1}
      \subseteq \addressesof{\allocsymmap_1'}$
      meaning that $\addressesof{\allocsymmap_1} \cup \{N_2\} < \allocstate_1$
      per the induction hypothesis and $\allocstate_1 = \allocstate_1$,
      since free does change the allocator state.
    \end{pfproof}
  \end{pfproof}
  \qed
\end{pfproof}
We now proceed to show well-formedness of the bump allocator strategy.
\begin{theorem}[Bump allocator strategy $\allocbump$ is well-formed]%
  \label{prop:bump:alloc:well-formed}
  The bump allocator strategy $\allocbump$ is well-formed as per
  \Cref{def:allocator:heap:well-formedness}.
\end{theorem}
\begin{pfproof}\pf\
  Let heaps $H, H_0, H_1$,
  reserved memory $\reservedmem$,
  allocator states of $\allocbump$ $\allocstate_0,\allocstate_1$,
  allocation map $\allocsymmap_1$,
  symbolic sequence $\symbseq$,
  and client update sequence $\clientupdseq_1$
  be given such that:
  \begin{enumerate}
  \item $\reservedmem$ is a contiguous memory interval $[N_1, N_2)$
  \item $\reservedmem \subseteq \dom{H}$
  \item $(H_0,\allocstate_0) = \bumpinit{H}$
  \item $\invseqrel{\allocbump, \reservedmem}%
    \ffrel%
    {\emptyset}%
    {H_0,\allocstate_0}%
    {\clientupdseq_1,\symbseq}%
    {H_1, \allocstate_1}%
    {\allocsymmap_1}$
  \end{enumerate}
  The proof now follows by proving each well-formedness property
  for $\allocbump$.
  \STEP[case-basic-1]{\hyperref[wf:disjoint:alloc]{
      \emph{allocated regions are disjoint:}}}
  \begin{pfproof}
    By induction in the $\ff$ relation on
    \(%
    \invseqrel{\allocbump, \reservedmem}%
    \ffrel%
    {\emptyset}%
    {H_0,\allocstate_0}%
    {\clientupdseq_1,\symbseq}%
    {H_1, \allocstate_1}%
    {\allocsymmap_1}%
    \).
    \STEP[case-basic-1-base]{\textsc{$\ff$-empty} (\(%
    \invseqrel{\allocbump, \reservedmem}%
    \ffrel%
    {\emptyset}%
    {H_0,\allocstate_0}%
    {\clientupdseq_1,\emptytr}%
    {H_0, \allocstate_0}%
    {\emptyset}%
    \)):}
    \begin{pfproof}
      Trivial since $\allocsymmap_1 = \emptyset$.
    \end{pfproof}
    \STEP[case-basic-1-step]{\textsc{$\ff$-step} (\(%
      \invseqrel{\allocbump, \reservedmem}%
      \ffrel%
      {\emptyset}%
      {H_0,\allocstate_0}%
      {\clientupdseq_1' \cdot \clientupdmeta,\symbseq_{\text{pre}} \cdot \symbevent}%
      {H_1, \allocstate_1}%
      {\allocsymmap_1}%
      \)):}
    \begin{pfproof}
      Let $H_1', H_1'', \allocstate_1', \allocsymmap_1'$
      be given such that \[%
        \invseqrel{\allocbump, \reservedmem}%
        \ffrel%
        {\emptyset}%
        {H_0,\allocstate_0}%
        {\clientupdseq_1',\symbseq_{\text{pre}}}%
        {H_1', \allocstate_1'}%
        {\allocsymmap_1'}%
      \]
      and \[%
        \invseqrel{\allocbump}%
        \playrel%
        {\allocsymmap_1'}%
        {H_1'', \allocstate_1'}%
        {\symbseq_{\text{pre}} \cdot \symbevent}%
        {H_1, \allocstate_1}%
        {\allocsymmap_1}%
      \]
      where \(H_1'' = \clientupd{H_1'}{\addressesof{\allocsymmap_1'} \cup \reservedmem}\).
      The statement follows by case analysis on
      \(\invseqrel{\allocbump}%
      \playrel%
      {\allocsymmap_1'}%
      {H_1'', \allocstate_1'}%
      {\symbseq_{\text{pre}} \cdot \symbevent}%
      {H_1, \allocstate_1}%
      {\allocsymmap_1}%
      \).
      \STEP[case-basic-1-step-malloc-ok]{\textsc{$\play$-malloc-ok}:}
      \begin{pfproof}
        For this case we have $\symbevent = \symbmalloc{s}$ for some $s$.
        In addition, we know that
        $\allocsymmap_1 = \allocsymmap_1' \cup \{(\allocstate_1',s;\lenof{\symbseq_\text{pre}} + 1)\}$.
        By induction hypothesis we know that \[
          \forall (a,k;i),(a',k';i') \in \allocsymmap_1'.~
          i \neq i' \implies [a, a+k) \cap [a', a'+k') = \emptyset
        \]
        meaning we just need to show that \[
          \forall (a,k;i) \in \allocsymmap_1'.~ [a, a+k) \cap [\allocstate_1', \allocstate_1' + s) = \emptyset
        \]
        By \Cref{lem:bump:alloc:invariant} we know $\allocstate_1'$
        is strictly greater than any address
        in $\addressesof{\allocsymmap_1'}$.
        But in that means $\addressesof{\allocsymmap_1'} \cap
        [\allocstate_1', \allocstate_1' + s) = \emptyset$,
        which in turn means $[\allocstate_1', \allocstate_1' + s)$
        is disjoint from any other allocated memory.
      \end{pfproof}
      \STEP[case-basic-1-step-malloc-fail]{\textsc{$\play$-malloc-fail}:}
      \begin{pfproof}
        Follows directly from induction hypothesis since
        $\allocsymmap_1' = \allocsymmap_1$.
      \end{pfproof}
      \STEP[case-basic-1-step-free]{\textsc{$\play$-free}:}
      \begin{pfproof}
        Since $\allocsymmap_1 \subseteq \allocsymmap_1'$,
        the statement follows from induction hypothesis.
      \end{pfproof}
    \end{pfproof}
  \end{pfproof}
  \STEP[case-basic-2]{\hyperref[wf:client:memory:final:heaps]{
      \emph{client-accessible memory is in the final heaps:}}}
  \begin{pfproof}
    By induction in the $\ff$ relation on
    \(%
    \invseqrel{\allocbump, \reservedmem}%
    \ffrel%
    {\emptyset}%
    {H_0,\allocstate_0}%
    {\clientupdseq_1,\symbseq}%
    {H_1, \allocstate_1}%
    {\allocsymmap_1}%
    \).
    \STEP[case-basic-2-base]{\textsc{$\ff$-empty} (\(%
      \invseqrel{\allocbump, \reservedmem}%
      \ffrel%
      {\emptyset}%
      {H_0,\allocstate_0}%
      {\clientupdseq_1,\emptytr}%
      {H_0, \allocstate_0}%
      {\emptyset}%
      \)):}
    \begin{pfproof}
      Since $\allocsymmap_1 = \emptyset$,
      it suffices to show that $\reservedmem \subseteq \dom{H_0}$.
      Per our assumptions we have that $\reservedmem \subseteq \dom{H}$,
      and since $\bumpinit{}$ keeps any memory in
      the heap, except $N_2$ which is not in $\reservedmem$,
      it must hold that $\reservedmem \subseteq \dom{H_0}$.
    \end{pfproof}
    \STEP[case-basic-2-step]{\textsc{$\ff$-step} (\(%
      \invseqrel{\allocbump, \reservedmem}%
      \ffrel%
      {\emptyset}%
      {H_0,\allocstate_0}%
      {\clientupdseq_1' \cdot \clientupdmeta,\symbseq_{\text{pre}} \cdot \symbevent}%
      {H_1, \allocstate_1}%
      {\allocsymmap_1}%
      \)):}
    \begin{pfproof}
      Let $H_1', H_1'', \allocstate_1', \allocsymmap_1'$
      be given such that \[%
        \invseqrel{\allocbump, \reservedmem}%
        \ffrel%
        {\emptyset}%
        {H_0,\allocstate_0}%
        {\clientupdseq_1',\symbseq_{\text{pre}}}%
        {H_1', \allocstate_1'}%
        {\allocsymmap_1'}%
      \]
      and \[%
        \invseqrel{\allocbump}%
        \playrel%
        {\allocsymmap_1'}%
        {H_1'', \allocstate_1'}%
        {\symbseq_{\text{pre}} \cdot \symbevent}%
        {H_1, \allocstate_1}%
        {\allocsymmap_1}%
      \]
      where \(H_1'' = \clientupd{H_1'}{\addressesof{\allocsymmap_1'} \cup \reservedmem}\).
      Per our induction hypothesis, we know that
      $\addressesof{\allocsymmap_1'} \cup \reservedmem \subseteq \dom{H_1'}$.
      Since client updates cannot modify the domain of the heap,
      this must in turn mean that $\addressesof{\allocsymmap_1'} \cup \reservedmem \subseteq \dom{H_1''}$.
      Now consider that neither malloc nor free of $\allocbump$
      changes the domain of the heap.
      The statement then follows by case analysis on the $\play$ relation.
    \end{pfproof}
  \end{pfproof}
  \STEP[case-basic-3]{\hyperref[wf:init:no:modify:client]{
      \emph{allocator initialization does not update reserved memory:}}}
  \begin{pfproof}
    Per our assumptions we have that $\reservedmem \subseteq \dom{H}$,
    and since $\bumpinit{}$ does not modify any memory in $H$
    it must necessarily hold that $H \heapeq{\reservedmem} H_0$.
  \end{pfproof}
  \STEP[case-basic-4]{\hyperref[wf:alloc:no:modify:client]{
      \emph{allocator does not modify client-accessible memory:}}}
  \begin{pfproof}
    Case analysis on $\symbevent$ and realize that neither malloc nor free
    modify the contents of the heap.
  \end{pfproof}
  \STEP[case-basic-5]{\hyperref[wf:alloc:no:overlap:reserved]{
      \emph{allocated addresses do not overlap with reserved memory:}}}
  \begin{pfproof}
    Induction in the $\ff$ relation.
    Base case is trivial since $\allocsymmap_1 = \emptyset$.
    For the inductive step, we have the induction hypothesis:
    $\addressesof{\allocsymmap_1'} \cap \reservedmem$.
    Then by case analysis on the $\play$ relation
    together with \Cref{lem:bump:alloc:invariant}
    we see that any new allocation would also be disjoint
    from $\reservedmem$.
  \end{pfproof}
  \STEP[case-basic-6]{\hyperref[wf:null:not:accessible]{
      \emph{null is not client-accessible:}}}
  \begin{pfproof}
    Induction in the $\ff$ relation.
    \STEP[case-basic-6-base]{\textsc{$\ff$-empty} (\(%
      \invseqrel{\allocbump, \reservedmem}%
      \ffrel%
      {\emptyset}%
      {H_0,\allocstate_0}%
      {\clientupdseq_1,\emptytr}%
      {H_0, \allocstate_0}%
      {\emptyset}%
      \)):}
    \begin{pfproof}
      Since $\allocsymmap_1 = \emptyset$,
      it suffices to show that $\bumpnull \notin \reservedmem$.
      But this is the same as showing $N_2 \notin [N_1,N_2)$
      which is true per definition.
    \end{pfproof}
    \STEP[case-basic-6-step]{\textsc{$\ff$-step} (\(%
      \invseqrel{\allocbump, \reservedmem}%
      \ffrel%
      {\emptyset}%
      {H_0,\allocstate_0}%
      {\clientupdseq_1' \cdot \clientupdmeta,\symbseq_{\text{pre}} \cdot \symbevent}%
      {H_1, \allocstate_1}%
      {\allocsymmap_1}%
      \)):}
    \begin{pfproof}
      Let $H_1', H_1'', \allocstate_1', \allocsymmap_1'$
      be given such that \[%
        \invseqrel{\allocbump, \reservedmem}%
        \ffrel%
        {\emptyset}%
        {H_0,\allocstate_0}%
        {\clientupdseq_1',\symbseq_{\text{pre}}}%
        {H_1', \allocstate_1'}%
        {\allocsymmap_1'}%
      \]
      and \[%
        \invseqrel{\allocbump}%
        \playrel%
        {\allocsymmap_1'}%
        {H_1'', \allocstate_1'}%
        {\symbseq_{\text{pre}} \cdot \symbevent}%
        {H_1, \allocstate_1}%
        {\allocsymmap_1}%
      \]
      where \(H_1'' = \clientupd{H_1'}{\addressesof{\allocsymmap_1'} \cup \reservedmem}\).
      The induction hypothesis states  $\bumpnull \notin \addressesof{\allocsymmap_1} \cup \reservedmem$.
      By case analysis on the $\play$ relation,
      the only non-trivial case is a successful malloc,
      meaning $\symbevent = \symbmalloc{k}$ for some $k$.
      For simplicity assume that $k > 0$ since the case of $k = 0$
      behaves internally like $k = 1$.
      Using the induction hypothesis,
      this reduces to showing
      $\bumpnull \notin [\allocstate_1',\allocstate_1' +k)$.
      But per \Cref{lem:bump:alloc:invariant},
      we know that $\forall a \in [\allocstate_1', \allocstate_1' + k).~ \bumpnull < a$,
      and thus $\bumpnull \notin [\allocstate_1',\allocstate_1' +k)$,
      concluded the proof case.
    \end{pfproof}
  \end{pfproof}
  \STEP[case-zero-1]{\hyperref[wf:zero:no:overlap]{
      \emph{allocated addresses are not reused:}}}
  \begin{pfproof}
    By induction in the $\ff$ relation.
    Base case is trivial since $\allocsymmap_1 = \emptyset$.
    For the inductive step, we do case analysis on the $\play$ relation.
    Here \textsc{$\play$-malloc-fail} and \textsc{$\play$-free}
    follow directly by induction hypothesis.
    For the case of \textsc{$\play$-malloc-ok}
    the statement follows by using the induction hypothesis with
    \Cref{lem:bump:alloc:invariant}.
  \end{pfproof}
  \STEP[case-zero-2]{\hyperref[wf:zero:no:space]{
      \emph{zero-sized allocations are disjoint from the client-updateable memory:}}}
  \begin{pfproof}
    By induction in the $\ff$ relation.
    Base case is trivial since $\allocsymmap_1 = \emptyset$.
    Inductive step by case analysis on the $\play$ relation as above.
    Here \textsc{$\play$-malloc-fail} and \textsc{$\play$-free}
    follow directly by induction hypothesis.
    This leaves the case of a successful malloc.
    If the malloc is not zero-sized, then
    the statement follows by induction hypothesis.
    If the malloc is zero-sized, then
    the statement follows by the fact that
    the allocation will have the address $\allocstate_1'$
    together with \Cref{lem:bump:alloc:invariant}
    and the induction hypothesis.
  \end{pfproof}
  \STEP[case-relational]{Relational properties:}
  \begin{pfproof}
    Let some $\clientupdseq_2$ be given.
    We need to show there exists $H_2,\allocstate_2$, and $\allocsymmap_2$
    such that the relational properties of \Cref{def:allocator:heap:well-formedness} hold.
    Now select $\allocstate_2 = \allocstate_1$ and
    $\allocsymmap_2 = \allocsymmap_1$.
    \STEP[case-rel-1]{\hyperref[wf:update:influence]{
        \emph{client updates to the heap do not influence allocator decision:}}}
    \begin{pfproof}
      The proof follows by induction in the $\ff$ relation on
      \(%
      \invseqrel{\allocbump, \reservedmem}%
      \ffrel%
      {\emptyset}%
      {H_0,\allocstate_0}%
      {\clientupdseq_1,\symbseq}%
      {H_1, \allocstate_1}%
      {\allocsymmap_1}%
      \).
      \STEP[case-rel-1-base]{\textsc{$\ff$-empty} (\(%
        \invseqrel{\allocbump, \reservedmem}%
        \ffrel%
        {\emptyset}%
        {H_0,\allocstate_0}%
        {\clientupdseq_1,\emptytr}%
        {H_0, \allocstate_0}%
        {\emptyset}%
        \)):}
      \begin{pfproof}
        Select $H_2 = H_0$. Statement then follows by applying \textsc{$\ff$-empty}.
      \end{pfproof}
      \STEP[case-rel-1-step]{\textsc{$\ff$-step} (\(%
        \invseqrel{\allocbump, \reservedmem}%
        \ffrel%
        {\emptyset}%
        {H_0,\allocstate_0}%
        {\clientupdseq_1' \cdot \clientupdmeta_1,\symbseq_{\text{pre}} \cdot \symbevent}%
        {H_1, \allocstate_1}%
        {\allocsymmap_1}%
        \)):}
      \begin{pfproof}
        Let $H_1', H_1'', \allocstate_1', \allocsymmap_1'$
        be given such that \[%
          \invseqrel{\allocbump, \reservedmem}%
          \ffrel%
          {\emptyset}%
          {H_0,\allocstate_0}%
          {\clientupdseq_1',\symbseq_{\text{pre}}}%
          {H_1', \allocstate_1'}%
          {\allocsymmap_1'}%
        \]
        and \[%
          \invseqrel{\allocbump}%
          \playrel%
          {\allocsymmap_1'}%
          {H_1'', \allocstate_1'}%
          {\symbseq_{\text{pre}} \cdot \symbevent}%
          {H_1, \allocstate_1}%
          {\allocsymmap_1}%
        \]
        where \(H_1'' = \clientupdmeta_1(H_1',\addressesof{\allocsymmap_1'} \cup \reservedmem)\).
        Per our assumptions we know that $\lenof{\clientupdseq_2} =
        \lenof{\clientupdseq_1' \cdot \clientupdmeta} =
        \lenof{\clientupdseq_1'} + 1$.  Therefore, there must exist some
        $\clientupdseq_2'$ and $\clientupdmeta_2$ such that $\clientupdseq_2 =
        \clientupdseq_2' \cdot \clientupdmeta_2$.
        Per our induction hypothesis, we also know that there exists
        some $H_2'$ such that \[
          \invseqrel{\allocbump, \reservedmem}%
          \ffrel%
          {\emptyset}%
          {H_0,\allocstate_0}%
          {\clientupdseq_2',\symbseq_{\text{pre}}}%
          {H_2', \allocstate_1'}%
          {\allocsymmap_1'}%
        \]
        Select $H_2 = \clientupdmeta_2(H_2',\addressesof{\allocsymmap_1'} \cup \reservedmem)$.
        We now need to show that
        \[
          \invseqrel{\allocbump, \reservedmem}%
          \ffrel%
          {\emptyset}%
          {H_0,\allocstate_0}%
          {\clientupdseq_2,\symbseq_{\text{pre}}\cdot \symbevent}%
          {H_2, \allocstate_1}%
          {\allocsymmap_1}%
        \]
        By applying \textsc{$\ff$-step} together with our induction
        hypothesis (after instantiating $H_2'$),
        it suffices to show \[
          \invseqrel{\allocbump}%
          \playrel%
          {\allocsymmap_1'}%
          {H_2, \allocstate_1'}%
          {\symbseq_{\text{pre}} \cdot \symbevent}%
          {H_2, \allocstate_1}%
          {\allocsymmap_1}%
        \]
        This follows by case analysis on $\symbevent$,
        and noticing that neither malloc nor free of $\allocbump$,
        modified the heap,
        together with the fact that none of the decisions
        depend on the heap input.
      \end{pfproof}
    \end{pfproof}
    \STEP[case-rel-2]{\hyperref[wf:alloc:map:equiv]{
        \emph{final allocation maps match:}}}
    \begin{pfproof}
      True since $\allocsymmap_2$ is selected to be $\allocsymmap_1$
    \end{pfproof}
  \end{pfproof}
  \qed
\end{pfproof}

\newenvironment{myarray}[1][t]{\begin{array}[#1]{rcll@{}}}{\end{array}}
\newenvironment{explbox}{\begin{tabular}[t]{@{}l@{}}}{\end{tabular}}
\newcommand{\expl}[1]{\mbox{\begin{explbox}#1\end{explbox}}}
\newcommand{\redbox}[1]{{\color{red}\framebox{\color{red}#1}}}

\newcommand{\avail}{\ensuremath{\mathit{avail}}}
\newcommand{\availbot}{\ensuremath{\mathit{avail}_{\bot}}}

\subsection{Curious allocator revisited}

The allocator strategy $\alloccurio$ below formalizes the Curious
Allocator from Section~\ref{ex:curious}. We assume a heap of the form
$H = [0,H^{\max} + 1)$, that is split into three non-overlapping parts
for some $m > 0$ thus
$H = \intvL{0,H^{\max} + 1} = \intv{0,2^{m-1}}\cup
\intv{2^{m-1}+1,2^m} \cup \intv{2^m+1,H^{\max}}$. For ease of
reference, we define $\mathtt{UPPER\_MAX} = 2^m$ and
$\mathtt{LOWER\_MAX} = 2^{m-1}$. Furthermore, we set $R = \emptyset$,
let the allocator fail on zero sized allocations, and initialised
newly allocated memory to zero. Furthermore, like the bump allocator,
the curious allocator does not support freeing of memory. The
allocator state is then defined as follows
\begin{displaymath}
  \stratapply{\alloccurio}{\domain{Alloc}} =
  \{\times\}
  \uplus
  (\domain{Addr} \times \domain{Size})
  \uplus
  (\domain{Addr} \times \domain{Addr})
\end{displaymath}
Here `$\times$' indicates that no (successful) allocations have been
made, a pair consisting of an address and an allocation size
$(a,k) \in \domain{Addr}\times\domain{Size}$ indicates that an initial
allocation has been made (of size $k$ at address $a$), and a pair of
addresses $(a_L,a_R) \in \domain{Addr}\times\domain{Addr}$ indicates
the bounds of the chosen memory half.

To simplify exposition, we define the following auxiliary functions.
The first determines the possible starting addresses (if any) for a
$k$-sized interval between the (inclusive) bounds $a_L$ and $a_R$ that
is free:
\begin{displaymath}
  \avail(H,a_L,a_R,k) =
  \{ a \,|\, \intvL{a,a+k} \subseteq \intv{a_L,a_R}, \intv{a_L,a_R}
  \cap\dom{H} = \emptyset \}
\end{displaymath}
and the second returns the minimal such address if it exists and
$\NULL$ otherwise
\begin{displaymath}
  \availbot(H,a_L,a_R,k) =
  \left\lbrace
    \begin{array}{ll}
      \min (\avail(H,a_L,a_R,k))
      & \mbox{if $\avail(H,a_L,a_R,k) \neq \emptyset$} \\
      \NULL & \mbox{otherwise}
    \end{array}
  \right.
\end{displaymath}
Allocator initialisation consists of marking all addresses as
unallocated, denoted by mapping all addresses to $\bot$ (extending the
update notation to intervals/sets of addresses
$\upd{H}{\intv{0,H^{\max}}}{\bot}$), and initializes the allocator
state to $\times$:
\[
  (H_{\bot},\times) =
  \stratapply{\alloccurio}{\INIT{H}}
\]
where $H_{\bot} = \upd{H}{\intv{0,H^{\max}}}{\bot}$. The zero address
(0) is chosen as the error address:
\[
  \stratapply{\alloccurio}{\NULL} = 0
\]
Allocation is defined as follows:
\[
  (H',\allocstate',a) =
  \stratapply{\alloccurio}{\MALLOC{H,\allocstate,k}}
\]
where
\begin{displaymath}
  a = 
  \left\lbrace
    \begin{array}{ll}
      \mathtt{UPPER\_MAX} + 1
      & \mbox{if $k > 0$, $\allocstate = \times$ and
        $\avail(H,\mathrm{UPPER\_MAX} + 1, H^{\max},k) \neq \emptyset$}
      \\
      \avail(H,a_L,a_R,k) & \mbox{if $k > 0$ and $\allocstate = (a_L,a_R)$}
      \\
      \NULL & \mbox{otherwise}
    \end{array}
  \right.
\end{displaymath}
and
\begin{displaymath}
  \allocstate' =
  \left\lbrace
    \begin{array}{ll}
      (a,k) & \mbox{if $k > 0$, $\allocstate = \times$, and $a\neq\NULL$} \\
      (\mathtt{LOWER\_MAX} + 1,\mathtt{UPPER\_})
      & \mbox{if $k > 0$, $\allocstate = (a',k')$ and $H'(a') > 0$} \\
      (1,\mathtt{HALF\_ADDR}_k)
      & \mbox{if $k > 0$, $\allocstate = (a',k')$ and $H'(a') \leq 0$} \\
      \allocstate & \mbox{otherwise}
    \end{array}
  \right.
\end{displaymath}
and
\begin{displaymath}
  H' =
  \left\lbrace
    \begin{array}{ll}
      H & \mbox{if $a = \NULL$} \\
      \upd{H}{\intvL{a,a+k}}{0} & \mbox{otherwise}
    \end{array}
  \right.
\end{displaymath}
Just like the bump allocator, memory can never be freed and thus
$\FREE{}$ is a no-op:
\begin{displaymath}
  (H, \allocstate) =
  \stratapply{\alloccurio}{\FREE{H,\allocstate,a}}
\end{displaymath}

\subsubsection{Well-formedness of the Curious Allocator}

\begin{lemma}
  The curious allocator strategy, $\alloccurio$, is well-formed.
\end{lemma}
\begin{proof}\mbox{}
  \begin{pfproof}
    \assume{
      \begin{enumerate}
      \item $H = \intvL{0,H^{\max}+1}$ (thus
        $\emptyset = R \subseteq \dom{H}$)
      \item $(H_0,\allocstate_0) = \stratapply{\alloccurio}{\INIT{H}}$
      \item $\clientupdseq_1$ is a client update sequence
      \item there exists $H_1, \allocstate_1, \allocsymmap_1$ such that
        \begin{math}
          \invseqrel{\alloccurio, \emptyset} \ffrel%
          {\emptyset}%
          {H_0,\allocstate_0}%
          {\clientupdseq_1,\symbseq}%
          {H_1, \allocstate_1}%
          {\allocsymmap_1}%
        \end{math}
      \end{enumerate}
    }%
    \prove{
      \begin{description}
      \item[Basic-1]
        $\forall (a,k;i), (a',k';i') \in \Phi_1\colon i \neq i'
        \implies \intvL{a,a+k} \cap \intvL{a',a'+k'} = \emptyset$
      \item[Basic-2]
        $\addressesof{\allocsymmap_1} \cup \reservedmem \subseteq
        \dom{H_1}$
      \item[Basic-3] $H \heapeq{\reservedmem} H_0$
      \item[Basic-4] if
        $\symbseq = \symbseq_{\text{pre}} \cdot \symbevent$ then for any
        $H', \allocstate', \allocsymmap'$, such that
        $ \invseqrel{\alloccurio,R}
        \ffrel{\emptyset}{H_0,A_0}{\clientupdseq,
          \symbseq_{\mathit{pre}}}{H',A'}{\allocsymmap'}$ and
        \begin{displaymath}
          \invseqrel{\alloccurio}%
          \playrel%
          {\allocsymmap'}%
          {H', \allocstate'}%
          {\symbseq_{\text{pre}} \cdot \symbevent}%
          {H_1, \allocstate_1}%
          {\allocsymmap_1}%
        \end{displaymath}
        then if $\symbevent = \symbfree{z}$ it holds that
        $H' \heapeq{\addressesof{\allocsymmap_1}\cup \reservedmem} H_1$;
        otherwise
        $H' \heapeq{\addressesof{\allocsymmap'}\cup \reservedmem} H_1$
      \item[Basic-5]
        $\addressesof{\allocsymmap_1} \cap \reservedmem = \emptyset$
      \item[Basic-6]
        $\stratapply{\alloccurio}{\NULL} \not \in
        \addressesof{\allocsymmap_1}\cup \reservedmem$
      \item[Zero-Alloc-1]
        $\forall (a,k;i), (a',k';i') \in \Phi_1\colon i \neq i' \implies a
        \neq a'$
      \item[Zero-Alloc-2]
        $\forall (a,k;i) \in \Phi_1\colon k = 0 \implies a
        \notin\addressesof{\Phi_1} \cup \reservedmem$
      \item[Relational]
        $\forall\clientupdseq_2\colon \lenof{\clientupdseq_1} =
        \lenof{\clientupdseq_2} \implies \exists H_2, \allocstate_2,
        \allocsymmap_2$:
        \begin{description}
        \item[Rel-1]
          \begin{math}
            \invseqrel{\alloccurio, \reservedmem} \ffrel%
            {\emptyset}%
            {H_0, \allocstate_0}%
            {\clientupdseq_2, \symbseq}%
            {H_2, \allocstate_2}%
            {\allocsymmap_2}%
          \end{math}
          
        \item[Rel-2]
          $\forall (a,k;i)\in \allocsymmap_1\colon \exists (a',k';i')\in
          \allocsymmap_2\colon k = k' \land i = i'$
        \end{description}
      \end{description}
    }
    \pf\ By induction on $\lenof{\symbseq}$.
    \STEP[dom-upd-inv]{$\dom{H} = \dom{\clientupd{H}{D}}$ for all
      $\clientupdmeta$ and $D \subseteq \dom{H}$}
    \begin{pfproof}
      \pf\ Follows from definition of client updates: only modifies
      the content of $H$ in the designated addresses $D \subseteq
      \dom{H}$.
    \end{pfproof}
    \STEP[curio-determ]{$\alloccurio$ is deterministic: if
      $(H',\allocstate',a') =
      \stratapply{\alloccurio}{\MALLOC{H,\allocstate,k}}$ and
      $(H'',\allocstate'',a'') =
      \stratapply{\alloccurio}{\MALLOC{H,\allocstate,k}}$ then
      $(H',\allocstate',a') = (H'',\allocstate'',a'')$ }
    \begin{pfproof}
      \pf\ Follows directly from the definition of $\alloccurio$,
      noting that all cases have non-overlapping conditions.
    \end{pfproof}
    \STEP{if
      $(H',\allocstate',a') =
      \stratapply{\alloccurio}{\MALLOC{H,\allocstate,k}}$ then
      $H(\intv{a',a'+k}) = \bot$ for $k > 0$}
    \begin{pfproof}
      \pflet{$k > 0$}%
      \pf\ It follows directly from expanding the definition of
      $\avail()$ that
      $\forall a\in \avail(H,a_L,a_R,k)\colon \intvL{a,a+k} \cap
      \dom{H} = \emptyset$.
    \end{pfproof}

    \STEP[basic-3]{\textbf{{Basic-3}}: $H \heapeq{\reservedmem} H_0$}
    \begin{pfproof}
      \pf\ Holds vacuously since $R = \emptyset$.
    \end{pfproof}

    \STEP[basic-5]{\textbf{{Basic-5}}:
      $\addressesof{\allocsymmap_1} \cap \reservedmem = \emptyset$}
    \begin{pfproof}
      \pf\ Trivial:
      $\addressesof{\allocsymmap_1} \cap R =
      \addressesof{\allocsymmap_1} \cap \emptyset = \emptyset$
    \end{pfproof}

    \STEP[h-init]{$(H_0,\allocstate_0) =
      (\upd{H}{\intv{0,H^{\max}}}{\bot},\times)$}
    \begin{pfproof}
      \pf\ Follows from definition of
      $\stratapply{\alloccurio}{\INIT{H}}$.
    \end{pfproof}

    \STEP[basic-5]{\textbf{Basic-6}:
      $\stratapply{\alloccurio}{\NULL} \not \in
      \addressesof{\allocsymmap_1}$}
    \begin{pfproof}
      \pf\ By definition of `$\play$' if $(a',0,i') \in
      \allocsymmap_1$ then it must come from a symbolic event
      $\symbmalloc{0}$ representing a successful zero sized
      allocation, which is not possible in $\alloccurio$. Furthermore,
      any successful allocation will be of the form $\intvL{a',a'+k}$
      where $a' > 0$ and thus $\forall a''\in
      \addressesof{(a',k',i')}\colon a'' > 0$.
    \end{pfproof}
    
    \pflet{
      \begin{enumerate}
      \item $\symbseq = \symbseq' \cdot \symbevent$
      \item $\clientupdseq_1 = \clientupdseq_1' \cdot \clientupdmeta_1$
      \end{enumerate}
    }
    \STEP[ff-dcons]{
      \begin{enumerate}
      \item
        \begin{math}
          \invseqrel{\alloccurio, \emptyset} \ffrel%
          {\emptyset}%
          {H_0,\allocstate_0}%
          {\clientupdseq_1',\symbseq'}%
          {H_0', \allocstate_0'}%
          {\allocsymmap_0'}%
        \end{math}
      \item
        $H_0'' =
        \clientupd[\clientupdmeta_1]{H_0'}{\addressesof{\allocsymmap_0'}}$
      \item
        $\invseqrel{\alloccurio}
        \playrel{\allocsymmap_0'}{H_0'',\allocstate_0'}{\symbseq\cdot
          \symbevent}{H_1,\allocstate_1}{\allocsymmap_1}$
      \end{enumerate}
    }
    \begin{pfproof}
      \pf\ Follows from assumptions and $\rulename{\ff-step}$.
    \end{pfproof}

    \STEP{\pfcase{$\lenof{\symbseq} = 1$ \textbf{(base case)}}}
    \begin{pfproof}
      \STEP{$\symbseq' = \emptytr$}
      \begin{pfproof}
        \pf\ Trivial.
      \end{pfproof}

      \STEP[step-simplify]{$H_0' = H_0$,
        $\allocstate_0' = \allocstate_0$, and
        $\allocsymmap_0' = \emptyset$}
      \begin{pfproof}
        \pf\ Follows from \stepref{ff-dcons} and $\rulename{\ff-empty}$
        since $\symbseq' = \emptytr$.
      \end{pfproof}

      \STEP[step-simp-h0]{$H_0'' = H_0$}
      \begin{pfproof}
        \pf\
        \begin{math}
          \begin{myarray}
            H_0''
            & = & \clientupd[\clientupdmeta_1]{H_0'}{\addressesof{\allocsymmap_0'}}
            \\
            & = & \clientupd[\clientupdmeta_1]{H_0}{\addressesof{\emptyset}}
            & \expl{(from \stepref{step-simplify})}
            \\
            & = & \clientupd[\clientupdmeta_1]{H_0}{\emptyset}
            & \expl{(def. $\addressesof{\cdot}$)}
            \\
            & = & H_0
            & \expl{(def. $\clientupd[\clientupdmeta_1]{\cdot}{\cdot}$)}
          \end{myarray}
        \end{math}
      \end{pfproof}

      \STEP[step-play-simp]{$\invseqrel{\alloccurio}
        \playrel{\emptyset}{H_0,\allocstate_0}{\emptytr\cdot
          \symbevent}{H_1,\allocstate_1}{\allocsymmap_1}$}
      \begin{pfproof}
        \pf\ From \stepref{ff-dcons}, \stepref{step-simplify} and
        \stepref{step-simp-h0}.
      \end{pfproof}

      \STEP[base-mfail]{\pfcase{$\symbevent = \symbfail{k}$}}
      \begin{pfproof}
        \STEP[mfail-step]{
          \begin{enumerate}
          \item
            $(H_1, \allocstate_1, a) =
            \stratapply{\alloccurio}{\MALLOC{H_0, \allocstate_0, k}}$
          \item $a = \stratapply{\alloccurio}{\NULL}$
          \item $\allocsymmap_1 = \emptyset$
          \end{enumerate}
        }%
        \begin{pfproof}
          \pf\ Follows from $\rulename{\play-malloc-fail}$ and
          \stepref{step-play-simp}.
        \end{pfproof}
        
        \STEP{$H_1 = H_0$ and $\allocstate_1 = \allocstate_0$}
        \begin{pfproof}
          \pf\ Follows from definition of $\alloccurio$, in particular
          the failure case where $\NULL$ is returned.
        \end{pfproof}

        \STEP{\textbf{Basic-1}:
          $\forall (a,k;i), (a',k';i') \in \allocsymmap_1\colon i \neq
          i' \implies \intvL{a,a+k} \cap \intvL{a',a'+k'} = \emptyset$}
        \begin{pfproof}
          \pf\ Holds vacuously, since $\allocsymmap_1 = \emptyset$.
        \end{pfproof}
        
        \STEP{\textbf{Basic-2}:
          $\addressesof{\allocsymmap_1} \subseteq \dom{H_1}$}
        \begin{pfproof}
          \pf\
          $\addressesof{\allocsymmap_1} = \addressesof{\emptyset} =
          \emptyset \subseteq \dom{H_1}$ from \stepref{mfail-step} and
          definition of $\addressesof{\cdot}$.
        \end{pfproof}

        \STEP{\textbf{Basic-4}: $\forall H', A', \allocsymmap'$, such
          that
          \begin{displaymath}
            \invseqrel{\alloccurio,\emptyset}
            \ffrel{\emptyset}{H_0,A_0}{\clientupdseq,
              \emptytr}{H',A'}{\allocsymmap'}
          \end{displaymath}
          and
          \begin{displaymath}
            \invseqrel{\alloccurio}%
            \playrel%
            {\allocsymmap'}%
            {H', \allocstate'}%
            {\emptytr \cdot \symbevent}%
            {H_1, \allocstate_1}%
            {\allocsymmap_1}%
          \end{displaymath}
          then
          $H' \heapeq{\addressesof{\allocsymmap'}} H_1$}
        \begin{pfproof}
          \assume{$\invseqrel{\alloccurio,\emptyset}
            \ffrel{\emptyset}{H_0,A_0}{\clientupdseq,
              \emptytr}{H',A'}{\allocsymmap'}$}
          \STEP{$\allocsymmap' = \emptyset$}
          \begin{pfproof}
            \pf\ Follows from $\rulename{\ff-empty}$.
          \end{pfproof}

          \STEP{$H' \heapeq{\addressesof{\allocsymmap'}} H_1$}
          \begin{pfproof}
            \pf\ Trivial, since $\allocsymmap' = \emptyset$.
          \end{pfproof}
        \end{pfproof}
        
        \STEP{\textbf{Zero-Alloc-1}:
          $\forall (a,k;i), (a',k';i') \in \allocsymmap_1\colon i \neq
          i' \implies a \neq a'$}
        \begin{pfproof}
          \pf\ Holds vacuously, since $\allocsymmap_1 = \emptyset$.
        \end{pfproof}

        \STEP{\textbf{Zero-Alloc-2}:
          $\forall (a,k;i) \in \allocsymmap_1\colon k = 0 \implies a
          \notin\addressesof{\allocsymmap_1}$}
        \begin{pfproof}
          \pf\ Holds vacuously, since $\allocsymmap_1 = \emptyset$.
        \end{pfproof}

        \STEP{For all
          $\clientupdseq_2\colon \lenof{\clientupdseq_1} =
          \lenof{\clientupdseq_2}$ there exists
          $H_2, \allocstate_2, \allocsymmap_2$ such that
          \begin{displaymath}
            \begin{myarray}
              & &
              \invseqrel{\alloccurio, \reservedmem} \ffrel%
              {\emptyset}%
              {H_0, \allocstate_0}%
              {\clientupdseq_2, \symbseq}%
              {H_2, \allocstate_2}%
              {\allocsymmap_2}%
              & \expl{(\textbf{Rel-1})}
              \\
              & \land &
              \forall (a,k;i)\in \allocsymmap_1\colon \exists (a',k';i')\in
              \allocsymmap_2\colon k = k' \land i = i'
              & \expl{(\textbf{Rel-2})}
            \end{myarray}
          \end{displaymath}}
        \begin{pfproof}
          \STEP{\textbf{Rel-1}:
            $\invseqrel{\alloccurio, \reservedmem} \ffrel%
            {\emptyset}%
            {H_0, \allocstate_0}%
            {\clientupdseq_2, \symbseq}%
            {H_2, \allocstate_2}%
            {\allocsymmap_2}$}%
          \begin{pfproof}
            \pf\ It follows from \stepref{step-simplify},
            \stepref{step-simp-h0}, and \stepref{step-play-simp} with
            $H_2 = H_0$, $\allocstate_2 = \allocstate_0$ and
            $\allocsymmap_2 = \emptyset$.
          \end{pfproof}

          \STEP{\textbf{Rel-2}:
            $\forall (a,k;i)\in \allocsymmap_1\colon \exists (a',k';i')\in
            \allocsymmap_2\colon k = k' \land i = i'$}
          \begin{pfproof}
            \pf\ Holds vacuously, since $\allocsymmap_1 = \emptyset$.
          \end{pfproof}
        \end{pfproof}

        \qedstep%
        \begin{pfproof}
          \pf\ The base case for $\symbevent = \symbfail{k}$ follows
          from \stepref{basic-3}, \stepref{basic-5},
          and the above.
        \end{pfproof}
      \end{pfproof}
      
      \STEP[base-malloc]{\pfcase{$\symbevent = \symbmalloc{k}$}}
      \begin{pfproof}
        \STEP[malloc-step]{
          \begin{enumerate}
          \item $k > 0$
          \item 
            $(H_1, \allocstate_1, a) =
            \stratapply{\alloccurio}{\MALLOC{H_0, \allocstate_0, k}}$
          \item $a \neq \stratapply{\alloccurio}{\NULL}$
          \item $i = \lenof{\emptytr} + 1 = 0 + 1 = 1$
          \item $\allocsymmap_1 = \{(a,k;1)\}$
          \end{enumerate}
        }%
        \begin{pfproof}
          \pf\ Follows from definition of $\alloccurio$ (zero sized
          allocations return $\NULL$), $\rulename{\play-malloc-ok}$,
          and \stepref{step-play-simp}.
        \end{pfproof}

        \STEP[tmp1]{$H_1 = \upd{H_0}{\intvL{a,a+k}}{0}$}
        \begin{pfproof}
          \pf\ Follows from definition of $\alloccurio$.
        \end{pfproof}

        \STEP[tmp2]{$\dom{H_1} = \intvL{a,a+k}$}
        \begin{pfproof}
          \pf\
          \begin{math}
            \begin{myarray}
              \dom{H_1}
              & = & \dom{\upd{H_0}{\intvL{a,a+k}}{0}}
              & \expl{(from \stepref{tmp1})}
              \\
              & = & \dom{\upd{H_{\bot}}{\intvL{a,a+k}}{0}}
              & \expl{(from \stepref{h-init})}
              \\
              & = & \intvL{a,a+k}
            \end{myarray}
          \end{math}
        \end{pfproof}

        \STEP[tmp3]{$\addressesof{\allocsymmap_1} = \intvL{a,a+k}$}
        \begin{pfproof}
          \pf\ 
          \begin{math}
            \begin{myarray}
              \addressesof{\allocsymmap_1}
              & = & \bigcup_{ (a',k'; i') \in
                \allocsymmap_1}\intvL{a',a'+k'}
              & \expl{(def.\ $\addressesof{\cdot}$)}
              \\
              & = & \bigcup_{ (a',k'; i') \in
                \{(a,k;1)\}}\intvL{a',a'+k'}
              & \expl{(from \stepref{malloc-step})}
              \\
              & = & \intvL{a,a+k}
            \end{myarray}
          \end{math}
        \end{pfproof}
        
        \STEP{\textbf{Basic-1}:
          $\forall (a,k;i), (a',k';i') \in \allocsymmap_1\colon i \neq
          i' \implies \intvL{a,a+k} \cap \intvL{a',a'+k'} = \emptyset$}
        \begin{pfproof}
          \pf\ Holds vacuously, since $\allocsymmap_1 = \{(a,k;1)\}$
          (from~\stepref{malloc-step}).
        \end{pfproof}
        
        \STEP{\textbf{Basic-2}:
          $\addressesof{\allocsymmap_1} \subseteq \dom{H_1}$}
        \begin{pfproof}
          \pf\ 
          \begin{math}
            \begin{myarray}
              \addressesof{\allocsymmap_1}
              & = & \intvL{a,a+k}
              & \expl{(from \stepref{tmp3})}
              \\
              & = & \dom{H_1}
              & \expl{(from \stepref{tmp2})}
            \end{myarray}
          \end{math}
        \end{pfproof}

        \STEP{\textbf{Basic-4}: $\forall H', A', \allocsymmap'$, such
          that
          \begin{displaymath}
            \invseqrel{\alloccurio,\emptyset}
            \ffrel{\emptyset}{H_0,A_0}{\clientupdseq,
              \emptytr}{H',A'}{\allocsymmap'}
          \end{displaymath}
          and
          \begin{displaymath}
            \invseqrel{\alloccurio}%
            \playrel%
            {\allocsymmap'}%
            {H', \allocstate'}%
            {\emptytr \cdot \symbevent}%
            {H_1, \allocstate_1}%
            {\allocsymmap_1}%
          \end{displaymath}
          then
          $H' \heapeq{\addressesof{\allocsymmap'}} H_1$}
        \begin{pfproof}
          \pf\ From $\rulename{\ff-empty}$ it follows that
          $\allocsymmap' = \emptyset$ and thus
          $H' \heapeq{\addressesof{\emptyset}} H_1$ holds vacuously.
        \end{pfproof}

        \STEP{\textbf{Zero-Alloc-1}:
          $\forall (a,k;i), (a',k';i') \in \allocsymmap_1\colon i \neq
          i' \implies a \neq a'$}
        \begin{pfproof}
          \pf\ Holds vacuously, since $\allocsymmap_1 = \{(a,k;1)\}$
          (from~\stepref{malloc-step}).
        \end{pfproof}

        \STEP{\textbf{Zero-Alloc-2}:
          $\forall (a,k;i) \in \allocsymmap_1\colon k = 0 \implies a
          \notin\addressesof{\allocsymmap_1}$}
        \begin{pfproof}
          \pf\ Since $\allocsymmap_1 = \{(a,k;1)\}$
          (from~\stepref{malloc-step}) and $k > 0$ by
          \stepref{malloc-step}.
        \end{pfproof}

        \STEP{For all $\clientupdseq_2$ such that
          $\lenof{\clientupdseq_1} = \lenof{\clientupdseq_2}$ there
          exists $H_2, \allocstate_2, \allocsymmap_2$ such that
          \begin{displaymath}
            \begin{myarray}
              & &
              \invseqrel{\alloccurio, \reservedmem} \ffrel%
              {\emptyset}%
              {H_0, \allocstate_0}%
              {\clientupdseq_2, \symbseq}%
              {H_2, \allocstate_2}%
              {\allocsymmap_2}%
              & \expl{(\textbf{Rel-1})}
              \\
              & \land &
              \forall (a,k;i)\in \allocsymmap_1\colon \exists (a',k';i')\in
              \allocsymmap_2\colon k = k' \land i = i'
              & \expl{(\textbf{Rel-2})}
            \end{myarray}
          \end{displaymath}}
        \begin{pfproof}
          \pflet{$\clientupdseq_2$ such that
            $\lenof{\clientupdseq_1} = \lenof{\clientupdseq_2}$}
          \STEP[rel1]{\textbf{Rel-1}:
            $\invseqrel{\alloccurio, \reservedmem} \ffrel%
            {\emptyset}%
            {H_0, \allocstate_0}%
            {\clientupdseq_2'\cdot\clientupdmeta_2, \emptytr \cdot
              \symbmalloc{k}}%
            {H_2, \allocstate_2}%
            {\allocsymmap_2}$}%
          \begin{pfproof}
            \STEP{$\invseqrel{\alloccurio, \reservedmem} \ffrel%
              {\emptyset}%
              {H_0, \allocstate_0}%
              {\clientupdseq_2, \emptytr}%
              {H_0, \allocstate_0}%
              {\emptyset}$}
            \begin{pfproof}
              \pf\ Follows from $\rulename{ff-empty}$.
            \end{pfproof}

            \STEP[tmp4]{$\invseqrel{\alloccurio}%
              \playrel%
              {\emptyset}%
              {H_0, \allocstate_0}%
              {\emptytr \cdot \symbmalloc{k}}%
              {H_2, \allocstate_2}%
              {\allocsymmap_2}$}
            \begin{pfproof}
              \pf\ Similar to \stepref{malloc-step}.
            \end{pfproof}
          \end{pfproof}

          \STEP[tmp5]{$H_2 = H_1$, $\allocstate_2 = \allocstate_1$, and
            $\allocsymmap_2 = \allocsymmap_1$}
          \begin{pfproof}
            \pf\ From \stepref{rel1} since $\alloccurio$ is
            deterministic.
          \end{pfproof}

          \STEP[rel2]{\textbf{Rel-2}:
            $\forall (a,k;i)\in \allocsymmap_1\colon \exists (a',k';i')\in
            \allocsymmap_2\colon k = k' \land i = i'$}
          \begin{pfproof}
            \pf\ Trivial, since $\allocsymmap_2 = \allocsymmap_1$ from
            \stepref{tmp5}.
          \end{pfproof}

          \qedstep
          \begin{pfproof}
            \pf\ Follows from \stepref{rel1} and \stepref{rel2}.
          \end{pfproof}
        \end{pfproof}

        \qedstep%
        \begin{pfproof}
          \pf\ The base case for $\symbevent = \symbmalloc{k}$ follows
          from \stepref{basic-3}, \stepref{basic-5},
          and the above.
        \end{pfproof}
      \end{pfproof}
      
      \STEP[base-free]{\pfcase{$\symbevent = \symbfree{z}$}}
      \begin{pfproof}
        \pf\ Holds vacuously, since a well-formed $\symbseq$ cannot
        contain an $\symbfree{z}$ event that has no matching (prior)
        $\symbmalloc{k}$ event (by $\rulename{\play-free}$).
      \end{pfproof}

      \qedstep%
      \begin{pfproof}
        \pf\ The base case $\lenof{\symbseq} = 1$ now follows from
        \stepref{base-mfail}, \stepref{base-malloc}, and
        \stepref{base-free}.
      \end{pfproof}
    \end{pfproof}

    \STEP{\pfcase{$\lenof{\symbseq} = n$}}
    \begin{pfproof}
      \assume{ \textbf{(IH)} $\alloccurio$ is \emph{well-formed} for
        all $\symbseq$ such that $\lenof{\symbseq} < n$.
      }

      \STEP{\pfcase{$\symbevent = \symbfail{k}$}}
      \begin{pfproof}
        \STEP[mfail-step]{
          \begin{enumerate}
          \item
            $(H_1, \allocstate_1, a) =
            \stratapply{\alloccurio}{\MALLOC{H_0'', \allocstate_0', k}}$
          \item $a = \stratapply{\alloccurio}{\NULL}$
          \item $\allocsymmap_1 = \allocsymmap_0'$
          \end{enumerate}
        }
        \begin{pfproof}
          \pf\ From \stepref{ff-dcons}.
        \end{pfproof}

        \STEP[h-simp]{$H_1 = H_0''$ and $\allocstate_1 = \allocstate_0'$}
        \begin{pfproof}
          \pf\ Follows from definition of $\alloccurio$, in particular
          the failure case where $\NULL$ is returned.
        \end{pfproof}
        
        \STEP{\textbf{Basic-1}:
          $\forall (a,k;i), (a',k';i') \in \allocsymmap_1\colon i \neq
          i' \implies \intvL{a,a+k} \cap \intvL{a',a'+k'} = \emptyset$}
        \begin{pfproof}
          \pf\ Follows from (IH) since
          $\allocsymmap_1 = \allocsymmap_0'$ (from
          \stepref{mfail-step}).
        \end{pfproof}

        \STEP{\textbf{Basic-2}:
          $\addressesof{\allocsymmap_1} \subseteq \dom{H_1}$}
        \begin{pfproof}
          \pf\
          \begin{math}
            \begin{myarray}
              \addressesof{\allocsymmap_1}
              & = & \addressesof{\allocsymmap_0'}
              & \expl{(by \stepref{mfail-step})}
              \\
              & \subseteq & \dom{H_0'}
              & \expl{(by (IH))}
              \\
              & = & \dom{H_0''}
              & \expl{(by \stepref{dom-upd-inv})}
              \\
              & = & \dom{H_1}
              & \expl{(by \stepref{h-simp})}
            \end{myarray}
          \end{math}
        \end{pfproof}

        \STEP{\textbf{Basic-4}: $\forall H', A', \allocsymmap'$, such
          that
          \begin{displaymath}
            \invseqrel{\alloccurio,\emptyset}
            \ffrel{\emptyset}{H_0,A_0}{\clientupdseq,
              \symbseq'}{H',A'}{\allocsymmap'}
          \end{displaymath}
          and
          \begin{displaymath}
            \invseqrel{\alloccurio}%
            \playrel%
            {\allocsymmap'}%
            {H', \allocstate'}%
            {\symbseq' \cdot \symbevent}%
            {H_1, \allocstate_1}%
            {\allocsymmap_1}%
          \end{displaymath}
          then
          $H' \heapeq{\addressesof{\allocsymmap'}} H_1$}
        \begin{pfproof}
          \pf\ Trivial, since $H_1 = H'$ (from the definition of
          $\alloccurio$ when failing to allocate).
        \end{pfproof}

      \STEP{\textbf{Zero-Alloc-1}:
        $\forall (a,k;i), (a',k';i') \in \allocsymmap_1\colon i \neq
        i' \implies a \neq a'$}
        \begin{pfproof}
          \pf\ Follows from (IH) and \stepref{mfail-step}.
        \end{pfproof}

        \STEP{\textbf{Zero-Alloc-2}:
          $\forall (a,k;i) \in \allocsymmap_1\colon k = 0 \implies a
          \notin\addressesof{\allocsymmap_1}$}
        \begin{pfproof}
          \pf\ Follows from (IH) and \stepref{mfail-step}.
        \end{pfproof}

        \STEP{For all $\clientupdseq_2$ such that
          $\lenof{\clientupdseq_1} = \lenof{\clientupdseq_2}$ there
          exists $H_2, \allocstate_2, \allocsymmap_2$ such that
          \begin{displaymath}
            \begin{myarray}
              & &
              \invseqrel{\alloccurio, \emptyset} \ffrel%
              {\emptyset}%
              {H_0, \allocstate_0}%
              {\clientupdseq_2, \symbseq}%
              {H_2, \allocstate_2}%
              {\allocsymmap_2}%
              & \expl{(\textbf{Rel-1})}
              \\
              & \land &
              \forall (a,k;i)\in \allocsymmap_1\colon \exists (a',k';i')\in
              \allocsymmap_2\colon k = k' \land i = i'
              & \expl{(\textbf{Rel-2})}
            \end{myarray}
          \end{displaymath}}
        \begin{pfproof}
          \pflet{$\clientupdseq_2 = \clientupdseq_2' \cdot
            \clientupdmeta_2$}%
          \STEP[tmp1]{$\exists H_0', \allocstate_0', \allocsymmap_0'$
            such that
            \begin{enumerate}
            \item $\invseqrel{\alloccurio, \emptyset} \ffrel%
              {\emptyset}%
              {H_0, \allocstate_0}%
              {\clientupdseq_2', \symbseq'}%
              {H_0', \allocstate_0'}%
              {\allocsymmap_0'}$
            \item
              $\forall (a,k;i)\in \allocsymmap_1\colon \exists
              (a',k';i')\in \allocsymmap_0'\colon k = k' \land i = i'$
            \end{enumerate}
          }%
          \begin{pfproof}
            \pf\ Follows from (IH).
          \end{pfproof}

          \STEP{\textbf{Rel-1}:
            $\invseqrel{\alloccurio, \emptyset} \ffrel%
            {\emptyset}%
            {H_0, \allocstate_0}%
            {\clientupdseq_2, \symbseq}%
            {H_2, \allocstate_2}%
            {\allocsymmap_2}$}%
          \begin{pfproof}
            \STEP{$\invseqrel{\alloccurio, \emptyset} \ffrel%
              {\emptyset}%
              {H_0, \allocstate_0}%
              {\clientupdseq_2', \symbseq'}%
              {H_0', \allocstate_0'}%
              {\allocsymmap_0'}$}
            \begin{pfproof}
              \pf\ Follows from \stepref{tmp1}.
            \end{pfproof}

            \pflet{
              \begin{enumerate}
              \item $\clientupdseq_2' = \clientupdseq_2'' \cdot
                \clientupdmeta_2'$ (since
                $\lenof{\clientupdseq_2} > 1$)
              \item
                $H_0'' =
                \clientupd[\clientupdmeta_2']{H_0'}{\allocsymmap_0'}$
              \end{enumerate}
            }
            \STEP{$ \invseqrel{\alloccurio}%
              \playrel%
              {\allocsymmap_0'}%
              {H_0'', \allocstate_0'}%
              {\symbseq' \cdot \symbfail{k}}%
              {H_2, \allocstate_2}%
              {\allocsymmap_2}$}%
            \begin{pfproof}
              \pf\ Follows from above with $H_2 = H_0''$,
              $\allocstate_2 = \allocstate_0'$, and
              $\allocsymmap_2 = \allocsymmap_0'$.
            \end{pfproof}

          \end{pfproof}

          \STEP{\textbf{Rel-2}:
            $\forall (a,k;i)\in \allocsymmap_1\colon \exists (a',k';i')\in
            \allocsymmap_2\colon k = k' \land i = i'$}
          \begin{pfproof}
            \pf\ Follows from (IH), \stepref{tmp1}, and above
            ($\allocsymmap_2 = \allocsymmap_0'$).
          \end{pfproof}
        \end{pfproof}
        \qedstep%
        \begin{pfproof}
          \pf\ The induction step for case $\symbevent = \symbfail{k}$
          follows from \stepref{basic-3}, \stepref{basic-5},
          and the above.
        \end{pfproof}
      \end{pfproof} 

      \STEP{\pfcase{$\symbevent = \symbmalloc{k}$}}
      \begin{pfproof}
        \STEP[malloc-step]{
          \begin{enumerate}
          \item
            $(H_1, \allocstate_1, a) =
            \stratapply{\alloccurio}{\MALLOC{H_0'', \allocstate_0', k}}$
          \item $a \neq \stratapply{\alloccurio}{\NULL}$
          \item $i = \lenof{\symbseq'} + 1 = \lenof{\symbseq} = n > 1$
          \item $\allocsymmap_1 = \allocsymmap_0' \cup \{ (a,k;i)\}$
          \end{enumerate}
        }
        \begin{pfproof}
          \pf\ From \stepref{ff-dcons} and
          $\rulename{\play-malloc-ok}$.
        \end{pfproof}

        \STEP[h1-init]{$H_1 = \upd{H_0''}{\intvL{a,a+k}}{0}$}
        \begin{pfproof}
          \pf\ Follows from \stepref{malloc-step} and the definiiotn
          of $\alloccurio$.
        \end{pfproof}

        \STEP[h1-dom]{$\dom{H_1} = \dom{H_0''} \cup \intvL{a,a+k}$}
        \begin{pfproof}
          \pf\ Follows from \stepref{malloc-step} and the definiiotn
          of $\alloccurio$.
        \end{pfproof}

        \STEP[h0-dom]{$\dom{H_0''} = \dom{H_0'}$}
        \begin{pfproof}
          \pf\ Follows from \stepref{dom-upd-inv}.
        \end{pfproof}
        
        \STEP[basic-1]{\textbf{Basic-1}:
          $\forall (a,k;i), (a',k';i') \in \allocsymmap_1\colon i \neq
          i' \implies \intvL{a,a+k} \cap \intvL{a',a'+k'} = \emptyset$}
        \begin{pfproof}
          \STEP{$\forall (a,k;i), (a',k';i') \in \allocsymmap_0'\colon
            i \neq i' \implies \intvL{a,a+k} \cap \intvL{a',a'+k'} =
            \emptyset$}
          \begin{pfproof}
            \pf\ Follows from (IH).
          \end{pfproof}

          \qedstep%
          \begin{pfproof}
            \pf\ By definition of $\alloccurio$: since no
            addresses are ever re-used (free is a no op) and an
            allocation only succeeds if there is no overlap between
            the requested interval and allocated memory.
          \end{pfproof}
        \end{pfproof}

        \STEP{\textbf{Basic-2}:
          $\addressesof{\allocsymmap_1} \subseteq \dom{H_1}$}
        \begin{pfproof}
          \pf\
          \begin{math}
            \begin{myarray}
              \addressesof{\allocsymmap_1}
              & = & \addressesof{\allocsymmap_0' \cup \{ (a,k;i)\}}
              & \expl{(by \stepref{malloc-step})}
              \\
              & = & \addressesof{\allocsymmap_0'} \cup \intvL{a,a+k}
              & \expl{(def.\ of $\addressesof{\cdot}$)}
              \\
              & \subseteq & \dom{H_0'} \cup \intvL{a,a+k}
              & \expl{(by (IH))}
              \\
              & = & \dom{H_0''} \cup \intvL{a,a+k}
              & \expl{(by \stepref{h0-dom})}
              \\
              & = & \dom{H_1}
              & \expl{(by \stepref{h1-dom})}
            \end{myarray}
          \end{math}
        \end{pfproof}

        \STEP{\textbf{Basic-4}: $\forall H', A', \allocsymmap'$, such
          that
          \begin{displaymath}
            \invseqrel{\alloccurio,\emptyset}
            \ffrel{\emptyset}{H_0,A_0}{\clientupdseq_1,
              \symbseq'}{H',A'}{\allocsymmap'}
          \end{displaymath}
          and
          \begin{displaymath}
            \invseqrel{\alloccurio}%
            \playrel%
            {\allocsymmap'}%
            {H', \allocstate'}%
            {\symbseq' \cdot \symbmalloc{k}}%
            {H_1, \allocstate_1}%
            {\allocsymmap_1}%
          \end{displaymath}
          then $H' \heapeq{\addressesof{\allocsymmap'}} H_1$}
        \begin{pfproof}
          \STEP[tmp-1]{$\addressesof{\allocsymmap'} \subseteq \dom{H'}$}
          \begin{pfproof}
            \pf\ Follows from (IH).
          \end{pfproof}

          \STEP[tmp-2]{$H_1 = \upd{H'}{\intvL{a,a+k}}{0}$ and
            $\isbot{H'}{\intvL{a,a+k}}$}
          \begin{pfproof}
            \pf\ Follows from definition of $\alloccurio$.
          \end{pfproof}

          \qedstep%
          \begin{pfproof}
            \pf\ From the definition of $H_1$ (by \stepref{tmp-2}) it
            follows that $a'\in\dom{H'}\colon H'(a') = H_1(a')$ and
            from \stepref{tmp-1} it follows that $H'
            \heapeq{\addressesof{\allocsymmap'}} H_1$.
          \end{pfproof}
        \end{pfproof}

        \STEP{\textbf{Zero-Alloc-1}:
          $\forall (a,k;i), (a',k';i') \in \allocsymmap_1\colon i \neq
          i' \implies a \neq a'$}
        \begin{pfproof}
          \pf\ Since $\alloccurio$ does not allow zero sized
          allocations, this follows from \stepref{basic-1}.
        \end{pfproof}

        \STEP{\textbf{Zero-Alloc-2}:
          $\forall (a,k;i) \in \allocsymmap_1\colon k = 0 \implies a
          \notin\addressesof{\allocsymmap_1}$}
        \begin{pfproof}
          \pf\ Trivial, since zero sized allocations are not allowed
          and results in a failed allocation (returning $\NULL$).
        \end{pfproof}

        \STEP{For all $\clientupdseq_2$ such that
          $\lenof{\clientupdseq_1} = \lenof{\clientupdseq_2}$ there
          exists $H_2, \allocstate_2, \allocsymmap_2$ such that
          \begin{displaymath}
            \begin{myarray}
              & &
              \invseqrel{\alloccurio, \emptyset} \ffrel%
              {\emptyset}%
              {H_0, \allocstate_0}%
              {\clientupdseq_2, \symbseq}%
              {H_2, \allocstate_2}%
              {\allocsymmap_2}%
              & \expl{(\textbf{Rel-1})}
              \\
              & \land &
              \forall (a,k;i)\in \allocsymmap_1\colon \exists (a',k';i')\in
              \allocsymmap_2\colon k = k' \land i = i'
              & \expl{(\textbf{Rel-2})}
            \end{myarray}
          \end{displaymath}}
        \begin{pfproof}
          \pflet{$\clientupdseq_2 = \clientupdseq_2' \cdot
            \clientupdmeta_2$}%

          \STEP[tmp-ih]{$\exists H'', \allocstate'', \allocsymmap''$
            such that
            \begin{enumerate}
            \item $\invseqrel{\alloccurio, \emptyset} \ffrel%
              {\emptyset}%
              {H_0, \allocstate_0}%
              {\clientupdseq_2', \symbseq'}%
              {H'', \allocstate''}%
              {\allocsymmap''}$
            \item
              $\forall (a,k;i)\in \allocsymmap_1\colon \exists
              (a',k';i')\in \allocsymmap''\colon k = k' \land i = i'$
            \end{enumerate}
          }%
          \begin{pfproof}
            \pf\ Follows from (IH).
          \end{pfproof}

          \STEP{\textbf{Rel-1}:
            $\invseqrel{\alloccurio, \emptyset} \ffrel%
            {\emptyset}%
            {H_0, \allocstate_0}%
            {\clientupdseq_2, \symbseq}%
            {H_2, \allocstate_2}%
            {\allocsymmap_2}$}
          \begin{pfproof}
            \STEP{$\invseqrel{\alloccurio, \emptyset} \ffrel%
              {\emptyset}%
              {H_0, \allocstate_0}%
              {\clientupdseq_2', \symbseq'}%
              {H'', \allocstate''}%
              {\allocsymmap''}$}%
            \begin{pfproof}
              \pf\ Follows from \stepref{tmp-ih}.
            \end{pfproof}

            \pflet{
              \begin{enumerate}
              \item
                $\clientupdseq_2' = \clientupdseq_2'' \cdot
                \clientupdmeta_2'$ (since
                $\lenof{\clientupdseq_2} > 1$)
              \item
                $H_{\clientupdmeta_2'}'' =
                \clientupd[\clientupdmeta_2']{H''}{\allocsymmap''}$
              \end{enumerate}
            }

            \STEP{$\invseqrel{\alloccurio}%
              \playrel%
              {\allocsymmap''}%
              {H_{\clientupdmeta_2'}'', \allocstate''}%
              {\symbseq' \cdot \symbmalloc{k}}%
              {H_2, \allocstate_2}%
              {\allocsymmap_2}$}
            \begin{pfproof}
              \pf\ Since $\alloccurio$ is deterministic and does
              not read from the client accessible heap, except for the
              initial allocation.
            \end{pfproof}
          \end{pfproof}

          \STEP{\textbf{Rel-2}:
            $\forall (a,k;i)\in \allocsymmap_1\colon \exists
            (a',k';i')\in \allocsymmap_2\colon k = k' \land i = i' $}
          \begin{pfproof}
            \pf\ Follows from \stepref{tmp-ih} and since
            $\allocsymmap_2 = \allocsymmap'' \cup \{ (a'',k,i)\}$.
          \end{pfproof}
        \end{pfproof}
      \end{pfproof}

      \STEP{\pfcase{$\symbevent = \symbfree{z}$}}
      \begin{pfproof}
        \STEP[free-step]{
          \begin{enumerate}
          \item
            $(H_1, \allocstate_1) =
            \stratapply{\alloccurio}{\FREE{H_0'',A_0',a}}$
          \item $\allocsymmap_1 = \allocsymmap_0' \setminus \{ (a,k;i)
            \}$
          \item $j = \lenof{\symbseq'} + 1 = n > 1$
          \item $H_1 = \upd{H_0''}{\intvL{a,a+k}}{\bot}$
          \end{enumerate}
        }
        \begin{pfproof}
          \pf\ From \stepref{ff-dcons} and
          $\rulename{\play-free}$.
        \end{pfproof}

        \STEP[dom-h0]{$\dom{H_1} = \dom{H_0''} \setminus
          \intvL{a,a+k}$}
        \begin{pfproof}
          \pf\
          $\dom{H_1} = \upd{H_0''}{\intvL{a,a+k}}{\bot} = \dom{H_0''}
          \setminus \intvL{a,a+k}$.
        \end{pfproof}

        \STEP[dom-h0h0]{$\dom{H_0'} = \dom{H_0''}$}
        \begin{pfproof}
          \pf\ Follows from \stepref{dom-upd-inv}.
        \end{pfproof}
        
        \STEP{\textbf{Basic-1}:
          $\forall (a,k;i), (a',k';i') \in \allocsymmap_1\colon i \neq
          i' \implies \intvL{a,a+k} \cap \intvL{a',a'+k'} =
          \emptyset$}
        \begin{pfproof}
          \pf\ Follows from (IH) and \stepref{free-step}.
        \end{pfproof}

        \STEP[basic-2]{\textbf{Basic-2}:
          $\addressesof{\allocsymmap_1} \subseteq \dom{H_1}$}
        \begin{pfproof}
          \pf\
          \begin{math}
            \begin{myarray}
              \addressesof{\allocsymmap_1}
              & = & \addressesof{\allocsymmap_0' \setminus
                \{(a,k;i)\}}
              & \expl{(by \stepref{free-step})}
              \\
              & = & \addressesof{\allocsymmap_0'} \setminus
              \intvL{a,a+k}
              & \expl{(by def.\ $\addressesof{\cdot}$)}
              \\
              & \subseteq & \dom{H_0'} \setminus \intvL{a,a+k}
              & \expl{(by (IH))}
              \\
              & = & \dom{H_0''} \setminus \intvL{a,a+k}
              & \expl{(by \stepref{dom-h0h0})}
              \\
              & = & \dom{H_1}
              & \expl{(by \stepref{dom-h0})}
            \end{myarray}
          \end{math}
        \end{pfproof}

        \STEP{\textbf{Basic-4}: For all $H', A', \allocsymmap'$, such
          that
          \begin{displaymath}
            \invseqrel{\alloccurio,\emptyset}
            \ffrel{\emptyset}{H_0,A_0}{\clientupdseq_1',
              \symbseq'}{H',A'}{\allocsymmap'}
          \end{displaymath}
          and
          \begin{displaymath}
            \invseqrel{\alloccurio, \emptyset}%
            \playrel%
            {\allocsymmap'}%
            {H', \allocstate'}%
            {\symbseq' \cdot \symbfree{z}}%
            {H_1, \allocstate_1}%
            {\allocsymmap_1}%
          \end{displaymath}
          then $H' \heapeq{\addressesof{\allocsymmap_1}} H_1$}
        \begin{pfproof}
          \pf\ Follows from $H_1 = \upd{H'}{\intvL{a,a+k}}{\bot}$ (by
          \stepref{free-step}) and $\addressesof{\allocsymmap_1}
          \subseteq \dom{H_1}$ (by \stepref{basic-2}).
        \end{pfproof}

        \STEP{\textbf{Zero-Alloc-1}:
          $\forall (a,k;i), (a',k';i') \in \allocsymmap_1\colon i \neq
          i' \implies a \neq a'$}
        \begin{pfproof}
          \pf\ Follows from (IH) and \stepref{free-step}.
        \end{pfproof}

        \STEP{\textbf{Zero-Alloc-2}:
          $\forall (a,k;i) \in \allocsymmap_1\colon k = 0 \implies a
          \notin\addressesof{\allocsymmap_1}$}
        \begin{pfproof}
          \pf\ Follows from (IH) and \stepref{free-step}.
        \end{pfproof}
        
        \STEP{For all $\clientupdseq_2$ such that
          $\lenof{\clientupdseq_1} = \lenof{\clientupdseq_2}$ there
          exists $H_2, \allocstate_2, \allocsymmap_2$ such that
          \begin{displaymath}
            \begin{myarray}
              & &
              \invseqrel{\alloccurio, \emptyset} \ffrel%
              {\emptyset}%
              {H_0, \allocstate_0}%
              {\clientupdseq_2, \symbseq}%
              {H_2, \allocstate_2}%
              {\allocsymmap_2}%
              & \expl{(\textbf{Rel-1})}
              \\
              & \land &
              \forall (a,k;i)\in \allocsymmap_1\colon \exists (a',k';i')\in
              \allocsymmap_2\colon k = k' \land i = i'
              & \expl{(\textbf{Rel-2})}
            \end{myarray}
          \end{displaymath}}
        \begin{pfproof}
          \STEP[tmp-ih]{$\exists H'', \allocstate'', \allocsymmap''$
            such that
            \begin{enumerate}
            \item $\invseqrel{\alloccurio, \emptyset} \ffrel%
              {\emptyset}%
              {H_0, \allocstate_0}%
              {\clientupdseq_2', \symbseq'}%
              {H'', \allocstate''}%
              {\allocsymmap''}$
            \item
              $\forall (a,k;i)\in \allocsymmap_1\colon \exists
              (a',k';i')\in \allocsymmap''\colon k = k' \land i = i'$
            \end{enumerate}
          }%
          \begin{pfproof}
            \pf\ Follows from (IH).
          \end{pfproof}

          \STEP{\textbf{Rel-1}:
            $\invseqrel{\alloccurio, \emptyset} \ffrel%
            {\emptyset}%
            {H_0, \allocstate_0}%
            {\clientupdseq_2, \symbseq}%
            {H_2, \allocstate_2}%
            {\allocsymmap_2}$}%
          \begin{pfproof}
            \STEP{$\invseqrel{\alloccurio, \emptyset} \ffrel%
              {\emptyset}%
              {H_0, \allocstate_0}%
              {\clientupdseq_2', \symbseq'}%
              {H_0', \allocstate_0'}%
              {\allocsymmap_0'}$}
            \begin{pfproof}
              \pf\ Follows from \stepref{tmp-ih}.
            \end{pfproof}

            \pflet{
              \begin{enumerate}
              \item $\clientupdseq_2' = \clientupdseq_2'' \cdot
                \clientupdmeta_2'$ (since
                $\lenof{\clientupdseq_2} > 1$)
              \item
                $H_0'' =
                \clientupd[\clientupdmeta_2']{H_0'}{\allocsymmap_0'}$
              \end{enumerate}
            }

            \STEP{$ \invseqrel{\alloccurio}%
              \playrel%
              {\allocsymmap_0'}%
              {H_0'', \allocstate_0'}%
              {\symbseq' \cdot \symbfree{z}}%
              {H_2, \allocstate_2}%
              {\allocsymmap_2}$}%
            \begin{pfproof}
              \pflet{$j = \lenof{\symbseq} + 1$}
              \STEP{$i \mallocfreerel{\symbseq' \cdot \symbfree{z}} j$
                and $\symbseqlookup{\symbseq'}{i} = \symbmalloc{k}$}
              \begin{pfproof}
                \pf\ Follows from (IH) and \stepref{ff-dcons}.
              \end{pfproof}

              \qedstep%
              \begin{pfproof}
                \pf\ Follows by letting
                $(H_2, A_2) =
                \stratapply{\allocstrategy}{\FREE{H_0'',A_0',a}}$ and
                $\allocsymmap_2 = \allocsymmap_1 \setminus \{ (a,k;i)
                \}$
              \end{pfproof}
            \end{pfproof}

          \end{pfproof}

          \STEP{\textbf{Rel-2}:
            $\forall (a,k;i)\in \allocsymmap_1\colon \exists
            (a',k';i')\in \allocsymmap_2\colon k = k' \land i = i' $}
          \begin{pfproof}
            \pf\ Follows from \stepref{tmp-ih} and
            \stepref{free-step}.
          \end{pfproof}
        \end{pfproof}

      \end{pfproof}

    \end{pfproof}
  \end{pfproof}
\end{proof}

\subsection{Null allocator}%
\label{sec:alloc:null}%
A null allocator always returns null on allocation,
which is to say allocation always fails.
\subsubsection{Null allocator strategy}
The null allocator strategy does need to track state,
so the allocator state will be represented by a singleton set:
\[
  A \in \{ \bot \}
\]
\paragraph{$\stratapply{\allocstrategy}{\INIT{H}}$}
On initialization, the null allocator strategy picks
a single unreserved byte of memory to be used as a null value.
Denote this address $\mathsf{null\_ptr}$.
\[
  \stratapply{\allocstrategy}{\INIT{H}} = (H', \bot)
\]
\paragraph{$\stratapply{\allocstrategy}{\NULL}$}
The $\NULL$ value of the strategy is the unreserved address
determined during initialization.
\[
  \stratapply{\allocstrategy}{\NULL} = \mathsf{null\_ptr}
\]
\paragraph{$\stratapply{\allocstrategy}{\MALLOC{H,A,s}}$}
The null allocator always return null on allocation.
\[
  \stratapply{\allocstrategy}{\MALLOC{H,A,s}} = (H,A,\NULL)
\]
\paragraph{$\stratapply{\allocstrategy}{\FREE{H,A,a}}$}
Freeing for the null allocator strategy is a no-op:
\[
  \stratapply{\allocstrategy}{\FREE{H,A,a}} = (H, A)
\]
\subsubsection{Real world null allocators}
\paragraph{Odin}
The core library of the Odin programming language includes 
a null allocator: \texttt{nil\_allocator}.
The stated reason for the existence of this allocator:
``This type of allocator can be used in scenarios where memory doesn't
need to be allocated, but an attempt to allocate memory is not an
error''.
\paragraph{D}
The D programming language includes a null allocator:
\texttt{NullAllocator}.  This allocator is explained as follows:
``\texttt{NullAllocator} is an emphatically empty implementation of
the allocator interface. Although it has no direct use, it is useful
as a `terminator' in composite allocators''.

\subsection{No zero allocator}
The no zero allocator strategy takes a strategy
that is well-formed for non zero-sized allocations,
and wraps it to handle zero-sized allocations in a
well-formed manner by failing on these.
\\[2mm]
Our definition of a well-formed allocator strategy
includes properties specifically concerned with
handling zero-sized allocations.
We may however still be interested in
allocation strategies for which
zero-sized allocation may be undefined behaviour.
In the Rust programming language,
zero-sized allocations are undefined behaviour
for global allocator implementations~\cite{rustGlobalAlloc}.
\subsubsection{Strategy}
Given an allocation strategy $\allocstrategy$,
we define the no zero allocation strategy of $\allocstrategy$.
\paragraph{$\INIT{H}$}
\[
  \INIT{H} = \stratapply{\allocstrategy}{\INIT{H}}
\]
\paragraph{$\NULL$}
\[
  \NULL = \stratapply{\allocstrategy}{\NULL}
\]
\paragraph{$\MALLOC{H,A,s}$}
\[
  \MALLOC{H,A,s} = \begin{cases}
    (H, A, \NULL), & \text{if }s = 0\\
    \stratapply{\allocstrategy}{\MALLOC{H,A,s}}, & \text{otherwise}
  \end{cases}
\]
\paragraph{$\FREE{H,A,a}$}
\[
  \FREE{H,A,a} = \stratapply{\allocstrategy}{\FREE{H,A,a}}
\]

\section{Examples}%
\label{sec:examples}
The following section contains additional Notac examples.
\subsection{XOR linked list}%
\label{appendix:example:xor:linked:list}
An XOR linked list is a doubly-linked list that instead of storing the
pointer to the previous and next nodes, stores the XOR of the two
pointers.  The benefit of this is that a node only has to store a
single word rather than two.  This comes at the cost of not being able
to delete or insert in the list if only given a single node, since the
previous or next address is needed to ``extract'' the other pointer
from the node.

As an example, consider three nodes $n_1, n_2, n_3$ linked together in
this fashion:
\[
  n_1 \leftrightarrow n_2 \leftrightarrow n_3
\]
In this scenario $n_2$ would store the XOR'ed value of the address of
$n_1$ and $n_3$: $\&n_1 \oplus \&n_3$.  To move back and forth in the
list, the next node address can then be extracted by computing the XOR
of the stored value and the previous node address.  So if coming from
$n_1$, we have that
\[
  \&n_1 \oplus (\&n_1 \oplus \&n_3) = \&n_3
\]
and similarly if coming from $n_3$
\[
  \&n_3 \oplus (\&n_1 \oplus \&n_3) = \&n_1
\]

Assuming the node addresses refer to allocated memory there are no
issues with using this strategy.  But note how this method makes heavy
use of XOR on pointer addresses, which presents a challenge when
arguing for the memory safety of an implementation.
\subsubsection{\thelanguage{} XOR linked list implementation}
The following section presents an implementation of an
XOR linked list in a version of \thelanguage{} extended with functions.

In the implementation, the XOR linked list consists of nodes of
memory size 2, where the first location contains the data and the
second location contains the XOR'ed addresses.  As an example, assume
that \texttt{node\_ptr} is a pointer to a node in the list.  With this
pointer, we can access the data by dereferencing the pointer by
\texttt{*node\_ptr}, and access the XOR'ed addresses by dereferencing
the address following the pointer: \texttt{*(node\_ptr + 1)}.  To
indicate the ends of the list, the implementation uses \texttt{NULL}
as a sentinel value.  The choice of \texttt{NULL} for the sentinel
value, serves two purposes.  Firstly, since the nodes stores the XOR
of its neighbors addresses, the implementation will need to know at
least one of the these to decode the addresses.  Secondly, for a
well-formed allocator, \texttt{NULL} has the particular property that
it will never be a used for allocated memory.  This is why
\texttt{NULL} is used over simply using the address \texttt{0}, since
if the \texttt{0} address is not used as \texttt{NULL}, then it could
potentially be used for a node in the list, meaning the implementation
would not be able to distinguish between the end of the list and a
node allocated at address \texttt{0}.
\paragraph{New}
\thelanguage{} function for creating a new XOR linked list can be seen
in \Cref{fig:xor:new}.  Assuming \texttt{elem} is in scope, create a
new XOR linked list containing the element \texttt{elem}.  Since the
list will only contain a single element, previous and next nodes are
null pointers, indicating the end of the list.  The implementation
stores the result in the \texttt{result} variable.

Concretely, the \texttt{new} function start by allocating space for a
node that stores the data and the XOR'ed neighbor addresses.  Since
the node will not have any neighbors, the XOR'ed addresses is simply
the XOR of \texttt{NULL} with itself, which is zero.
\begin{figure}[H]
\begin{lstlisting}
// Create a new xor-linked-list containing elem.
new(elem) {
  // Allocate space for the elem pointer and the XOR'ed value.
  node = malloc(2);
  if (node == NULL) return NULL; // Return to indicate error.
  *(node) = elem;
  // NULL xor NULL = 0
  *(node + 1) = 0;
  return node;
}
\end{lstlisting}
  \caption{\texttt{new} function of XOR linked list}\label{fig:xor:new}
\end{figure}
\paragraph{Push}
\Cref{fig:xor:push} shows a
\thelanguage{} implementation of a function pushing \texttt{elem} to
the front of the XOR linked list at pointer \texttt{ptr}.

Concretely, the implementation first allocates space for a new node in
the list.  It then stores the pushed data in the new node and sets the
XOR data of the new node to be \verb|NULL ^ ptr|, where we recall that
\texttt{ptr} is the head of the list we are pushing to.  On an
intuitive level, \verb|NULL ^ ptr| means there is no preceding
node (\texttt{NULL}) and the next node is \texttt{ptr}.  After
creating the new head of the list, the XOR data in the old head must
be updated to reflect that is has a new neighbor.  To do this, the
implementation first computes the address of the node following the
old head with \verb|NULL ^ *(ptr + 1)|.  Since the node at
\texttt{ptr} was the head before, we know that it does not have any
preceding nodes, meaning its XOR data (which is at address \texttt{ptr
+ 1}) is of the form \verb|NULL ^ next|.  As such, the implementation
decodes the \texttt{next} value and updates the old heads XOR value to
be \texttt{newNode \^{} next} to reflect that it now has a preceding
node.
\begin{figure}[H]
\begin{lstlisting}
// Push elem to the front of list at ptr. 
push(ptr, elem) {
  list = *ptr;
  // Allocate a new node.
  newNode = malloc(2);
  if (newNode == NULL) return 1; // Error
  // Assign the elem value to the first location.
  *newNode = elem;
  // New node will be the head, so point to last is NULL 
  // and the point to next is the previous head.
  *(newNode + 1) = NULL ^ list;
  // Find the address of the second node.
  next = NULL ^ *(list + 1);
  // Store the xor of the new head and the second node
  // in the old head node.
  *(list + 1) = newNode ^ next;
  // Update the head of the list.
  *ptr = newNode;
  return 0;
}
\end{lstlisting}
  \caption{\texttt{pushFront} function of XOR linked list}\label{fig:xor:push}
\end{figure}

\paragraph{Pop}
\Cref{fig:xor:pop} shows a \thelanguage{} implementation of a function
for popping the first element of an XOR linked list with pointer
\texttt{ptr}, freeing the node afterwards.

The implementation first loads the stored value from the head of the
list.  It then finds the address of the next node in the list by
computing \verb|NULL ^ *(ptr + 1)|.  To avoid assigning to the null
pointer, the implementation first checks if the address of the next
node in the list is \texttt{NULL}, which indicates we have popped the
last element in the list.  If the popped element was not the last in
the list, the next node is updated to reflect that it is now the new
head of the list, which means the XOR data is updated to be
\verb|NULL ^ nextNext| instead of \verb|ptr ^ nextNext|.  Following
this, the memory of the old node is freed and the reference to the
list is updated to point to the new head before returning the popped
value.
\begin{figure}[H]
\begin{lstlisting}
// Pop the first element of the list at ptr.
pop(ptr) {
  list = *ptr;
  //if (list == NULL) return NULL;
  // Get the popped value.
  res = *list;
  // Get the pointer for the next node.
  next = NULL ^ *(list + 1);
  // If the next node is not NULL update the XOR value.
  if (next != NULL) *(next + 1) = NULL ^ list ^ *(next + 1);
  // Free the memory of the popped node.
  free(list);
  // Update the head of the list.
  *ptr = next;
  // Return the popped value.
  return res;
}
\end{lstlisting}
  \caption{\texttt{popFront} function of XOR linked list}\label{fig:xor:pop}
\end{figure}

\paragraph{Get}
\Cref{fig:xor:get} shows a \thelanguage{} implementation of a function
for retrieving the value stored at index \texttt{i} from the XOR
linked list with pointer \texttt{ptr}. 

To retrieve the element at index \verb|i| in the list, the program
uses a loop to iterate through the list.
In the loop, the program tracks the address of the
previous node in \verb|last| to decode the XOR data and
determine the address of a following node.
\begin{figure}[H]
\begin{lstlisting}
// Get element i in list at ptr
get(ptr, i) {
  if (ptr == NULL) return NULL;
  j = 0;
  node = *ptr;
  last = NULL;
  // Iterate through the list.
  while (j < i) {
    next = last ^ *(node + 1);
    // If next is NULL the index is out of bounds.
    if (next == NULL) return NULL;
    last = node;
    node = next;
    j = j + 1;
  }
  return *node;
}
\end{lstlisting}
  \caption{\texttt{get} function of XOR linked list}\label{fig:xor:get}
\end{figure}

\paragraph{Delete}
\Cref{fig:xor:delete} shows a \thelanguage{} function that deletes
the node at index \texttt{i} from XOR linked list at pointer \texttt{ptr}.

Implementation of delete for the XOR linked list, has uses a similar
structure to the code for \verb|get|. The difference is in what
happens when the node to be deleted is found.  At this point, the
implementation updates the XOR data in the neighbors to refer to each
other over the node that will be deleted.  Once the neighbors have
been updated, the removed node is freed.
\begin{figure}[H]
\begin{lstlisting}
// Delete element at index i in list at address ptr.
delete(ptr, i) {
  if (ptr == NULL) return 1;
  j = 0;
  node = *ptr;
  last = NULL;
  // Find the i'th node.
  while (j < i) {
    next = last ^ *(node + 1);
    if (next == NULL) return 1;
    last = node;
    node = next;
    j = j + 1;
  }
  next = last ^ *(node + 1);
  if (next != NULL) *(next + 1) = last ^ node ^ *(next + 1);
  if (last != NULL) *(last + 1) = *(last + 1) ^ node ^ next;
  free(node);
  return 0;
}
\end{lstlisting}
  \caption{\texttt{delete} function of XOR linked list}\label{fig:xor:delete}
\end{figure}

\paragraph{Insert}
\Cref{fig:xor:insert} shows a \thelanguage{} implementation for
inserting an element \texttt{elem} at index \texttt{i} in the XOR
linked list with pointer \texttt{ptr}.

Similar to \verb|get| and \verb|delete|, \verb|insert| starts by
finding the node of the corresponding index.  When the node at index
\verb|i| is found, the implementation proceeds allocates memory for
the node that will be inserted and initializes it with the element
that will be stored, and the XOR of the neighboring nodes.  The
neighboring nodes then have their XOR data updated to reflect that
they neighbor the new node instead of one another.
\begin{figure}[H]
\begin{lstlisting}
// Insert elem at position i in list at point ptr.
insert(ptr, i, elem) {
  j = 0;
  node = *ptr;
  last = NULL;
  // Find the i'th node.
  while (j < i) {
    next = last ^ *(node + 1);
    if (next == NULL) return 1;
    last = node;
    node = next;
    j = j + 1;
  }
  // Allocate a new node.
  newNode = malloc(2);
  // Store the value in the new node.
  *(newNode) = elem;
  // The new node will be between last and node.
  *(newNode + 1) = last ^ node;
  // Update node to refer to newNode instead of last.
  *(node + 1) = last ^ *(node + 1) ^ newNode;
  if (last != NULL) {
    // If last is not NULL, update it to refer to newNode.
    *(last + 1) = node ^ *(last + 1) ^ newNode;
  } else {
    // If last is NULL the insertion will be the new head of the list.
    *ptr = newNode;
  }
  return 0;
}
\end{lstlisting}
  \caption{\texttt{insert} function of XOR linked list}\label{fig:xor:insert}
\end{figure}

\section{Comparison to formalized allocators}
This section compares our definition of a well-formed allocator
to specifications of some formally verify allocators.
\subsection{Verified sequential malloc/free}
Consider the specifications of malloc and free from \citeauthor{appel:naumann2020verified:malloc:free}.\\
\textbf{Malloc spec:}
\[
  \left \{
    S(n) = i \land V(i) > 0 \land \text{rmm}(V)
  \right \}
\]
\[ \text{p} = \text{malloc(n);} \]
\[
  \left \{
    \text{rmm}(V[i \coloneq V(i) - 1]) \ast \text{mtok}(p,n) \ast p \mapsto_{(n)} \_
  \right \}
\]
\textbf{Free spec:}
\[
  \left \{ S(n) = i \land \text{rmm}(V) \ast p \mapsto_{(n)} \_  \right \}
\]
\[ 
  \text{free(p);}
\]
\[
  \left \{ \text{rmm}(V[i \coloneq V(i) + 1]) \right \}
\]
\subsection{Well-formedness of spec}
We consider if the specification given by \citeauthor{appel:naumann2020verified:malloc:free}
aligns with the definition of a well-formed allocator.
\paragraph{Basic-1}
With this spec, are the allocated regions disjoint?
This follows as a property of the separating conjunction $(\ast)$ in the malloc spec,
since this prevents the allocator from using memory that is already owned by the client.
\paragraph{Basic-2}
Is client-accessible memory in the final heaps?
The spec does not directly refer to this, but it arguably follows from
the use of separation logic, since the client has ownership of the client-accessible
memory in the heap.
\paragraph{Basic-3}
Does the spec guarantee the allocator does not update the client memory?
Again this follows from the separation logic, since the allocator does
not require memory access outside calls to free memory.
\paragraph{Basic-4}
Same as for \textbf{Basic-3}.
\paragraph{Basic-5}
The spec's use of separation logic prevents the allocated addresses from
overlapping with client memory.
\paragraph{Basic-6}
The paper presents two specs, one that is resource aware and one that is not.
The resource aware spec never returns NULL. The version that is not resource aware
can return NULL, but will not give the resource to the client.
\paragraph{Zero-Alloc-1}
Separation logic ensures this.
\paragraph{Zero-Alloc-2}
Separation logic ensures this.
\paragraph{Rel-1}
The spec does not depend on client updates. In fact, it cannot as the
allocator may not have ownership of the memory.
\paragraph{Rel-2}
As with above, the allocator cannot read from client memory,
so the property trivially holds.
\subsection{The curious allocator}
To see a well-formed allocator may not necessarily satisfy the specification of
\citeauthor{appel:naumann2020verified:malloc:free},
consider the curious allocator and recall the malloc specification:
\[
  \left \{
    S(n) = i \land V(i) > 0 \land \text{rmm}(V)
  \right \}
\]
\[ \text{p} = \text{malloc(n);} \]
\[
  \left \{
    \text{rmm}(V[i \coloneq V(i) - 1]) \ast \text{mtok}(p,n) \ast p \mapsto_{(n)} \_
  \right \}
\]
The precondition $\left \{ S(n) = i \land V(i) > 0 \land \text{rmm}(V)\right \}$
does not include access to any client-accessible resources, which would
be required of the curious allocator to decide on a semi-space.
The relational properties of a well-formed allocator allows reading
of client memory, but reading the memory is not allowed to influence the 

\section{Trace similarity equivalence relation}
This section shows the formalisation of the symbolic filter in
\Cref{fig:trace:filtering} and shows trace similarity, as defined in
\Cref{def:trace:similarity}, is an equivalence relation.

\begin{figure}
  \framebox{
    \begin{mathpar}
      \inferrule[Filter-Malloc-Pass]{
        i = \lenof{\symbseq_{\mathit{pre}}} + 1 \and
        \stepcorresponds{\allocsymmap[a \mapsto (k,i)]}{\tr}{ \symbseq_{\mathit{pre}} \cdot \symbmalloc{k}}{\symbseq}{\tr_r}
      }{
        \stepcorresponds{\allocsymmap}{\evmalloc{k}{a} \cdot  \tr }{\symbseq_{\mathit{pre}}}{ \symbmalloc{k}  \cdot \symbseq}{\tr_r}}
      \and
      \inferrule[Filter-Mfail-Pass]{
        \stepcorresponds{\allocsymmap}{\tr}{ \symbseq_{\mathit{pre}} \cdot \symbfail{k}}{\symbseq}{\tr_r}
      }{
        \stepcorresponds{\allocsymmap}{\evmallocfail{k}  \cdot  \tr }{\symbseq_{\mathit{pre}}}{ 
          \symbfail{k}  \cdot \symbseq}{\tr_r}}
      \\
      \inferrule[Filter-Free-Pass]{
        \hat\symbseq = \symbseq_{\mathit{pre}} \cdot \symbfree{z} \cdot \symbseq \and
        j = \lenof{\symbseq_{\mathit{pre}}} + 1 \and
        i \mallocfreerel{\hat\symbseq} j \\
        \symbseqlookup{\hat\symbseq}{i} = \symbmalloc{k} \\
        \allocsymmap(a) = (k,i)
        \and
        \stepcorresponds{\allocsymmap[a \mapsto \bot]}{\tr}{ \symbseq_{\mathit{pre}} \cdot \symbfree{z}}{\symbseq}{\tr_r}
      }{
        \stepcorresponds{\allocsymmap}{ \evfree{a} \cdot  \tr }{\symbseq_{\mathit{pre}}}{ \symbfree{z}  \cdot \symbseq}{\tr_r}
      }
      \\
      \inferrule[Filter-Free-Residue]{
        \allocsymmap(a) = \bot
        \and
        \stepcorresponds{\allocsymmap}{\tr }{\symbseq_{\mathit{pre}} }{ \symbseq}{\tr_r}
      }{
        \stepcorresponds{\allocsymmap}{ \evfree{a} \cdot \tr }{\symbseq_{\mathit{pre}} }{ \symbseq}{ \evfree{a} \cdot  \tr_r}
      }
      \\
      \inferrule[Filter-Obs-Cast-Residue]{
        \event \in \cup_{v} \{ \evcast{v}, \evobs{v} \} \and
        \stepcorresponds{\allocsymmap}{\tr }{\symbseq_{\mathit{pre}} }{ \symbseq}{\tr_r}
      }{
        \stepcorresponds{\allocsymmap}{\event \cdot \tr }{\symbseq_{\mathit{pre}} }{ \symbseq}{\event \cdot  \tr_r}
      }
      \and
      \inferrule[Filter-Empty]{
      }{
        \stepcorresponds{\allocsymmap}{\emptrace }{\symbseq_{\mathit{pre}} }{ \emptytr}{\emptytr}
      }
    \end{mathpar}
  }
  \caption{Symbolic filter}%
  \label{fig:trace:correspondence}%
  \label{fig:trace:filtering}
\end{figure}

\subsection{Existence characteristic filter and residue}
We start by showing that for all traces there exists
a characteristic filter and residue.
To show this, we prove a more general statement,
that the filtering symbolic allocation sequence exists
for any symbolic filter relation, as long as the
allocation map and symbolic ``pre-sequence'' are compatible.
\begin{definition}[Map-sequence compatibility]
  An allocation map $\allocsymmap$ and symbolic allocation sequence $\symbseq$
  are compatible if $\allocsymmap(a) = (k,i) \implies \symbseq[i] = \symbmalloc{k}$.
\end{definition}
\begin{lemma}[Symbolic filter exists]%
  \label[lemma]{lemma:symbolic:filter:exists}
  For any allocation map $\allocsymmap$, trace $\tr$,
  and symbolic allocation sequence $\symbseq_{\mathit{pre}}$,
  if $\allocsymmap$ and $\symbseq_{\mathit{pre}}$ are compatible,
  then there exists symbolic allocation sequence $\symbseq$ and
  residue $\tr_r$ such that
  \[
    \stepcorresponds{\allocsymmap}{\tr}{\symbseq_{\mathit{pre}}}{\symbseq}{\tr_r}
  \]
\end{lemma}
Existence of characteristic filter then follows as a corollary of this lemma.
\begin{pfproof}\pf\
  Let some trace $\tr$, allocation map $\allocsymmap$,
  and symbolic allocation sequence $\symbseq_{pre}$ be given,
  such that $\allocsymmap$ and $\symbseq_{\mathit{pre}}$ are compatible.
  The goal is to show that there exists well-formed $\symbseq$
  and $\tr_r$, such that $\stepcorresponds{\allocsymmap}{\tr}{\symbseq_{pre}}{\symbseq}{\tr_r}$
  is well-defined.\\
  The proof proceeds by induction in the trace $\tr$.
  \STEP[base-case]{$\tr = \emptytr$}
  \begin{pfproof}
    Select $\symbseq = \symbseqemp$ and $\tr_r = \emptytr$.
    Clearly $\symbseqemp$ is well-formed.
    The statement $\stepcorresponds{\allocsymmap}{\emptytr}
    {\symbseq_{pre}}{\emptytr}{\emptytr}$
    then holds by (\ruleref{Filter-Empty}).
  \end{pfproof}
  \STEP[step-case]{$\tr = \event \tracecons \tr'$}
  \begin{pfproof}
    Given some compatible $\symbseq_{pre}$ and $\allocsymmap$,
    we need to show there exists well-formed $\symbseq$ and $\tr_r$,
    such that $\stepcorresponds{\allocsymmap}{\event \tracecons \tr'}
    {\symbseq_{pre}}{\symbseq}{\tr_r}$.
    Induction hypothesis states that for all compatible
    $\symbseq_{\mathit{pre}}'$ and $\allocsymmap'$,
    there exists well-formed $\symbseq'$ and $\tr_r'$,
    such that $\stepcorresponds{\allocsymmap}{\tr'}
    {\symbseq_{\mathit{pre}}'}{\symbseq'}{\tr_r'}$.
    The proof the proceeds by case analysis on $\event$.
    \STEP[case-cast]{$\event = \evcast{v}$}
    \begin{pfproof}
      Select $\symbseq = \symbseq'$ and $\tr_r = \evcast{v} \tracecons \tr_r'$.
      The goal is now to show
      $\stepcorresponds{\allocsymmap}{\evcast{v} \tracecons \tr'}%
      {\symbseq_{\mathit{pre}}}{\symbseq'}{\evcast{v} \tracecons \tr_r'}$.
      The statement then follows by (\ruleref{Filter-Obs-Cast-Residue})
      and the induction hypothesis,
      since $\symbseq$ and $\allocsymmap$ are compatible by assumption.
    \end{pfproof}
    \STEP[case-obs]{$\event = \evobs{v}$}
    \begin{pfproof}
      Analogous to proof of $\event = \evcast{v}$ (\stepref{case-cast}).
    \end{pfproof}
    \STEP[case-free]{$\event = \evfree{a}$}
    \begin{pfproof}
      Consider the two cases $\allocsymmap(a) = \bot$ and $\allocsymmap(a) = (k,i)$
      for some $k$ and $i$.
      \STEP[case-free-filter-residue]{$\allocsymmap(a) = \bot$}
      \begin{pfproof}
        Select $\symbseq = \symbseq'$ and $\tr_r = \evfree{a} \tracecons \tr_r'$.
        The goal is to show
        $\stepcorresponds{\allocsymmap}{\evfree{a} \tracecons \tr'}%
        {\symbseq_{\mathit{pre}}}{\symbseq'}{\evfree{a} \tracecons \tr_r'}$.
        The statement then follows by (\ruleref{Filter-Free-Residue})
        and the induction hypothesis.
      \end{pfproof}
      \STEP[case-free-filter-pass]{$\allocsymmap(a) = (k,i)$}
      \begin{pfproof}
        Given $\allocsymmap$ and $\symbseq_{\mathit{pre}}$ are compatible,
        we know that $\symbseq_{\mathit{pre}}[i] = \symbmalloc{k}$.
        Now let $z$ be the number of symbolic allocation events in
        $\symbseq_{\mathit{pre}}$ with index higher than $i$.
        Clearly for this $z$ when
        $j = \lenof{\symbseq_{\mathit{pre}}} + 1$ and 
        $\hat{\symbseq} = \symbseq_{\mathit{pre}} \symbcons \symbfree{z} \symbcons \symbseq'$,
        it holds that $i \mallocfreerel{\hat{\symbseq}} j$.
        Now select $\symbseq = \symbfree{z} \symbcons \symbseq'$ and $\tr_r = \tr_r'$.
        The goal is then to show
        $\stepcorresponds{\allocsymmap}{\evfree{v} \tracecons \tr'}%
        {\symbseq_{\mathit{pre}}}{\symbfree{z} \symbcons \symbseq'}{\tr_r'}$.
        By the (\ruleref{Filter-Free-Pass}) rule,
        the induction hypothesis, and how we have selected $z$,
        it suffices to show that $\allocsymmap[a \mapsto \bot]$ and
        $\symbseq_{\mathit{pre}} \symbcons \symbfree{z}$ are compatible.
        This statement clearly follows from the assumption that
        $\allocsymmap$ and $\symbseq_{\mathit{pre}}$ are compatible.
      \end{pfproof}
    \end{pfproof}
    \STEP[case-mfail]{$\tr = \evmallocfail{k} \tracecons \tr'$}
    \begin{pfproof}
      Select $\symbseq = \symbfail{k} \symbcons \symbseq'$ and $\tr_r = \tr_r'$.
      For the statement
      $\stepcorresponds{\allocsymmap}{\evmallocfail{k} \tracecons \tr'}%
      {\symbseq_{pre}}{\symbfail{k} \symbcons \symbseq'}{\tr_r'}$
      with the (\ruleref{Filter-Mfail-Pass}) rule and
      the induction hypothesis,
      it suffices to show that $\allocsymmap$ and $\symbseq_{\mathit{pre}} \symbcons \symbfail{k}$ are compatible.
      Since adding an element to the end of $\symbseq_{\mathit{pre}}$
      does not change the indices of the other elements,
      $\allocsymmap$ and $\symbseq_{\mathit{pre}} \symbcons \symbfail{k}$ are compatible.
    \end{pfproof}
    \STEP[case-malloc]{$\tr = \evmalloc{k}{a}$}
    \begin{pfproof}
      Select $\symbseq = \symbmalloc{k} \symbcons \symbseq'$ and $\tr_r = \tr_r'$.
      The goal is to show
      $\stepcorresponds{\allocsymmap}{\evmalloc{k}{a} \tracecons \tr'}%
      {\symbseq_{pre}}{\symbmalloc{k} \symbcons \symbseq'}{\tr_r'}$.
      By the (\ruleref{Filter-Malloc-Pass}) rule and the induction
      hypothesis, it suffices to show that
      $\allocsymmap[a \mapsto (k, \lenof{\symbseq_{\mathit{pre}}} + 1)]$
      and $\symbseq_{\mathit{pre}} \symbcons \symbmalloc{k}$ are compatible.
      The compatibility follows from the fact that
      $\allocsymmap[a \mapsto (k, \lenof{\symbseq_{\mathit{pre}}} + 1)](a) = \allocsymmap[a \mapsto (k, \lenof{\symbseq_{\mathit{pre}}} + 1)$
      matches $\symbseq_{\mathit{pre}} \symbcons \symbmalloc{k}[\lenof{\symbseq_{\mathit{pre}}} + 1] = \symbmalloc{k}$
      with $\allocsymmap$ and $\symbseq_{\mathit{pre}}$ being compatible per our assumptions.
    \end{pfproof}
  \end{pfproof}
  \qed
\end{pfproof}

\begin{corollary}[Characteristic filter exists]%
  \label[corollary]{corollary:characteristic:filter:exists}
  For every trace $\tr$ there exists a characteristic filter $\trchar{\tr}$.
\end{corollary}
\begin{pfproof}\pf\
  Let some trace $\tr$ be given. Showing that $\trchar{\tr}$ exists
  means we need to show there exists a symbolic allocation sequence $\symbseq$ and
  residue $\tr_r$ such that
  $\stepcorresponds{\emptyset}{\tr}{\symbseqemp}{\symbseq}{\tr_r}$.
  Since $\emptyset$ and $\symbseqemp$ are compatible, the statement
  follows directly from \Cref{lemma:symbolic:filter:exists}.
  \qed{}
\end{pfproof}
\subsection{Uniqueness of characteristic filter and residue}
We now show that the characteristic filter and residue of
a trace are uniquely determined.
To do this, we show more generally that for a given trace,
allocation map and symbolic pre-sequence, the
filtering symbolic allocation sequence and residue are
unique.
\begin{lemma}[Symbolic filter and residue are unique]%
  \label[lemma]{lemma:symbolic:filter:residue:unique}
  For any trace $\tr$, allocation map $\allocsymmap$,
  and symbolic allocation sequence $\symbseq_{\mathit{pre}}$,
  if $\stepcorresponds{\allocsymmap}{\tr}{\symbseq_{\mathit{pre}}}{\symbseq}{\tr_r}$,
  then $\symbseq$ and $\tr_r$ are unique.
  More precisely the following statement holds:
  \[
    \forall \allocsymmap, \tr, \symbseq_{\mathit{pre}}, \symbseq, \symbseq',
    \tr_r, \tr_r'.~
    \stepcorresponds{\allocsymmap}{\tr}{\symbseq_{\mathit{pre}}}{\symbseq}{\tr_r}
    \land
    \stepcorresponds{\allocsymmap}{\tr}{\symbseq_{\mathit{pre}}}{\symbseq'}{\tr_r'}
    \implies
    \symbseq = \symbseq' \land \tr_r = \tr_r'
  \]
\end{lemma}
\begin{pfproof}\pf\
  Let some trace $\tr$, allocation map $\allocsymmap$,
  symbolic allocation sequences $\symbseq_{pre}, \symbseq, \symbseq'$,
  and residues $\tr_r, \tr_r'$  be given.
  The goal is to show:
  \[
    \stepcorresponds{\allocsymmap}{\tr}{\symbseq_{\mathit{pre}}}{\symbseq}{\tr_r}
    \land
    \stepcorresponds{\allocsymmap}{\tr}{\symbseq_{\mathit{pre}}}{\symbseq'}{\tr_r'}
    \implies
    \symbseq = \symbseq' \land \tr_r = \tr_r'
  \]
  The proof follows by induction in the relation
  $\stepcorresponds{\allocsymmap}%
  {\tr}%
  {\symbseq_{pre}}%
  {\symbseq}%
  {\tr_r}$.
  \STEP[case-filter-empty]{(\ruleref{Filter-Empty})}
  \begin{pfproof}
    Per our assumptions we know   \(
    \stepcorresponds{\allocsymmap}{\emptytr}{\symbseq_{pre}}{\symbseq'}{\tr_r'}
    \).
    The goal is to show $\symbseq' = \symbseqemp$ and $\tr_r' = \emptytr$.
    Now consider the only way the statement 
    \(
    \stepcorresponds{\allocsymmap}{\emptytr}{\symbseq_{pre}}{\symbseq'}{\tr_r'}
    \)
    holds is if $\symbseq' = \symbseqemp$ and $\tr_r' = \emptytr$,
    which concludes the case.
  \end{pfproof}
  \STEP[case-filter-obs-cast-residue]{(\ruleref{Filter-Obs-Cast-Residue})}
  \begin{pfproof}
    Let $\event = \evobs{v} \lor \event = \evcast{v}$.
    Per assumption, it holds that
    \begin{itemize}
    \item\(
      \stepcorresponds{\allocsymmap}{\tr}{\symbseq_{pre}}{\symbseq}{\tr_r}
      \)
    \item\(
      \stepcorresponds{\allocsymmap}{\event \tracecons \tr}{\symbseq_{pre}}{\symbseq}{\event \tracecons \tr_r}
      \)
    \item\(
      \stepcorresponds{\allocsymmap}{\event \tracecons \tr}{\symbseq_{pre}}{\symbseq'}{\tr_r'}
      \)
    \end{itemize}
    Induction hypothesis states:
    \[
      \forall \symbseq'', \tr_r''.~
      \stepcorresponds{\allocsymmap}{\tr}{\symbseq_{pre}}{\symbseq''}{\tr_r''}
      \implies
      \symbseq = \symbseq'' \land \tr_r = \tr_r''
    \]
    The goal is to prove $\symbseq = \symbseq'$ and $\event \tracecons \tr_r = \tr_r'$.

    The only way \(\stepcorresponds{\allocsymmap}{\event \tracecons \tr}
    {\symbseq_{pre}}{\symbseq'}{\tr_r'} \) can hold is
    if it way formed by the (\ruleref{Filter-Obs-Cast-Residue}) rule,
    meaning $\tr_r'$ must be of the form $\event \tracecons \tr_r'''$,
    for some $\tr_r'''$, and that
    \(\stepcorresponds{\allocsymmap}{\tr}{\symbseq_{pre}}{\symbseq'}{\tr_r'''} \).
    This means it suffices to show 
    $\symbseq = \symbseq'$ and $\tr_r = \tr_r'''$,
    since $\tr_r = \tr_r''' \implies \symbseq = \symbseq' \land \event \tracecons \tr_r = \event \tracecons \tr_r'$.
    The case then follows by the induction hypothesis and \(\stepcorresponds{\allocsymmap}{\tr}{\symbseq_{pre}}{\symbseq'}{\tr_r'''} \).
  \end{pfproof}
  \STEP[case-filter-free-residue]{(\ruleref{Filter-Free-Residue})}
  \begin{pfproof}
    Per assumption, it holds that
    \begin{itemize}
    \item\(
      \stepcorresponds{\allocsymmap}{\tr}{\symbseq_{pre}}{\symbseq}{\tr_r}
      \)
    \item\(
      \stepcorresponds{\allocsymmap}{\evfree{a} \tracecons \tr}{\symbseq_{pre}}{\symbseq}{\evfree{a} \tracecons \tr_r}
      \)
    \item\( \allocsymmap(a) = \bot \)
    \item\(
      \stepcorresponds{\allocsymmap}{\evfree{a} \tracecons \tr}{\symbseq_{pre}}{\symbseq'}{\tr_r'}
      \)
    \end{itemize}
    We need to show that $\symbseq = \symbseq'$ and $\evfree{a} \tracecons \tr_r = \tr_r'$.
    First we consider the only way \(\stepcorresponds{\allocsymmap}
    {\evfree{a} \tracecons \tr}{\symbseq_{pre}}{\symbseq'}{\tr_r'}\)
    can be true is if formed by the (\ruleref{Filter-Free-Residue}) rule.
    If it was formed by (\ruleref{Filter-Free-Pass}) it would
    contradict the assumption that $\allocsymmap(a) = \bot$.
    As such, it suffices to show $\symbseq = \symbseq'$ and
    $\evfree{a} \tracecons \tr_r = \evfree{a} \tracecons \tr_r'''$.
    The proof from here is analogous to the proof for (\ruleref{Filter-Obs-Cast-Residue})
    in case \stepref{case-filter-obs-cast-residue}.
  \end{pfproof}
  \STEP[case-filter-free-pass]{(\ruleref{Filter-Free-Pass})}
  \begin{pfproof}
    Per assumption, it holds that
    \begin{itemize}
    \item\(
      \stepcorresponds{\allocsymmap[a \mapsto \bot]}{\tr}{\symbseq_{pre} \symbcons \symbfree{z}}{\symbseq}{\tr_r}
      \)
    \item\(
      \stepcorresponds{\allocsymmap}{\evfree{a} \tracecons \tr}{\symbseq_{pre}}{\symbfree{z} \symbcons \symbseq}{\tr_r}
      \)
    \item\( \allocsymmap(a) = (k,i) \)
    \item\( j = \lenof{\symbseq_{\mathit{pre}}} + 1 \)
    \item\( \hat\symbseq = \symbseq_{\mathit{pre}} \symbcons \symbfree{z} \symbcons \symbseq \)
    \item\( \hat{\symbseq}[i] = \symbmalloc{k} \)
    \item\( i \mallocfreerel{\hat{\symbseq}} j \)
    \item\(
      \stepcorresponds{\allocsymmap}{\evfree{a} \tracecons \tr}{\symbseq_{pre}}{\symbseq'}{\tr_r'}
      \)
    \end{itemize}
    Induction hypothesis states:
    \[
      \forall \symbseq'', \tr_r''.~
      \stepcorresponds{\allocsymmap[a \mapsto \bot]}{\tr}{\symbseq_{pre}\symbcons \symbfree{z}}{\symbseq''}{\tr_r''}
      \implies
      \symbseq = \symbseq'' \land \tr_r = \tr_r''
    \]
    We need to show $\symbfree{z} \symbcons \symbseq = \symbseq'$ and
    $\tr_r = \tr_r'$.
    Start by considering how the statement \(
    \stepcorresponds{\allocsymmap}{\evfree{a} \tracecons \tr}
    {\symbseq_{pre}}{\symbseq'}{\tr_r'}\)
    can be true.
    Clearly, it has to be formed either by (\ruleref{Filter-Free-Residue})
    or (\ruleref{Filter-Free-Pass}), since only these are valid for the trace
    $\evfree{a} \tracecons \tr$.
    We can however also rule out (\ruleref{Filter-Free-Residue}),
    as this would require $\allocsymmap(a) = \bot$,
    which contradicts the assumption that $\allocsymmap(a) = (k,i)$.
    As such, we there must exist some $\symbseq'''$ and $z'$ such that
    \(\stepcorresponds{\allocsymmap}{\evfree{a} \tracecons \tr}
    {\symbseq_{pre}}{\evfree{z'} \symbcons \symbseq'''}{\tr_r'}\),
    where for $\hat{\symbseq}' =
    \symbseq_{\mathit{pre}} \symbcons \evfree{z'} \symbcons \symbseq'''$
    it holds that $\hat{\symbseq}'[i] = \symbmalloc{k}$ and
    $i \mallocfreerel{\hat{\symbseq}'} j$.

    Now consider that $i \mallocfreerel{\hat{\symbseq}} j$
    means $i < j$ by definition,
    and since $j = \lenof{\symbseq_{\mathit{pre}}} + 1$
    it must be the case that the value of $z$ and $z'$ only depend
    on $\symbseq_{\mathit{pre}}$. Therefore, $z'$ must be equal to $z$,
    since both refer to the number of symbolic malloc events
    after index $i$ in $\symbseq_{\mathit{pre}}$.
    Due to this, it suffices to show that
    $\symbfree{z} \symbcons \symbseq = \symbfree{z} \symbcons \symbseq'''$
    and $\tr_r = \tr_r'$,
    which reduces to showing $\symbseq = \symbseq'''$ and $\tr_r = \tr_r'$.
    This statement follow directly from our induction hypothesis and assumptions.
  \end{pfproof}
  \STEP[case-filter-mfail-pass]{(\ruleref{Filter-Mfail-Pass})}
  \begin{pfproof}
    Per assumption, it holds that
    \begin{itemize}
    \item\(
      \stepcorresponds{\allocsymmap}{\tr}{\symbseq_{pre} \symbcons \symbfail{k}}{\symbseq}{\tr_r}
      \)
    \item\(
      \stepcorresponds{\allocsymmap}{\evmallocfail{k} \tracecons \tr}{\symbseq_{pre}}{\symbfail{k} \symbcons \symbseq}{\tr_r}
      \)
    \item\(
      \stepcorresponds{\allocsymmap}{\evmallocfail{k} \tracecons \tr}{\symbseq_{pre}}{\symbseq'}{\tr_r'}
      \)
    \end{itemize}
    Induction hypothesis states:
    \[
      \forall \symbseq'', \tr_r''.~
      \stepcorresponds{\allocsymmap}{\tr}{\symbseq_{pre}\symbcons \symbfail{k}}{\symbseq''}{\tr_r''}
      \implies
      \symbseq = \symbseq'' \land \tr_r = \tr_r''
    \]
    The goal is then to show that $\symbfail{k} \symbcons \symbseq = \symbseq'$
    and $\tr_r = \tr_r'$.

    As in other cases, we consider how the statement \(
    \stepcorresponds{\allocsymmap}
    {\evmallocfail{k} \tracecons \tr}{\symbseq_{pre}}{\symbseq'}{\tr_r'}\)
    can be formed. This can only be true if
    $\symbseq' = \symbfail{k} \symbcons \symbseq'''$ for some $\symbseq'''$.
    We therefore need to show that $\symbfail{k} \symbcons \symbseq =
    \symbfail{k} \tracecons \symbseq'''$
    and $\tr_r = \tr_r'$, which reduces to showing
    $\symbseq = \symbseq'''$ and $\tr_r = \tr_r'$.
    The case then follows directly from the induction hypothesis with the assumption
    \(
      \stepcorresponds{\allocsymmap}{\tr}{\symbseq_{pre} \symbcons \symbfail{k}}{\symbseq}{\tr_r}
      \).
  \end{pfproof}
  \STEP[case-filter-malloc-pass]{(\ruleref{Filter-Malloc-Pass})}
  \begin{pfproof}
    Per assumption, it holds that
    \begin{itemize}
    \item\(
      \stepcorresponds{\allocsymmap[a \mapsto (k, i)]}{\tr}
      {\symbseq_{pre} \symbcons \symbmalloc{k}}{\symbseq}{\tr_r}
      \)
    \item\(
      \stepcorresponds{\allocsymmap}{\evmalloc{k}{a} \tracecons \tr}{\symbseq_{pre}}
      {\symbmalloc{k} \symbcons \symbseq}{\tr_r}
      \)
    \item\( i = \lenof{\symbseq_{\mathit{pre}}} + 1 \)
    \item\(
      \stepcorresponds{\allocsymmap}{\evmalloc{k}{a} \tracecons \tr}{\symbseq_{pre}}{\symbseq'}{\tr_r'}
      \)
    \end{itemize}
    Induction hypothesis states:
    \[
      \forall \symbseq'', \tr_r''.~
      \stepcorresponds{\allocsymmap[a \mapsto (k,i)}{\tr}{\symbseq_{pre}\symbcons \symbmalloc{k}}{\symbseq''}{\tr_r''}
      \implies
      \symbseq = \symbseq'' \land \tr_r = \tr_r''
    \]
    The goal is to show $\symbmalloc{k} \symbcons \symbseq = \symbseq'$ and
    $\tr_r = \tr_r'$.
    Starting by realizing the statement \(
    \stepcorresponds{\allocsymmap}{\evmalloc{k}{a} \tracecons \tr}
    {\symbseq_{pre}}{\symbseq'}{\tr_r'} \)
    can only have been formed by the (\ruleref{Filter-Malloc-Pass}) rule.
    As such, it suffices to show $\symbmalloc{k} \symbcons \symbseq =
    \symbmalloc{k} \symbcons \symbseq'''$ and $\tr_r = \tr_r'$,
    which reduces to showing $\symbseq = \symbseq'''$ and $\tr_r = \tr_r'$.
    That this statement is true follows from the induction
    hypothesis with our assumptions.
  \end{pfproof}
  \qed
\end{pfproof}

\begin{corollary}[Characteristic filter and residue are unique]%
  \label[corollary]{corollary:characteristic:filter:residue:unique}
  For every trace $\tr$, the characteristic filter $\trchar{\tr}$ and
  characteristic residue $\symfilter{\tr}{\trchar{\tr}}$ are unique.
\end{corollary}
\begin{pfproof}\pf\
  Let some trace $\tr$ be given. Per definition
  $\trchar{\tr}$ is a symbolic allocation sequence
  and $\symfilter{\tr}{\trchar{\tr}}$ is a residue,
  such that 
  \[
    \stepcorresponds{\emptyset}{\tr}{\symbseqemp}{\trchar{\tr}}
    {\symfilter{\tr}{\trchar{\tr}}}
  \]
  That $\trchar{\tr}$ and $\symfilter{\tr}{\trchar{\tr}}$ are
  unique is then a direct consequence of \Cref{lemma:symbolic:filter:residue:unique}.
  \qed{}
\end{pfproof}
\subsection{Trace similarity is an equivalence relation}
We now show that trace similarity is an equivalence relation,
meaning it is \emph{reflexive}, \emph{symmetric}, and \emph{transitive}.
\begin{lemma}[Trace similarity is reflexive]%
  \label[lemma]{lemma:trace:similarity:reflexive}
  For any trace $\tr$, it holds that $\tracesim{\tr}{\tr}$.
\end{lemma}
\begin{pfproof}\pf\
  Let some trace $\tr$ be given.
  By \Cref{corollary:characteristic:filter:exists} we know that
  there exists $\trchar{\tr}$ and $\symfilter{\tr}{\trchar{\tr}}$.
  Per \Cref{corollary:characteristic:filter:residue:unique} we also know
  that both of these are uniquely determined.
  It therefore follows that $\trchar{\tr} = \trchar{\tr}$ and
  $\symfilter{\tr}{\trchar{\tr}} = \symfilter{\tr}{\trchar{\tr}}$,
  which means $\tracesim{\tr}{\tr}$ concluding the proof.
  \qed{}
\end{pfproof}
\begin{lemma}[Trace similarity is symmetric]%
  \label[lemma]{lemma:trace:similarity:symmetric}
  For any two trace $\tr_1$ and $\tr_2$,
  if $\tracesim{\tr_1}{\tr_2}$, it also holds that
  $\tracesim{\tr_2}{\tr_1}$.
\end{lemma}
\begin{pfproof}\pf\
  Let traces $\tr_1$ and $\tr_2$ be given and assume $\tracesim{\tr_1}{\tr_2}$.
  This, by definition, means that $\trchar{\tr_1} = \trchar{\tr_2}$
  and $\symfilter{\tr_1}{\trchar{\tr_1}} = \symfilter{\tr_2}{\trchar{\tr_2}}$.
  But then it must also be true that $\trchar{\tr_2} = \trchar{\tr_1}$
  and $\symfilter{\tr_2}{\trchar{\tr_2}} = \symfilter{\tr_1}{\trchar{\tr_1}}$,
  meaning $\tracesim{\tr_2}{\tr_1}$, which concludes the proof.
  \qed{}
\end{pfproof}
\begin{lemma}[Trace similarity is transitive]%
  \label[lemma]{lemma:trace:similarity:transitive}
  For any three traces $\tr_1$, $\tr_2$, and $\tr_3$,
  if $\tracesim{\tr_1}{\tr_2}$ and $\tracesim{\tr_2}{\tr_3}$,
  then $\tracesim{\tr_1}{\tr_3}$.
\end{lemma}
\begin{pfproof}\pf\
  Let traces $\tr_1$, $\tr_2$, and $\tr_3$ be given
  and assume $\tracesim{\tr_1}{\tr_2}$ and $\tracesim{\tr_2}{\tr_3}$,
  meaning $\trchar{\tr_1} = \trchar{\tr_2}$,
  $\symfilter{\tr_1}{\trchar{\tr_1}} = \symfilter{\tr_2}{\trchar{\tr_2}}$,
  $\trchar{\tr_2} = \trchar{\tr_3}$,
  and $\symfilter{\tr_2}{\trchar{\tr_2}} = \symfilter{\tr_3}{\trchar{\tr_3}}$.
  Per \Cref{corollary:characteristic:filter:residue:unique},
  all of these filters and residues are uniquely determined.
  This means $\trchar{\tr_1} = \trchar{\tr_2} = \trchar{\tr_3}$
  and
  $\symfilter{\tr_1}{\trchar{\tr_1}} = \symfilter{\tr_2}{\trchar{\tr_2}} = \symfilter{\tr_3}{\trchar{\tr_3}}$,
  meaning $\tracesim{\tr_1}{\tr_3}$,
  which concludes the proof.
  \qed{}
\end{pfproof}
\begin{theorem}[Trace similarity is an equivalence relation]%
  \label{theorem:trace:similarity:equivalence:relation}
  Trace similarity, as defined in \Cref{def:trace:similarity}, is an
  equivalence relation.
\end{theorem}
\begin{pfproof}\pf\
  Trace similarity is an equivalence relation because it is:
  \emph{reflexive} by \Cref{lemma:trace:similarity:reflexive},
  \emph{symmetric} by \Cref{lemma:trace:similarity:symmetric},
  and \emph{transitive} by \Cref{lemma:trace:similarity:transitive}.
  \qed{}
\end{pfproof}

\fi
\end{document}